\documentclass[longauth]{aa} 

\usepackage{graphicx}
\usepackage{natbib}

\usepackage{booktabs}  
\usepackage[colorlinks=true,linkcolor=blue,citecolor=blue]{hyperref}
\usepackage{txfonts}

\graphicspath{{./Pictures/}} 

\providecommand{\abs}[1]{\lvert#1\rvert} 

\usepackage[dvipsnames]{xcolor} 

\usepackage{multirow}
\usepackage{longtable}
\usepackage{lscape}

\begin{document}

   \title{A multi-planetary system orbiting the early-M dwarf TOI-1238}

   \author{E.~Gonz\'alez-\'Alvarez\inst{\ref{inst:CSIC-INTA1}}
   \and M.\,R.~Zapatero Osorio\inst{\ref{inst:CSIC-INTA1}}
   \and J.~Sanz-Forcada\inst{\ref{inst:CSIC-INTA1}}
   \and J.\,A.~Caballero\inst{\ref{inst:CSIC-INTA1}}
    \and S.~Reffert \inst{\ref{inst:zah_lsw}}
    \and V.\,J.\,S.~B\'ejar\inst{\ref{inst:IAC},\ref{inst:ULL}}
    \and A.\,P.~Hatzes \inst{\ref{inst:TLS}} 
    \and E.~Herrero \inst{\ref{inst:IEEC-CSIC},\ref{inst:IEEC}}
    \and S.\,V.~Jeffers \inst{\ref{inst:MP_inst}}
    \and J.~Kemmer \inst{\ref{inst:zah_lsw}}
   \and M.\,J.~L\'opez-Gonz\'alez \inst{\ref{inst:IAA}}
   \and R.~Luque \inst{\ref{inst:IAA}}
   \and K.~Molaverdikhani \inst{\ref{inst:zah_lsw},\ref{inst:MPI_heil}, \ref{inst:Uni_Munchen}}
   \and G.~Morello \inst{\ref{inst:IAC},\ref{inst:ULL}}
   \and E.~Nagel \inst{\ref{inst:Hamb}, \ref{inst:TLS}}
   \and A.~Quirrenbach \inst{\ref{inst:zah_lsw}}
   \and E.~Rodr\'iguez\inst{\ref{inst:IAA}}
   \and C.~Rodr\'iguez-L\'opez\inst{\ref{inst:IAA}}
    \and M.~Schlecker \inst{\ref{inst:MPI_heil}}
   \and A.~Schweitzer\inst{\ref{inst:Hamb}}
   \and S.~Stock \inst{\ref{inst:zah_lsw}}
   \and V.\,M.~Passegger\inst{\ref{inst:Hamb},\ref{inst:USA}}
   \and T.~Trifonov \inst{\ref{inst:MPI_heil}}
   \and P.\,J.~Amado\inst{\ref{inst:IAA}}
   \and D.~Baker \inst{\ref{inst:TX}}
   \and P.\,T.~Boyd \inst{\ref{inst:NASA_greenbelt}}
   \and C.~Cadieux \inst{\ref{inst:Canada}}
   \and D.~Charbonneau \inst{\ref{inst:Harv&Smith}}
   \and K.\,A.~Collins \inst{\ref{inst:Harv&Smith}}
   \and R.~Doyon \inst{\ref{inst:Canada}}
   \and S.~Dreizler \inst{\ref{inst:IAG_goett}}
   \and N.~Espinoza \inst{\ref{inst:STSI_baltimore}}
   \and G.~F\H{u}r\'esz \inst{\ref{inst:Kavli_inst_MA}}
   \and E.~Furlan \inst{\ref{inst:NASA_inst}}
   \and K.~Hesse \inst{\ref{inst:wesleyan_univ}}
   \and S.\,B.~Howell \inst{\ref{inst:NASA_CA}}
   \and J.\,M.~Jenkins \inst{\ref{inst:NASA_CA}}
   \and R.\,C.~Kidwell \inst{\ref{inst:STSI_baltimore}}
   \and D.\,W.~Latham \inst{\ref{inst:Harv&Smith}}
   \and K.\,K.~McLeod \inst{\ref{inst:Wellesley}}
   \and D.~Montes \inst{\ref{inst:UCM}}
   \and J.\,C.~Morales\inst{\ref{inst:IEEC-CSIC},\ref{inst:IEEC}}
   \and T.~O'Dwyer \inst{\ref{inst:TX}}
   \and E.~Pall\'e \inst{\ref{inst:IAC},\ref{inst:ULL}}
   \and S.~Pedraz \inst{\ref{inst:CAHA}}
   \and A.~Reiners\inst{\ref{inst:IAG_goett}}
   \and I.~Ribas \inst{\ref{inst:IEEC-CSIC},\ref{inst:IEEC}}
   \and S.\,N.~Quinn \inst{\ref{inst:Harv&Smith}}
   \and C.~Schnaible \inst{\ref{inst:TX}}
   \and S.~Seager \inst{ \ref{inst:Kavli_inst_MA},\ref{inst:Massachusetts}, \ref{inst:MIT}}
   \and B.~Skinner \inst{\ref{inst:TX}}
    \and J.\,C.~Smith \inst{ \ref{inst:NASA_CA}}
   \and R.\,P.~Schwarz \inst{\ref{inst:Patashnick}}
   \and A.~Shporer \inst{\ref{inst:Kavli_inst_MA}}
   \and R.~Vanderspek \inst{\ref{inst:Kavli_inst_MA}}
   \and J.\,N.~Winn \inst{\ref{inst:Princeston}}
          }
 
                        \institute{Centro de Astrobiolog\'ia (CSIC-INTA), Carretera de Ajalvir km 4, E-28850 Torrej\'on de Ardoz, Madrid, Spain \label{inst:CSIC-INTA1}   
            \and Landessternwarte, Zentrum f\"ur Astronomie der Universit\"at Heidelberg, K\"onigstuhl 12, D-69117 Heidelberg, Germany\label{inst:zah_lsw}
            \and Instituto de Astrof\'isica de Canarias, Avenida V\'ia L\'actea s/n, E-38205 La Laguna, Tenerife, Spain\label{inst:IAC}
            \and Departamento de Astrof\'isica, Universidad de La Laguna, E-38206 La Laguna, Tenerife, Spain\label{inst:ULL}  
            \and Th\"uringer Landessternwarte Tautenburg, Sternwarte 5, D-07778 Tautenburg, Germany\label{inst:TLS}
            \and Institut de Ci\`encies de l'Espai (IEEC-CSIC), Campus UAB, Carrer de Can Magrans s/n, E-08193, Bellaterra, Spain \label{inst:IEEC-CSIC} 
            \and Institut d'Estudis Espacials de Catalunya (IEEC), E-08034 Barcelona, Spain  \label{inst:IEEC} 
            \and Max-Planck-Institut f\"ur Sonnensystemforschung, D-37077 G{\"o}ttingen, Germany \label{inst:MP_inst}
            \and Instituto de Astrof\'isica de Andaluc\'ia (IAA-CSIC), Glorieta de la Astronom\'ia s/n, E-18008 Granada, Spain \label{inst:IAA}
                        \and Max-Planck-Institut f\"ur Astronomie, K\"onigstuhl 17, D-69117 Heidelberg, Germany \label{inst:MPI_heil}
                    \and Universit\"ats-Sternwarte,Ludwig-Maximilians-Universit\"at M\"unchen, Scheinerstrasse 1, D-81679 M\"unchen, Germany \label{inst:Uni_Munchen}
                        \and Hamburger Sternwarte, Universit\"at Hamburg, Gojenbergsweg 112,
                        D-21029 Hamburg, Germany \label{inst:Hamb}  
                        \and Homer L. Dodge Department of Physics and Astronomy, University
                        of Oklahoma, 440 West Brooks Street, Norman, OK 73019, USA \label{inst:USA}
                        \and Physics Department, Austin College, Sherman, TX 75090, USA \label{inst:TX}
                        \and Astrophysics Science Division, NASA Goddard Space Flight Center, Greenbelt, MD 20771, USA \label{inst:NASA_greenbelt}
                        \and Universit\'e de Montr\'eal, D\'epartement de Physique, C.P.~6128 Succ. Centre-ville, Montr\'eal, QC H3C~3J7, Canada \label{inst:Canada}
                        \and Center for Astrophysics \textbar \ Harvard \& Smithsonian, 60 Garden Street, Cambridge, MA 02138, USA \label{inst:Harv&Smith}
                        \and Institut f\"ur Astrophysik, Georg-August-Universit\"at G\"ottingen, Friedrich-Hund-Platz 1, D-37077 G\"ottingen, Germany\label{inst:IAG_goett}
                        \and Space Telescope Science Institute, Baltimore, MD 21218, USA \label{inst:STSI_baltimore}
                        \and Department of Physics and Kavli Institute for Astrophysics and Space Science, Massachusetts Institute of Technology, Cambridge, MA 02139, USA \label{inst:Kavli_inst_MA}
                        \and NASA Exoplanet Science Institute, Caltech/IPAC, Mail Code 100-22, 1200 E. California Blvd., Pasadena, CA 91125, USA \label{inst:NASA_inst}
                        \and Wesleyan University, Middletown, CT 06459, USA \label{inst:wesleyan_univ}
                        \and Intelligent Systems Division, NASA Ames Research Center, Moffett Field, CA 94035, USA \label{inst:NASA_CA}
                        \and Department of Astronomy, Wellesley College, Wellesley, MA 02481, USA \label{inst:Wellesley}
                        \and Departamento de F\'isica de la Tierra y Astrof\'isica \& IPARCOS-UCM (Instituto de F\'isica de Part\'iculas y del Cosmos de la UCM), Facultad de Ciencias F\'isicas, Universidad Complutense de Madrid, E-28040 Madrid, Spain \label{inst:UCM}
                        \and Centro Astron\'omico Hispano Alem\'an, Observatorio de Calar Alto, Sierra de los Filabres, E-04550, G\'ergal, Spain \label{inst:CAHA}
                        \and Department of Earth, Atmospheric, and Planetary Sciences, Massachusetts Institute of Technology, 54–1718, 77 Massachusetts Avenue Cambridge, MA 02139, USA \label{inst:Massachusetts}
                        \and Department of Aeronautics and Astronautics, Massachusetts Institute of Technology, 77 Massachusetts Avenue, Cambridge, MA 02139, USA \label{inst:MIT}
                        \and Patashnick Voorheesville Observatory, Voorheesville, NY 12186, USA \label{inst:Patashnick}
                        \and Princeton University, Cambridge, MA, USA \label{inst:Princeston}
                                }

   \offprints{Esther~Gonz\'alez-\'Alvarez \\ 
   \href{mailto:egonzalez@cab.inta-csic.es}{egonzalez@cab.inta-csic.es}
   }
   \date{Received 1 September 2021 / Accepted 16 November 2021}

 
  \abstract
   {The number of super-Earth and Earth-mass planet discoveries has increased significantly in the last two decades thanks to the Doppler radial velocity and planetary transit observing techniques. Either technique can detect planet candidates on its own, but the power of a combined photometric and spectroscopic analysis is unique for an insightful characterization of the planets, which in turn has repercussions for our understanding of the architecture of planetary systems and, therefore, their formation and evolution.}
   {Two transiting planet candidates with super-Earth radii around the nearby ($d$ = 70.64 $\pm$ 0.06\,pc) K7--M0 dwarf star TOI-1238 were announced by NASA's Transiting Exoplanet Survey Satellite (\textit{TESS}), which observed the field of TOI-1238 in four different sectors. We aim to validate their planetary nature using precise radial velocities taken with the CARMENES spectrograph.}
   {We obtained 55 CARMENES radial velocity measurements that span the 11 months between 9 May 2020 and 5 April 2021. For a better characterization of the parent star's activity, we also collected contemporaneous optical photometric observations at the Joan Or\'o and Sierra Nevada observatories and retrieved archival photometry from the literature. We performed a combined \textit{TESS}$+$CARMENES photometric and spectroscopic analysis by including Gaussian processes and Keplerian orbits to account for the stellar activity and planetary signals simultaneously. 
   }
   {We estimate that TOI-1238 has a rotation period of {40\,$\pm$\,5\,d} based on photometric and spectroscopic data. The combined analysis confirms the discovery of two transiting planets, TOI-1238 b and c, with orbital periods of $0.764597^{+0.000013}_{-0.000011}$\,d and $3.294736^{+0.000034}_{-0.000036}$\,d, masses of 3.76$^{+1.15}_{-1.07}$\,M$_{\oplus}$ and 8.32$^{+1.90}_{-1.88}$\,M$_{\oplus}$, and radii of $1.21^{+0.11}_{-0.10}$\,R$_{\oplus}$ and $2.11^{+0.14}_{-0.14}$\,R$_{\oplus}$. They orbit their parent star at semimajor axes of 0.0137$\pm$0.0004\,au and 0.036$\pm$0.001\,au, respectively. The two planets are placed on opposite sides of the radius valley for M dwarfs and lie between the star and the inner border of TOI-1238's habitable zone. The inner super-Earth TOI-1238\,b is one of the densest ultra-short-period planets ever discovered ($\rho=11.7^{+4.2}_{-3.4}$\,g\,$\rm cm^{-3}$). The CARMENES data also reveal the presence of an outer, non-transiting, more massive companion with an orbital period and radial velocity amplitude of $\geq$600\,d and $\geq$70\,m\,s$^{-1}$, which implies a likely mass of $M \geq 2 \sqrt{1-e^2}$\,M$_{\rm Jup}$ and a separation $\geq$1.1\,au from its parent star.}
   {}
 

\maketitle
%

\section{Introduction}
\label{Introduction}

The number of discovered exoplanets has grown tremendously over the last two decades thanks to forefront space- and ground-based instruments and the combination of successful techniques such as the transit photometry and the spectroscopic radial velocity (RV) methods. Transiting planetary systems with bright parent stars offer a unique opportunity to determine the bulk density of the planets and to explore the atmospheres of these planets (\citealt{2019ARA&A..57..617M}). Spectrographs delivering high-precision RV measurements combined with the superb quality of photometric light curves (LCs) obtained from space-based missions are now approaching the golden objective of finding and characterizing planets the size of the Earth (e.g., \citealt{2019A&A...628A..39L, 2020A&A...642A.121L}).

In recent years, the search for new exoplanets has been focused on low-mass stars because they are the most numerous stars in the Galaxy and allow lower-mass planets to be identified for a given level of observational precision \citep{2013ApJ...767...95D, 2015ApJ...807...45D,2021A&A...650A.201R}. At present, about 800 M-type stars are known to harbor planetary systems, accounting for more than 1000 exoplanets. Considering planetary systems with transiting planets that have a robust mass and radius determination, this number is reduced to about 40--50 exoplanets orbiting M-type dwarfs \citep{2019ApJ...887..261M, 2021Sci...371.1038T}, 26 of which are in nine different multi-planetary systems. These planets typically have masses in the Earth to super-Earth mass regime in rather circular orbits around their low-mass parent stars with short orbital periods. Very interestingly, a few of these systems ($\sim$1\,\%) also harbor additional, more massive planets in the Neptune- to Jupiter-mass regime that pose a challenge to the core accretion theory of planet formation \citep{2004ApJ...612L..73L, 2010Sci...330..653H, Ida_2013, 2019Sci...365.1441M}.

Here, we present the confirmation of two transiting planets orbiting the early-M dwarf TOI-1238, whose transit signals were first detected by NASA's Transiting Exoplanet Survey Satellite \citep[\textit{TESS;}][]{2015JATIS...1a4003R}. Their masses and radii are derived through a joint modeling of the \textit{TESS} photometry and follow-up RV observations with the Calar Alto high-Resolution search for M dwarfs with Exoearths with Near-infrared and optical \'Echelle Spectrographs \citep[CARMENES;][]{2016SPIE.9908E..12Q, 2018SPIE10702E..0WQ, 2018A&A...609L...5R}, together with extensive ground-based photometric follow-up observations with telescopes at the Joan Or\'o and Sierra Nevada observatories. TOI-1238 thus becomes the tenth known transiting multi-planetary system around an M dwarf. The signal of a third companion, with no transit detected in the \textit{TESS} LC, is evident from the CARMENES RV time series. This candidate is located $>$30 times farther away from its parent star than the outer of the two transiting planets. This work is part of the \textit{TESS} follow-up program within the CARMENES guaranteed time observations (GTO) survey, which aims to validate the planetary nature of transit events detected around M dwarfs \citep{2019A&A...628A..39L, 2020A&A...642A.236K,2020A&A...642A.173N, 2020A&A...644A.127D,2020A&A...639A.132B, 2021A&A...650A..78B,2021A&A...649A.144S}.

In Sect.~\ref{sec:observations} we present all space- and ground-based observations. The properties of the host star, TOI-1238, are introduced in Sect.~\ref{sec:TOI-1238}. With the rotation period of the parent star being constrained by our data, we proceeded to analyze the photometric and spectroscopic observations with the aim of characterizing the transiting planetary candidates and searching for new planet candidates. The properties of the newly confirmed and discovered planets orbiting TOI-1238 are given in Sect.~\ref{section4}. A brief discussion on the implications of these findings is presented in Sect.~\ref{discussion}, and the conclusions appear in Sect. \ref{sec:summary_discussion}.

\section{Observations}
\label{sec:observations}

\subsection{\textit{TESS} photometry}
\label{subsec:TESS photometric time serie}

\begin{small}
\begin{table}[]
\centering
\caption{\textit{TESS} observations of TOI-1238.}
\label{tab:tess_observ_TOI-1238}
\begin{tabular}{c c c c  }
\hline
\hline
\noalign{\smallskip}
Sector  & Camera  & Start date & End date\\
\noalign{\smallskip}    
\hline  
\noalign{\smallskip}
14 & 4 &   18 Jul 2019 & 15 Aug 2019\\
15 & 4 &   15 Aug 2019 & 11 Sep 2019\\
21 & 3 &   21 Jan 2020 & 18 Feb 2020\\
22 & 3 &   18 Feb 2020 & 18 Mar 2020 \\

\noalign{\smallskip}
\hline
\end{tabular}
\end{table}
\end{small}

\textit{TESS} is an all-sky photometric transit survey with the main objective of finding planets smaller than Neptune orbiting bright stars that are amenable to follow-up observations leading to the determination of planetary masses and atmospheric compositions. In its primary mission, \textit{TESS} has conducted high-precision photometry of more than 200,000 stars over two years of observations until 4 July 2020. All targets were made available to the community as target pixel files (TPFs) and calibrated LCs. \textit{TESS} LC files include the timestamps, simple aperture photometry (SAP) fluxes, and pre-search data conditioned simple aperture photometry (PDCSAP) fluxes \citep{2012PASP..124.1000S,Stumpe2012,Stumpe2014}. The SAP flux is the flux after summing the calibrated pixels within the \textit{TESS} optimal photometric aperture, while the PDCSAP flux corresponds to the SAP flux values corrected for instrumental variations and for crowding. The optimal photometric aperture is defined such that the stellar signal has a high signal-to-noise ratio, with minimal contamination from the background. The second extension header data unit (HDU) consists of an image that contains the registered pixels for the target star and marks the pixels used for the optimal aperture photometry. The \textit{TESS} detector bandpass spans from 600 to 1000\,nm and is centered on the traditional Cousins $I$ band (786.5\,nm). This wide red optical bandpass was chosen to reduce photon-counting noise and to increase the sensitivity to small planets transiting cool, red stars.

TOI-1238 (TIC\,153951307) was observed by \textit{TESS} at the 2\,min short-cadence integrations in sectors 14, 15, 21, and 22 during the \textit{TESS} primary mission (see Table~\ref{tab:tess_observ_TOI-1238}). All sectors were processed by the Science Processing Operations Center (SPOC) pipeline \citep{2016SPIE.9913E..3EJ} and searched for transiting planet signatures with an adaptive, wavelet-based transit detection algorithm \citep{2002ApJ...575..493J,2010SPIE.7740E..0DJ}. In a first analysis of sectors 14 and 15, conducted separately {by the SPOC pipeline}, a planet candidate was identified at an orbital period of 3.294\,d, and TOI-1238 was announced as a \textit{TESS} object of interest (TOI) via the dedicated MIT \textit{TESS} data alerts public website\footnote{\url{https://tess.mit.edu/toi-releases/}} on 17 October 2019. In a subsequent joint analysis of sectors 14 and 15, a second transiting planet candidate at a shorter separation from its parent star with orbital period of 0.764\,d was announced on 15 November 2019. Both planet candidates were fitted with limb-darkening transit models \citep{Li:DVmodelFit2019} by SPOC and successfully passed all the diagnostic tests \citep{Twicken:DVdiagnostics2018} performed on them: {comparison of even and odd transits to screen against eclipsing binaries and ghost diagnostic tests to help rule out scattered light or background eclipsing binaries. From the difference image centroiding test, the transit source is coincident with the core of the stellar point spread function, so it is clear that the transit events are associated with the target and not with nearby bright stars}. The SPOC analysis located the source of the transit signatures to within 1.5$\pm$2.3\,arcsec, and 4.0$\pm$4.0\,arcsec, respectively. No further transiting planet signatures were detected in the residual LC by the SPOC. The candidates have estimated radii of 2.29$\pm$0.29\,R$_{\oplus}$ and 1.3$\pm$3.5\,R$_{\oplus}$ for the outer and inner objects, respectively. The later \textit{TESS} observations in sectors 21 and 22 did confirm the presence of shallow {transit-like signatures} at these two different periodicities.

\begin{figure*}[]
\centering
\includegraphics[width=0.24\textwidth]{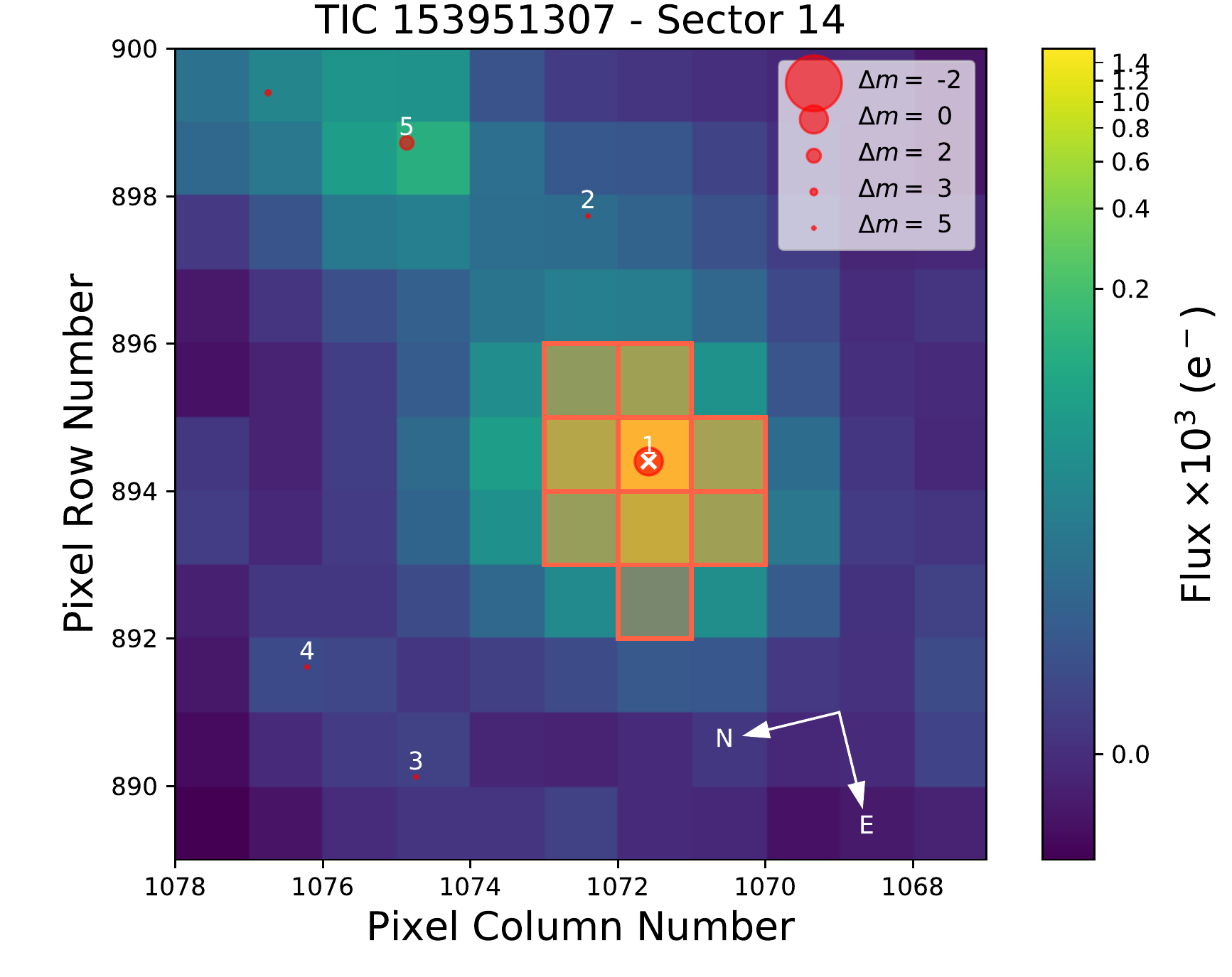}
\includegraphics[width=0.24\textwidth]{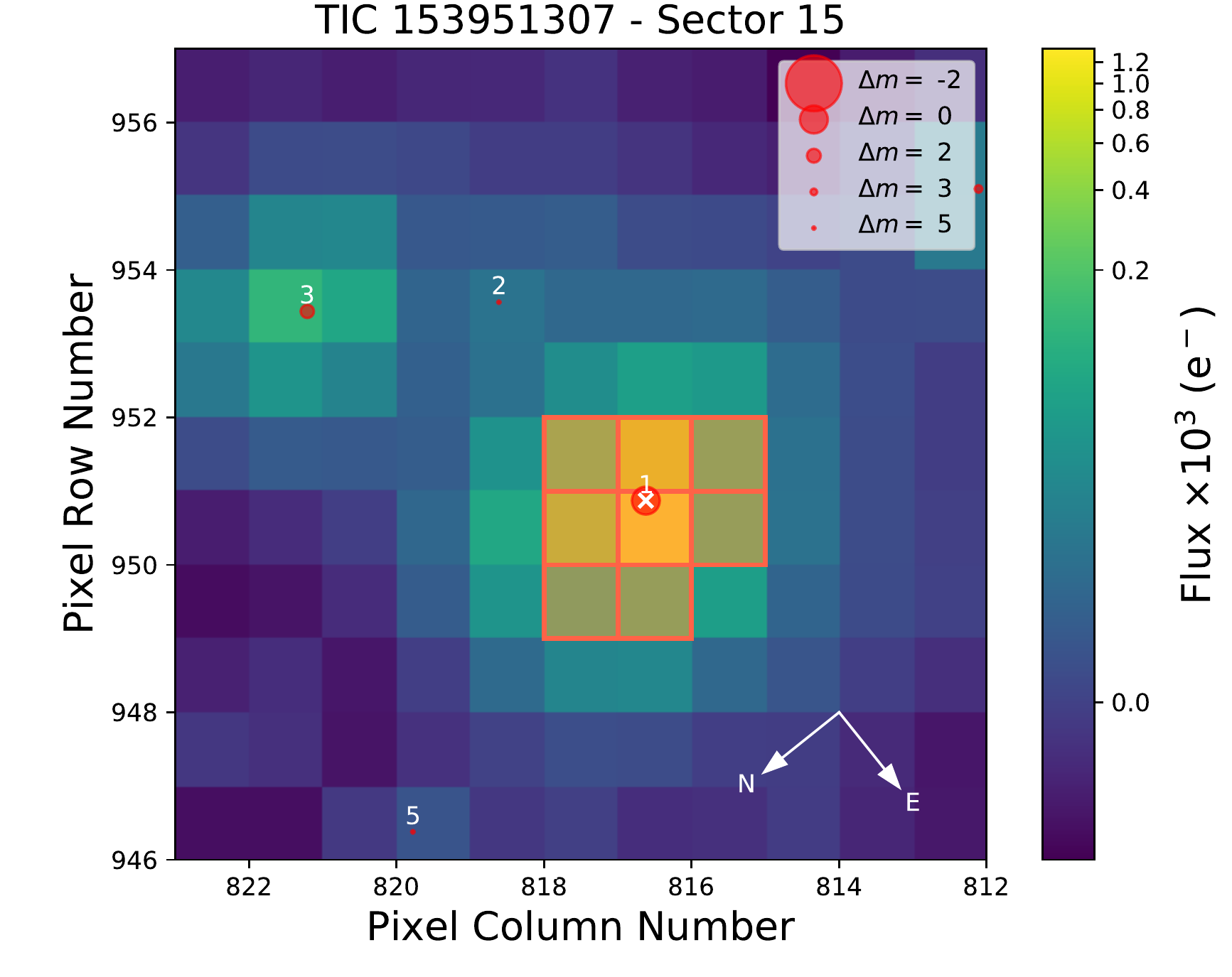}
\includegraphics[width=0.24\textwidth]{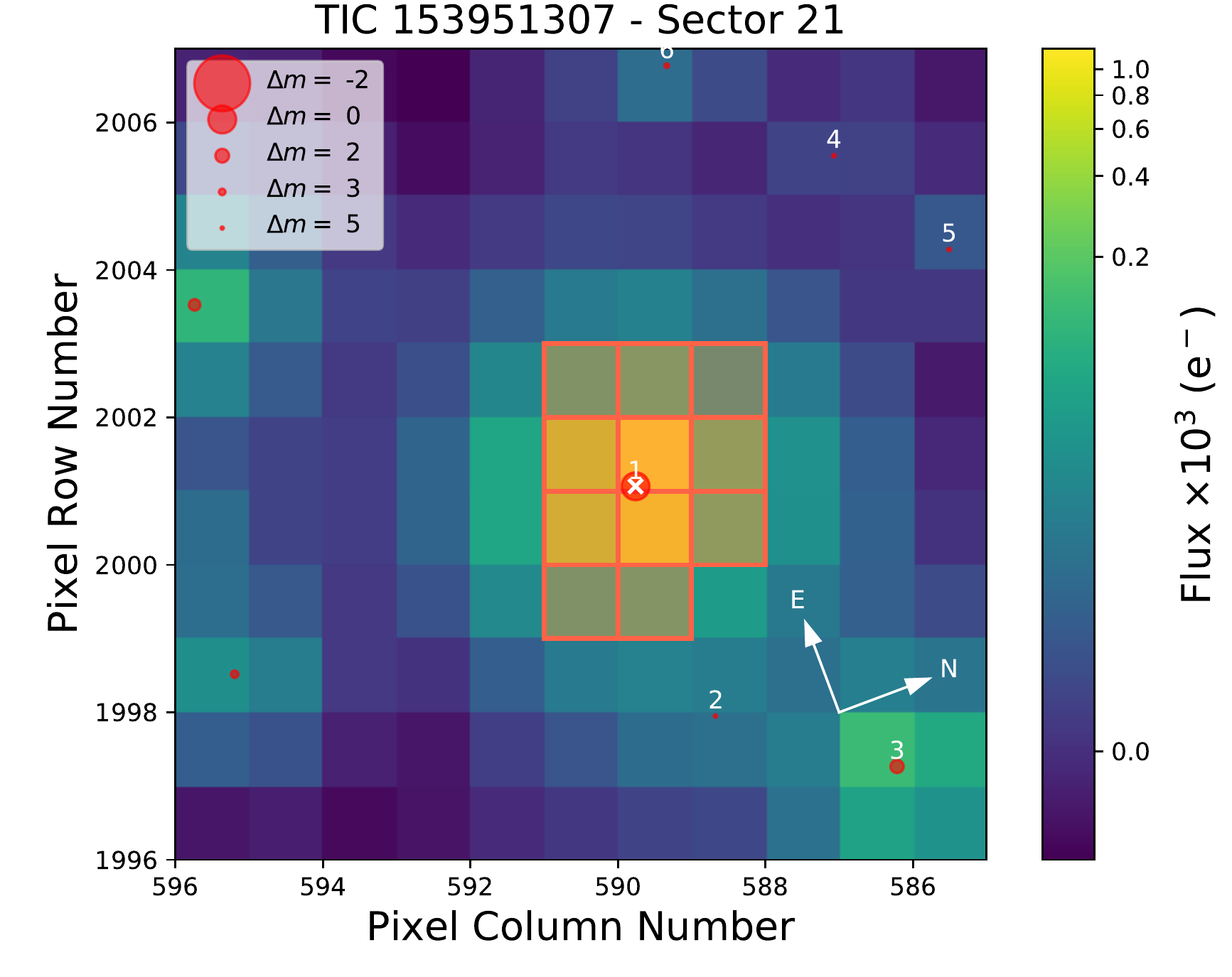}
\includegraphics[width=0.24\textwidth]{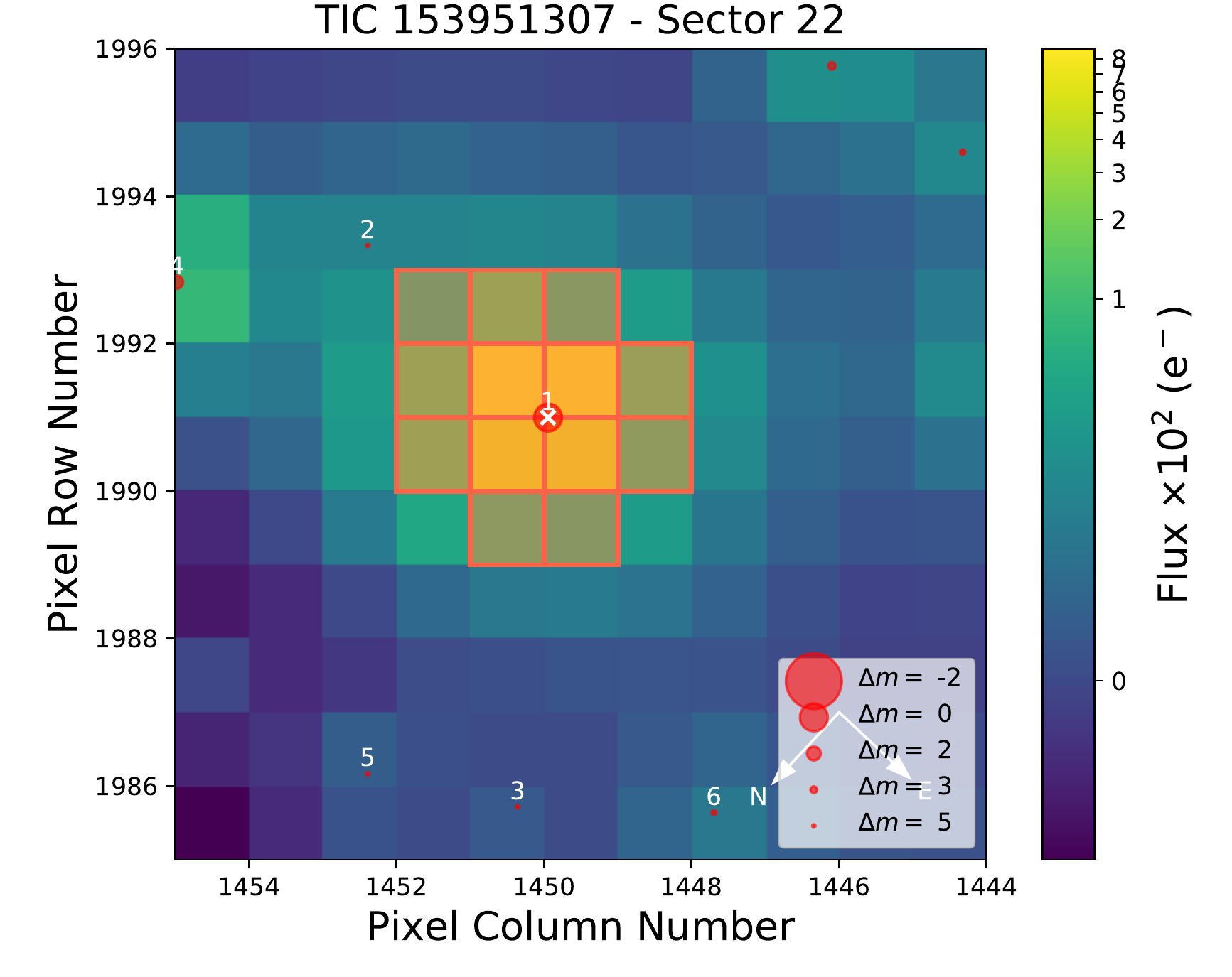}
\caption{TPFs of TOI-1238 (cross symbol) for \textit{TESS} sectors 14, 15, 21, and 22 (left to right). The electron counts are color-coded. The \textit{TESS} optimal photometric aperture
 per sector used to obtain the SAP fluxes is marked with red squares. The \textit{Gaia} DR2 objects with $G$-band magnitudes down to 6 mag fainter than TOI-1238 are labeled with numbers (TOI-1238 corresponds to number 1), and their scaled brightness, based on \textit{Gaia} magnitudes, is shown by red circles of different sizes (see figure inset). The pixel scale is 21\,arcsec\,pixel$^{-1}$.}
\label{fig:apertures}
\end{figure*}

\begin{figure*}[] 
\centering
\includegraphics[width=0.99\textwidth]{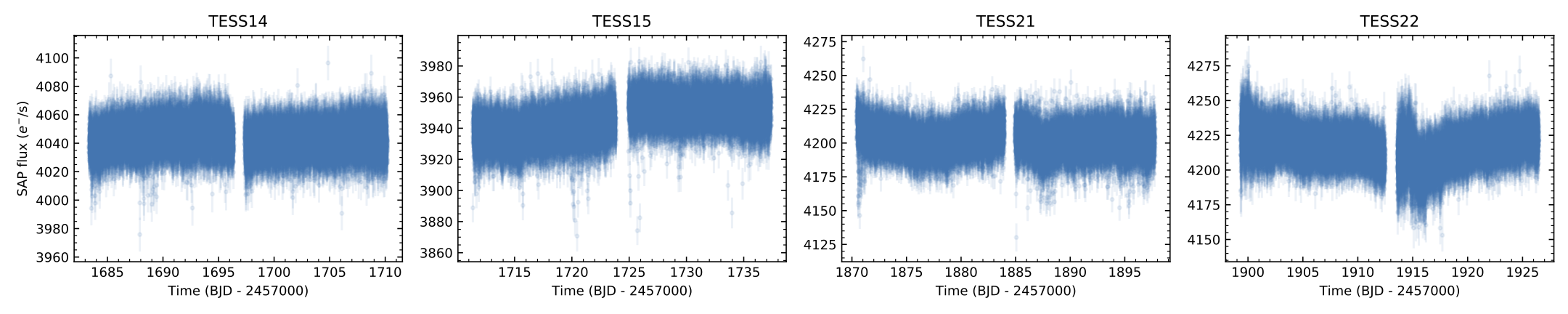}
\includegraphics[width=0.99\textwidth]{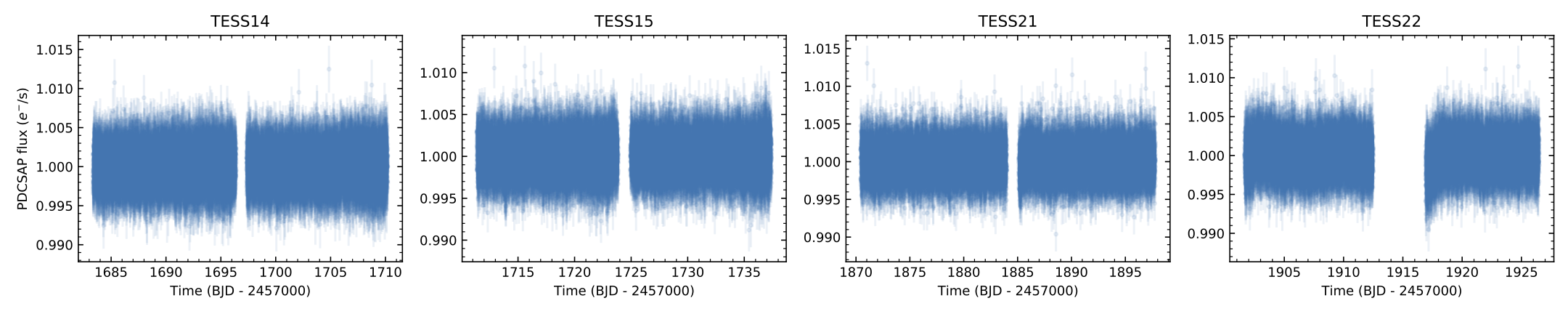}
\includegraphics[width=0.99\textwidth]{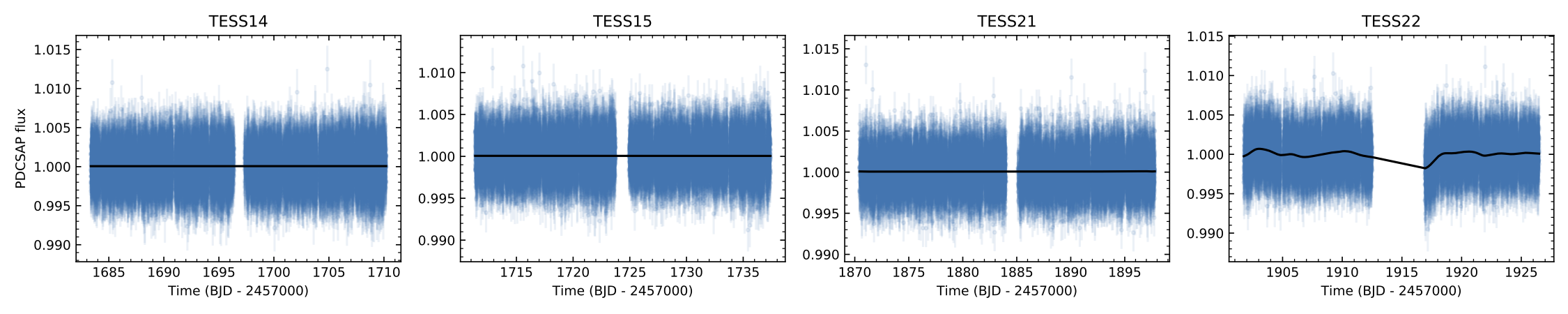}
\caption{\textit{TESS} LCs. {\sl Top and middle panels:} SAP and PDCSAP LCs of TOI-1238 as obtained by \textit{TESS} in sectors 14, 15, 21, and 22 and processed by SPOC (left to right). {\sl Bottom panel:} PDCSAP LCs with the detrending GP models overplotted (black line). The GP models were used to remove any residual instrumental effect (after masking the position of all transits). As illustrated, only the data of sector 22 required this additional correction to produce a flat LC. }
\label{fig:SAP_and_PDCSAP}
\end{figure*}

All LCs and TPF files were downloaded from the Mikulski Archive for Space Telescopes, which is a NASA funded project. Our first step was to verify that the SAP and PDCSAP fluxes automatically computed by the pipeline are useful for scientific studies by confirming that no additional bright source is contaminating the aperture photometry. Figure~\ref{fig:apertures} displays the TPFs of TOI-1238 for all four sectors using the publicly available {\tt tpfplotter}\footnote{ \url{https://github.com/jlillo/tpfplotter}} code \citep{2020A&A...635A.128A}. {This code} overplots {all sources from the} \textit{Gaia} Data Release 2 (DR2) catalog \citep{2018A&A...616A...1G} {with magnitude contrast up to $\Delta m$=6\,mag} on top of the \textit{TESS} TPFs. {There are no additional \textit{Gaia} sources within the photometric aperture around TOI-1238 automatically selected by the pipeline. Therefore, we considered the extracted \textit{TESS} LC to be free of contamination from nearby stars.} The SAP and PDCSAP LCs of TOI-1238 are illustrated in the top and middle panels of Fig.~\ref{fig:SAP_and_PDCSAP}.

\subsection{ASAS-SN photometry}

\begin{figure}[]
\centering
\includegraphics[width=0.5\textwidth]{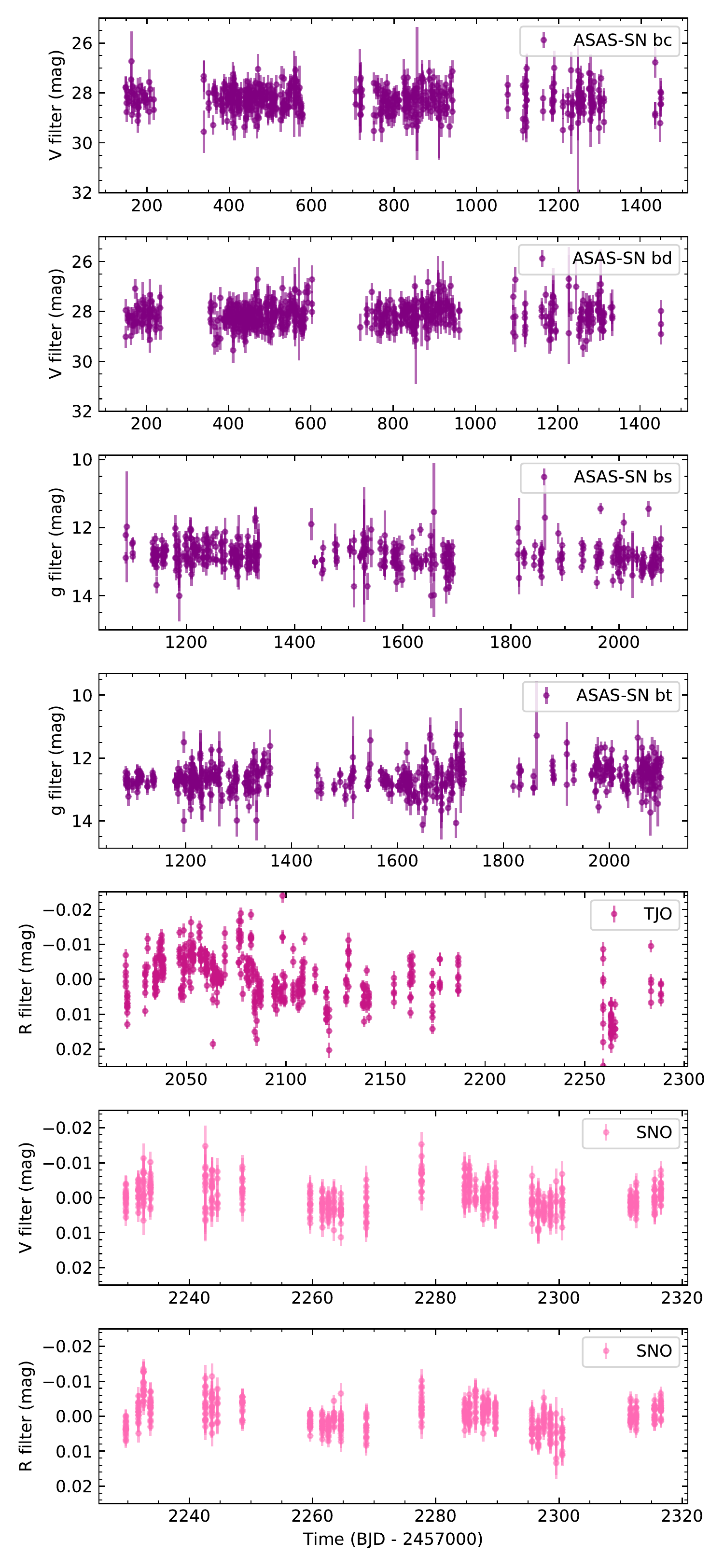}
\caption{Photometric time series of TOI-1238 using ASAS-SN (aperture photometry), TJO (differential photometry), and SNO (differential photometry) data.}
\label{fig:ASAS-SN_phot}
\end{figure}

TOI-1238 was photometrically observed by the All-Sky Automated Survey for Supernovae (ASAS-SN) project \citep{2014ApJ...788...48S, 2017PASP..129j4502K}. ASAS-SN currently consists of 24 telescopes distributed around the globe. These telescopes are used to survey the entire visible sky every night down to 18\,mag using robotic telescopes with a diameter of 14\,cm; the large number of telescopes minimizes the impact of poor weather. The ASAS-SN data of TOI-1238 were taken with two different passbands: {$V$ (cameras ``bc'' and ``bd'') and $g'$ (cameras ``bt'' and ``bs''). A total of 1065 and 1018 measurements in the $V$ and $g'$ filters were obtained between 6 May 2015 and 28 November 2018, and between 29 November 2018 and 5 September 2020, respectively. TOI-1238 was typically observed on two to four occasions per filter and per day.} The ASAS-SN aperture photometry is calibrated using the AAVSO Photometric All-Sky Survey (APASS) catalog \citep{2015AAS...22533616H}. The ASAS-SN detectors have a pixel scale of $8\arcsec$ projected onto the sky; the images of the stars have a typical full-width-half-maximum of about $15\arcsec$, so there could be some stellar blending particularly for crowded fields. We downloaded the ASAS-SN $V$- and $g'$-band photometry of TOI-1238. The different cameras were treated separately in order to minimize any possible systematics. The original photometry contained several outliers and we applied a 2.5$\sigma$-clipping algorithm to clean the various ASAS-SN LCs. The root mean square ($rms$) of the data is $\sim$0.4\,mag. Both $V$- and $g'$-band photometric time-series are shown in the four top panels of Fig.~\ref{fig:ASAS-SN_phot}.

\subsection{TJO photometry}

We observed TOI-1238 with the 0.8\,m Joan Or\'o {Telescope} \citep[TJO;][]{2010SPIE.7740E..3KC} at the {Montsec Astronomical Observatory} in Lleida, Spain, from 19 June 2020 through 19 February 2021. A total of 507 images {(5--10 images per night)} were obtained using the Johnson $R$ filter, an exposure time of 60\,s, and the LAIA\footnote{Large Area Imager for Astronomy} imager, a 4k$\times$4k charge-coupled device (CCD) with a field of view of 30$\arcmin$ and a plate scale of 0.4$\arcsec$\,$\rm pixel^{-1}$. Raw frames were corrected for dark current and bias and were flat-fielded using the ICAT\footnote{The IEEC Calibration and Analysis Tool} pipeline \citep{2006IAUSS...6E..11C} of the TJO. The aperture photometry of TOI-1238 was extracted with the {\tt AstroImageJ} software \citep{2017AJ....153...77C} by using an optimal aperture size that minimized the $rms$ of the resulting relative fluxes. To derive the differential photometry of TOI-1238, we selected the 30 brightest comparison stars in the field that did not show any variability. Then, we employed our own pipelines to remove outliers and measurements affected by poor observing conditions or low signal-to-noise ratio. The $rms$ of the TJO differential photometry after the removal of outliers is 8\,mmag. The TJO $R$-band LC is presented in the fifth panel of Fig.~\ref{fig:ASAS-SN_phot}.

\subsection{SNO photometry}

We also monitored TOI-1238 using the 90-cm T90 telescope at the Observatorio de Sierra Nevada (SNO), Spain. The T90 telescope is equipped with a VersArray 2k$\times$2k CCD camera, which delivers images with a field of view of 13.2\arcmin$\times$13.2{\arcmin} \citep{2021A&A...650A.188A}. The observations were collected with both Johnson $V$ and $R$ filters at a total of 31 observing epochs over 88 days in the period January--April 2021. Per observing epoch, we typically obtained 20 individual measurements per filter, each with an integration time of 20\,s. We obtained aperture photometry from the unbinned frames, which were bias subtracted and properly flat-fielded beforehand. We explored different aperture sizes and the circular aperture with radius of 12 pixels produced the highest-quality photometric LC for both filters. We identified six stars in the field that were non-variable in the time interval of the SNO observations and could act as reference stars to produce the differential photometry of TOI-1238. In Fig.~\ref{fig:SNO_field}, the target and the reference stars are marked. These reference stars are 0.2--2.5\,mag fainter than TOI-1238. From the final SNO photometric time series, we removed low-quality data obtained under poor observing conditions or at very high air masses. The final SNO LCs have an $rms$ of 4\,mmag for the two filters. The $V$- and $R$-band SNO photometric time series are shown in the bottom panels of Fig.~\ref{fig:ASAS-SN_phot}.

\subsection{Gemini speckle imaging}
\label{subsec:speckle}

\begin{figure}
\centering
\includegraphics[width=0.49\textwidth]{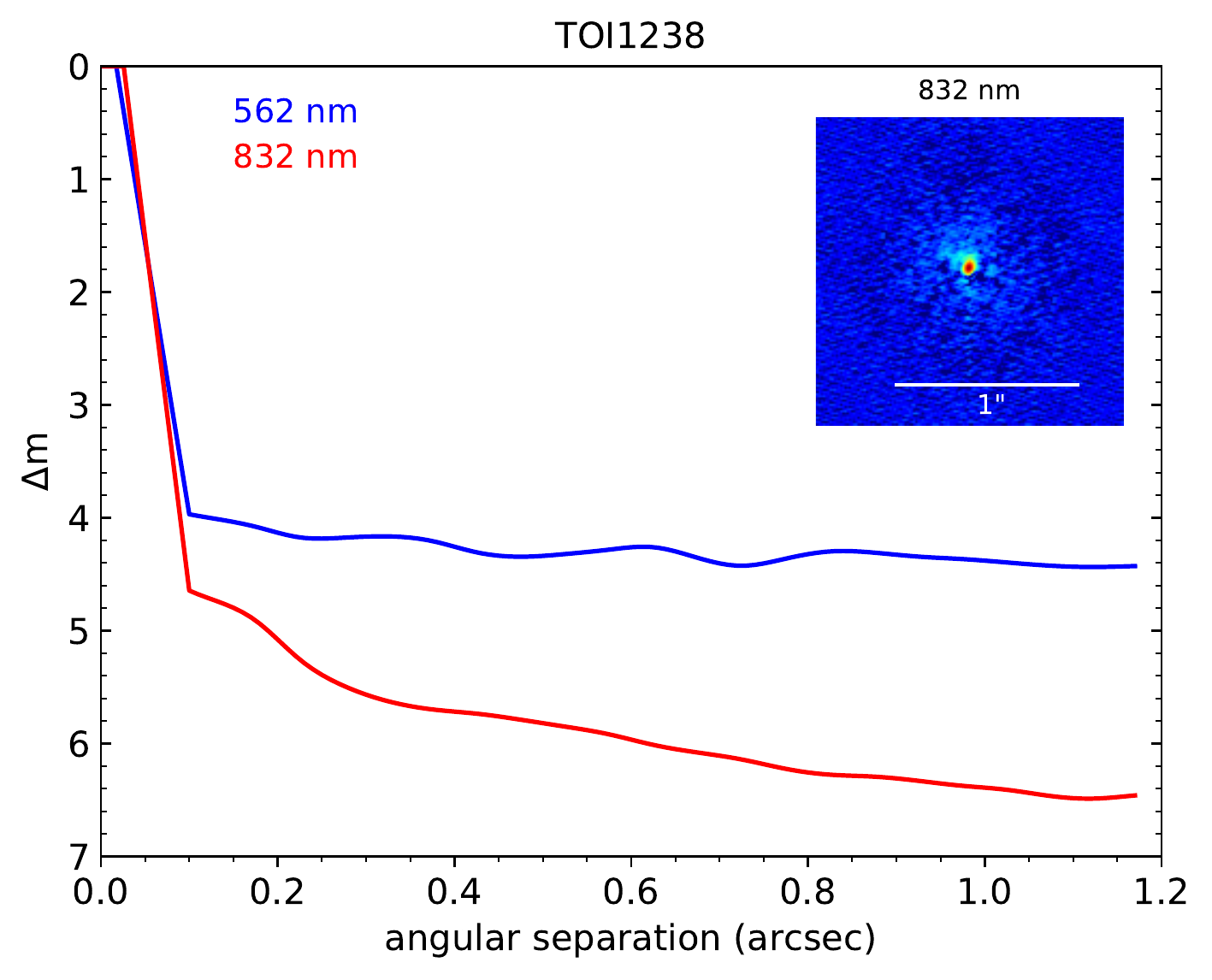}
\caption{Relative 5$\sigma$ detection limit magnitude as a function of the angular separation for Gemini (\textit{`Alopeke}) at 562\,nm (blue line) and 832\,nm (red line). The inset figure shows the reconstructed \textit{`Alopeke} high-contrast, high-resolution speckle image at 832\,nm.
}
\label{fig:Gemini-N_image}
\end{figure}

TOI-1238 was observed on 17 February 2020 using the \textit{`Alopeke} speckle instrument on the Gemini North 8.1\,m telescope\footnote{\url{https://www.gemini.edu/sciops/instruments/alopeke-zorro/}}. \textit{`Alopeke} provides simultaneous speckle imaging in two bands (562\,nm and 832\,nm) with output data products including a reconstructed image with robust contrast limits on companion detections \citep[e.g.,][]{2016ApJ...829L...2H}. Five sets of 1000$\times$0.06\,s exposures were collected and subjected to Fourier analysis in our standard reduction pipeline \citep[see][]{2011AJ....142...19H}. Figure \ref{fig:Gemini-N_image} shows our final {5-$\sigma$} contrast curves and the 832\,nm reconstructed speckle image.

\subsection{CARMENES spectroscopy \label{carmenes_spectroscopy}}

\begin{figure}[]
\centering
\includegraphics[width=0.5\textwidth]{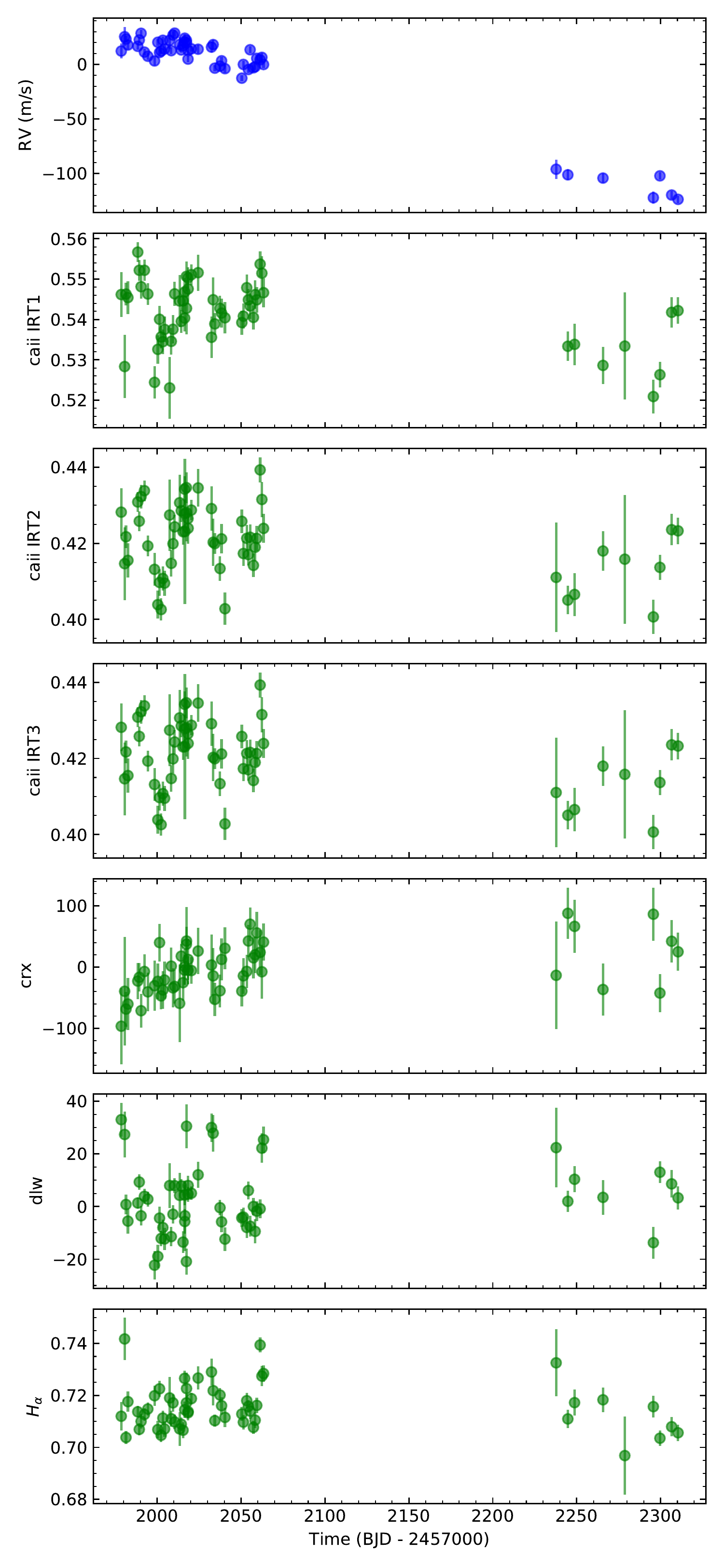}
\caption{CARMENES VIS RV time series (blue dots) and various spectroscopic activity indicators (green dots) of TOI-1238. }
\label{fig:toi1238_activity_time}
\end{figure}

After the announcement of TOI-1238 as a \textit{TESS} object of interest, we obtained a total of 55 high-resolution spectra using the fiber-fed-\'echelle spectrograph CARMENES from 9 May 2020 through 5 April 2021. {The first observing season (9 May 2020 through 1 August 2020) has the largest amount of CARMENES data (48 spectra) and was mainly used to characterize the transiting planets. On three different nights, we acquired two CARMENES spectra per night to address the aliasing issues described in Sects.~\ref{spec-indicators} and \ref{sec:aliasing}. The second CARMENES season (23 January 2021 through 5 April 2021) has seven spectra and was intended to monitor the RV long-term trend of TOI-1238.} The TJO and SNO photometry and the CARMENES data were thus contemporaneous. CARMENES is installed at the 3.5\,m telescope at the Calar Alto Observatory in Almer\'ia (Spain). It was specifically designed to deliver high-resolution spectra at optical (resolving power $R$ = 94,600) and near-infrared ($R$ = 80,500) wavelengths covering the wavelength range from 520 to 1710\,nm without any significant gap. CARMENES has two different channels, one for the optical (the VIS channel) and one for the near-infrared (the NIR channel) with the break at 960\,nm \citep{2016SPIE.9908E..12Q}. All data were acquired with integration times of 1800\,s (which is the maximum exposure employed for precise RV measurements) and followed the data flow of the CARMENES GTO program. {Only the spectrum taken on BJD = 2459278.65 was acquired with a lower exposure time, 1200\,s.} In this work, we used only the CARMENES VIS data because the NIR channel delivers relative RVs with less precision than the VIS channel \citep{2018A&A...612A..49R,2020A&A...640A..50B}, and the expected RV amplitudes of the transiting planets {($\le 5$\,m\,s$^{-1}$)} are {notably} lower than the median RV precision obtained in the NIR {(mean error bar of 24\,m\,s$^{-1}$)}. Nevertheless, the CARMENES NIR RVs of TOI-1238 are analyzed briefly in Sect.~\ref{external-planet} in relation to an outer companion candidate with large RV amplitude. From now on, when we mention CARMENES data (without the ``VIS'') we refer to the VIS channel, unless specified otherwise.

CARMENES raw data are automatically reduced with the {\tt caracal} pipeline \citep{2016SPIE.9910E..0EC}. Relative RVs are extracted separately for the VIS and NIR channels using the {\tt serval} software \citep{2018A&A...609A..12Z}, where each \'echelle order is fitted separately. The final RV per epoch is computed as the weighted RV mean over all \'echelle orders. Our first step was to remove RV outliers from the CARMENES time series by applying a 2.5$\sigma$-clipping algorithm and to remove data points with very large error bars (more than three times the mean error bar size, 3.93\,m\,s$^{-1}$). The final CARMENES data set has 53 RV measurements {out of a total of 55 RVs, and it} is displayed in the top panel of Fig.~\ref{fig:toi1238_activity_time}. {The 53 individual CARMENES relative RVs used in our analysis and their associated uncertainties are listed in Table~\ref{tab:toi1238_rv_act_data_used}. For completeness, we also give in Table~\ref{tab:toi1238_rv_act_data_rejected} the deviating velocity measurements not used here}. The CARMENES RVs show a marked downward trend, which will be discussed in the next sections. 

The CARMENES {\tt serval} pipeline also provides measurements for a number of spectral features that are considered indicators of stellar activity: pseudo-equivalent widths of He\,{\sc i}\,D$_3$, He\,{\sc i} $\lambda$10833\,\AA, H$\alpha$, Pa$_\beta$, Na\,{\sc i}\,D doublet, and the Ca\,{\sc ii} infrared triplet (IRT), all of which may have a chromospheric component in active M dwarfs. The pipeline also provides measurements of the strength of two photospheric molecular bands: TiO and VO \citep[see also][]{2018A&A...609A..12Z, 2019A&A...623A..44S}. At high spectral resolution, the profile of the stellar lines may change due to photospheric and chromospheric activity, which has an impact on accurate RV measurements. It is important to disentangle the effects of stellar activity from the Keplerian signals. We used the following CARMENES spectral indices: H${\alpha}$, Ca\,{\sc ii} IRT $\lambda\lambda\lambda$8498, 8542, 8662\,\AA, the differential line width (dLW), and the chromatic index (CRX), all of which were defined by \cite{2018A&A...609A..12Z}. The CRX determines the RV--$\log \lambda$ correlation, and it is used as an indicator of the presence of stellar active regions. The measured CARMENES activity indicators of TOI-1238 are {given in Table~\ref{tab:toi1238_rv_act_data_used} and are} shown in Fig.~\ref{fig:toi1238_activity_time}, where data points deviating significantly (more than 2.5 $\sigma$) from the sequence or data with very large error bars have been removed from the time series. We used {these $\sigma$-clipped} activity indices to analyze the properties of the parent star. {Typically, one to three data points were rejected per CARMENES activity indicator. These rejected data correspond to the short-integration-time spectrum and the spectra acquired at high air masses. For completeness, they are provided in Table~\ref{tab:toi1238_rv_act_data_rejected}}.

\begin{figure}[] 
\centering
\includegraphics[width=0.5\textwidth]{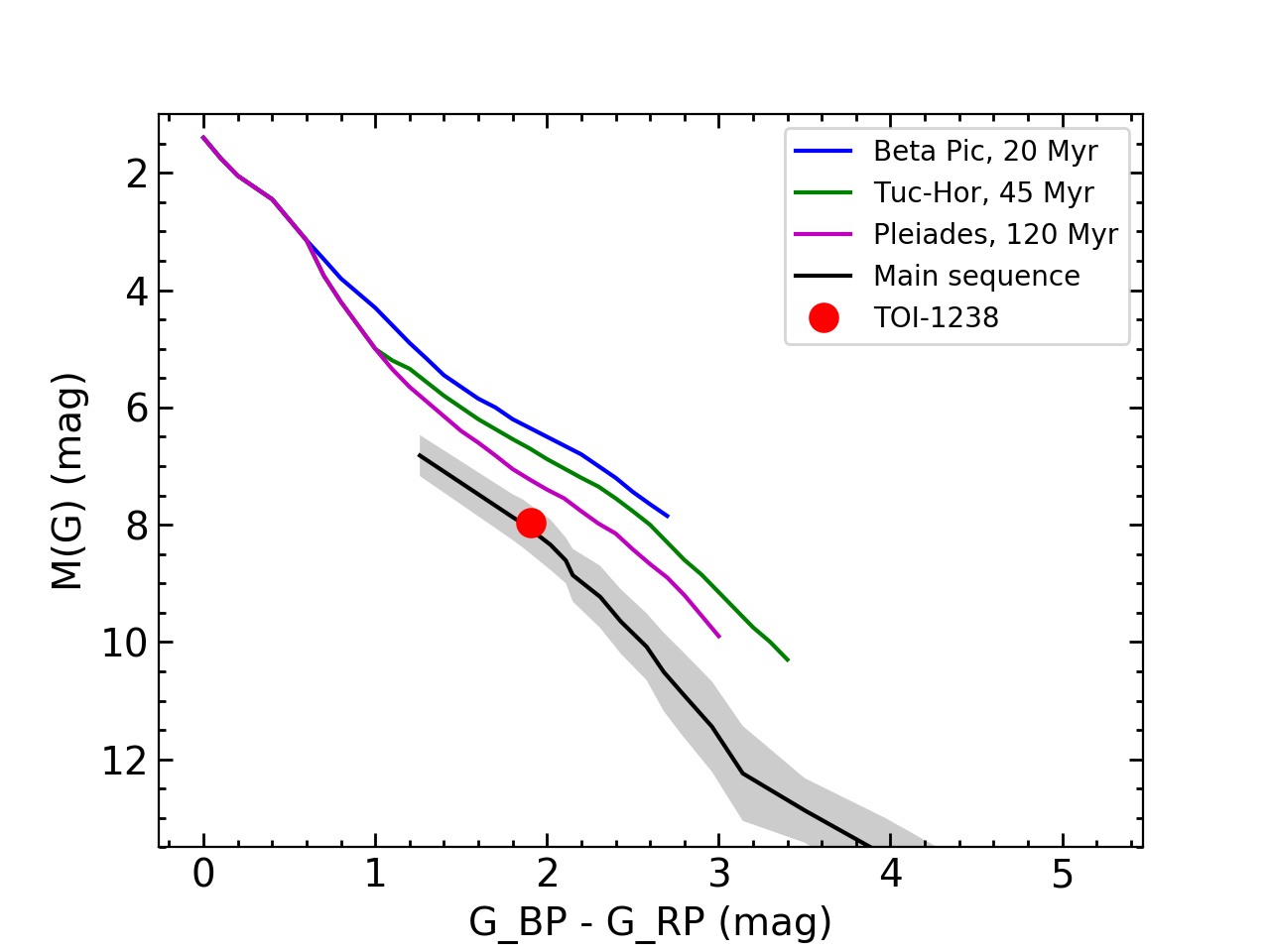}
\caption{Location of TOI-1238 (red circle) in the \textit{Gaia} color-magnitude diagram. The young stellar sequences of the $\beta$~Pic (blue) and Tucana-Horologium (green) moving groups and the Pleiades star cluster (purple) are taken from \citet{2018AJ....156..271L} and the main sequence of field stars (black) from \citet{2020A&A...642A.115C}. The gray area represents the dispersion observed among stars on the main sequence.
}
\label{fig:main_sequence}
\end{figure}

\section{TOI-1238}
\label{sec:TOI-1238}

\subsection{Stellar properties}
TOI-1238 (2MASS J13253177$+$6850106) is not an especially well-studied star, and does not appear very often in the literature except in large catalogs. In Table~\ref{tab:stellar_properties_TOI-1238}, we provide all optical, near- and mid-infrared photometry extracted from the \textit{Gaia}, 2MASS\footnote{Two Micron All-Sky Survey}, and AllWISE\footnote{Wide-field Infrared Survey Explorer} catalogs \citep{2016A&A...595A...1G, 2006AJ....131.1163S, 2010AJ....140.1868W}, together with the \textit{Gaia} Early Data Release 3 (EDR3) trigonometric parallax, proper motion, and equatorial coordinates \citep{2021A&A...649A...1G}. TOI-1238 is located at a distance of 70.642 $\pm$ 0.061\,pc from the Sun and has optical and infrared colors typical of K7--M0 type stars with no obvious evidence of extinction at short wavelengths. The location of TOI-1238 in the \textit{Gaia} color-magnitude diagram is shown in Fig.~\ref{fig:main_sequence}; it is obvious that this star is not overluminous (i.e., it is not young) and has absolute magnitudes compatible with main sequence K7--M0 star. All stellar sequences shown in the figure were built using \textit{Gaia} photometry and parallaxes (see \citealt{2018AJ....156..271L} for the young isochrones and \citealt{2020A&A...642A.115C} for the main sequence). 

\begin{figure}[] 
\centering
\includegraphics[width=0.49\textwidth]{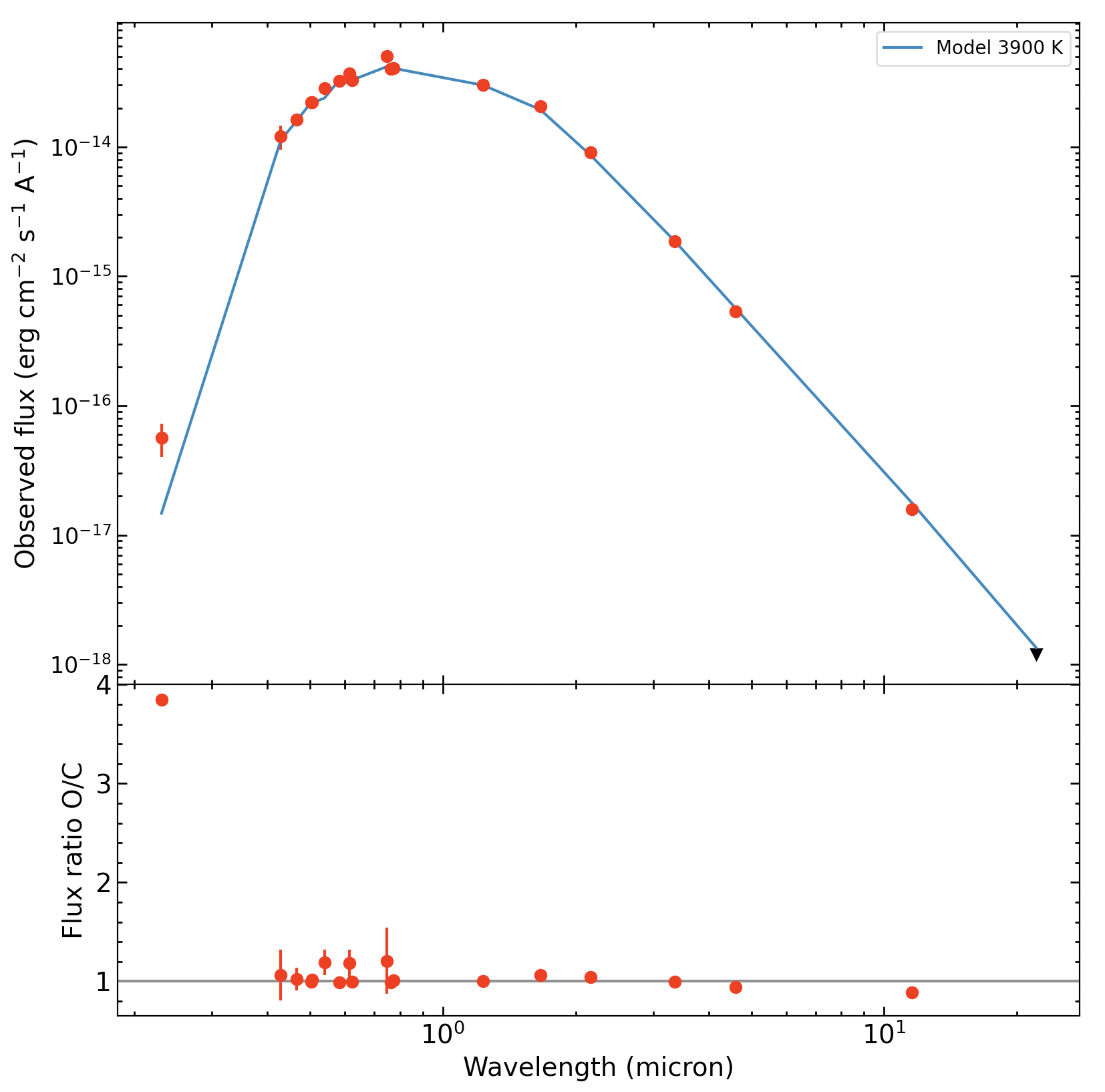}
\caption{Photometric spectral energy distribution of TOI-1238 (red dots) and photometry of the best BT-Settl model ($T_{\rm eff}$ = 3900 $\pm$ 100\,K, log\,$g$ = 5.5 $\pm$ 0.5 dex, and [Fe/H] = 0.0--0.5 dex, \citealt{2012RSPTA.370.2765A}). The reddest data point ($W4$) is an upper limit.
}
\label{fig:sed}
\end{figure}

We first estimated the physical parameters of TOI-1238 empirically by fitting the observational spectral energy distribution constructed from cataloged broadband photometric magnitudes, following the prescription of the VOSA\footnote{Virtual Observatory SED analyzer} tool \citep{2008A&A...492..277B}. The available photometry includes the near-ultraviolet (NUV) magnitude from \textit{GALEX}\footnote{Galaxy Evolution Explorer} \citep{2017ApJS..230...24B}, the \textit{Gaia} $BP$ and $RP$ magnitudes, the $gri$ magnitudes from the Sloan Digital Sky Survey \citep{2017AJ....154...28B}, the $BV$ magnitudes of the APASS catalog \citep{2015AAS...22533616H}, the $JHK_s$ magnitudes from 2MASS, and the $W$1, $W$2, and $W$3 magnitudes from AllWISE \citep{2010AJ....140.1868W, 2011ApJ...731...53M}. We employed the BT-Settl models \citep{2012RSPTA.370.2765A}, which are computed for a wide range of effective temperatures ($T_{\rm eff}$), surface gravities and atmospheric metallicities with steps of 100\,K, 0.05\,dex and 0.05\,dex, respectively. The photometric spectral energy distribution of TOI-1238 covers the wavelength interval 0.23--22.09 $\mu$m. We assumed no interstellar extinction due to the proximity of the star and the presence of no obvious reddening in the stellar colors at short wavelengths. From this analysis, we obtained $T_{\rm eff}$ = 3900 $\pm$ 100\,K, log\,$g$ = 5.5 $\pm$ 0.5\,dex, and [Fe/H] = 0.0--0.5\,dex. The high surface gravity is consistent with a main sequence star. The observed spectral energy distribution of TOI-1238 and its best fit are shown in Fig.~\ref{fig:sed}. The \textit{GALEX} NUV emission of the star at 230\,nm is higher than the expected photospheric emission, which is a signpost of the presence of a stellar chromosphere/corona  \citep{2020A&A...642A.115C}. From our fit we derived a stellar luminosity of $\log$\,$L$/$\rm L_\odot$ = $-1.0824 \pm 0.0024$ dex (Table~\ref{tab:stellar_properties_TOI-1238}).

We also calculated the stellar parameters of TOI-1238 using the CARMENES VIS and NIR spectra and the grid of PHOENIX-ACES synthetic models, following the method of \cite{2019A&A...627A.161P}. The resulting values (listed in Table~\ref{tab:stellar_properties_TOI-1238}) are $T_{\rm eff}$ = 4089 $\pm$ 54\,K, log\,$g$ = 4.63 $\pm$ 0.06\,dex, and [Fe/H] = +0.31 $\pm$ 0.19\,dex; at the 2-$\sigma$ level, they are consistent with those derived from the photometric spectral energy distribution and indicate that the star is likely more metal-rich than the Sun. We used these values (and an age $>$ 0.8\,Gyr; see below) to determine the radius and mass of TOI-1238 at $R = 0.58 \pm 0.02$ R$_\odot$ and $M = 0.59 \pm 0.02$ M$_\odot$, following the mass--radius relation of \cite{2019A&A...625A..68S}, which was based on eclipsing binary stars. From the spectral fitting, we also derived an upper limit on the projected rotational velocity of the star at $v$\,sin\,$i \leq$ 2\,km\,s$^{-1}$.

The Galactic space velocities $U V W$ of TOI-1238 were derived using the \textit{Gaia} coordinates, proper motion, and RV with the formulation developed by \citet{1987AJ.....93..864J}. The $U V W$ components in the directions of the Galactic center, Galactic rotation, and north Galactic pole, respectively, are given in Table~\ref{tab:stellar_properties_TOI-1238}.  We note that the right-handed system is used and that we did not subtract the solar motion from our calculations. The uncertainties associated with each space velocity component were obtained from the observational quantities and their error bars. TOI-1238 has kinematics typical of ``the field'' (it does not appear to belong to any known young stellar moving group) indicating a likely age of $>$ 0.8\,Gyr. {This result was also confirmed by using the BANYAN $\Sigma$ code \citep{2018ApJ...856...23G}.}

\begin{small}
\begin{table}[]
\centering
\caption{Stellar parameters of TOI-1238.}
\label{tab:stellar_properties_TOI-1238}
\begin{tabular}{l c l }
\hline
\hline
\noalign{\smallskip}
Parameters  & Value & Ref.$^{a}$\\
\noalign{\smallskip}    
\hline  
\noalign{\smallskip}

TIC                                                                     & 153951307 & Stas18\\
Karm                                                            & J13255+688 & Cab16 \\
2MASS                                                   & J13253177+6850106 & 2MASS \\
$\rm \alpha$ (hh:mm:ss)             &   13:25:31.76  &  \textit{Gaia} EDR3\\
$\rm \delta$ (dd:mm:ss)         &   +68:50:09.8  &  \textit{Gaia} EDR3\\

$V$ (mag) & $12.79 \pm 0.0005$ &  Stas18\\
$G$ (mag) & $12.2139     \pm 0.0003$ &  \textit{Gaia} EDR3\\
$J$ (mag) & $10.039      \pm 0.020$ &  2MASS\\
$H$ (mag) & $9.348\pm 0.019$ &  2MASS\\
$K_s$ (mag) & $9.184     \pm 0.014$ &  2MASS\\
$W1$ (mag) & $9.106 \pm 0.023$ & AllWISE\\
$W2$ (mag) & $9.037 \pm 0.020$ & AllWISE\\
$W3$ (mag) & $9.037 \pm 0.027$ & AllWISE\\
$W4$ (mag) & $>9.0$ & AllWISE\\
$\pi$ (mas) & $14.1558   \pm 0.0123$ &  \textit{Gaia} EDR3\\
$d$ (pc) & $\rm 70.6424 \pm 0.0614$  & \\
$\mu _{\alpha} \cos \delta$ ($\rm mas\,yr^{-1}$) & $-4.887 \pm 0.016$ & \textit{Gaia} EDR3\\
$\mu _{\delta}$ ($\rm mas\,yr^{-1}$) & $-45.886 \pm 0.015$ & \textit{Gaia} EDR3\\

RV ($\rm km\,s^{-1}$)                   & $\rm -17.49   \pm0.85$ & \textit{Gaia} DR2\\
$U$ ($\rm km\,s^{-1}$)  &               $12.30\pm0.27$          & This work\\
$V$ ($\rm km\,s^{-1}$)  &               $-19.65\pm0.50$ &  This work\\
$W$ ($\rm km\,s^{-1}$)  &               $-2.70\pm0.63$          &  This work\\

\noalign{\smallskip}
\hline
\noalign{\smallskip}

\noalign{\smallskip}
Spectral type & K7--M0 & This work\\
$T_{\rm eff}$ (K)  & $\rm 4089  \pm 54 $ & This work\\
$\log g$ (cgs) & $\rm 4.63 \pm 0.06$ & This work\\
$\rm [Fe/H]$ (dex) & $\rm +0.31 \pm 0.19$ & This work\\
$M$ ($\rm M_{\odot}$) & $\rm 0.59 \pm 0.02$ & This work\\
$R$ ($\rm R_{\odot}$) & $\rm 0.58 \pm 0.02$ & This work\\
$L$ ($\rm L_{\odot}$) & $\rm 0.0827 \pm 0.002$ & This work\\
$ v \sin i$ ($\rm km\,s^{-1}$) & $\leq$ 2 & This work\\
$P_{\rm rot}$ (d) & {$40 \pm 5$} & This work\\
Age (Gyr) & $>$0.8 &  This work\\
\hline

\end{tabular}
\tablefoot{$^{(a)}$ Stas18: \cite{2018AJ....156..102S}; Cab16: \cite{2016csss.confE.148C}; 2MASS: \cite{2006AJ....131.1163S}; \textit{Gaia} EDR3: \cite{2021A&A...649A...1G}, equinox: J2000.0; epoch: 2016.0\,yr; AllWISE: \cite{2010AJ....140.1868W}; \textit{Gaia} DR2: \cite{2018A&A...616A...1G}.
}
\end{table}
\end{small}


\begin{figure}[]
\centering
\includegraphics[width=0.53\textwidth]{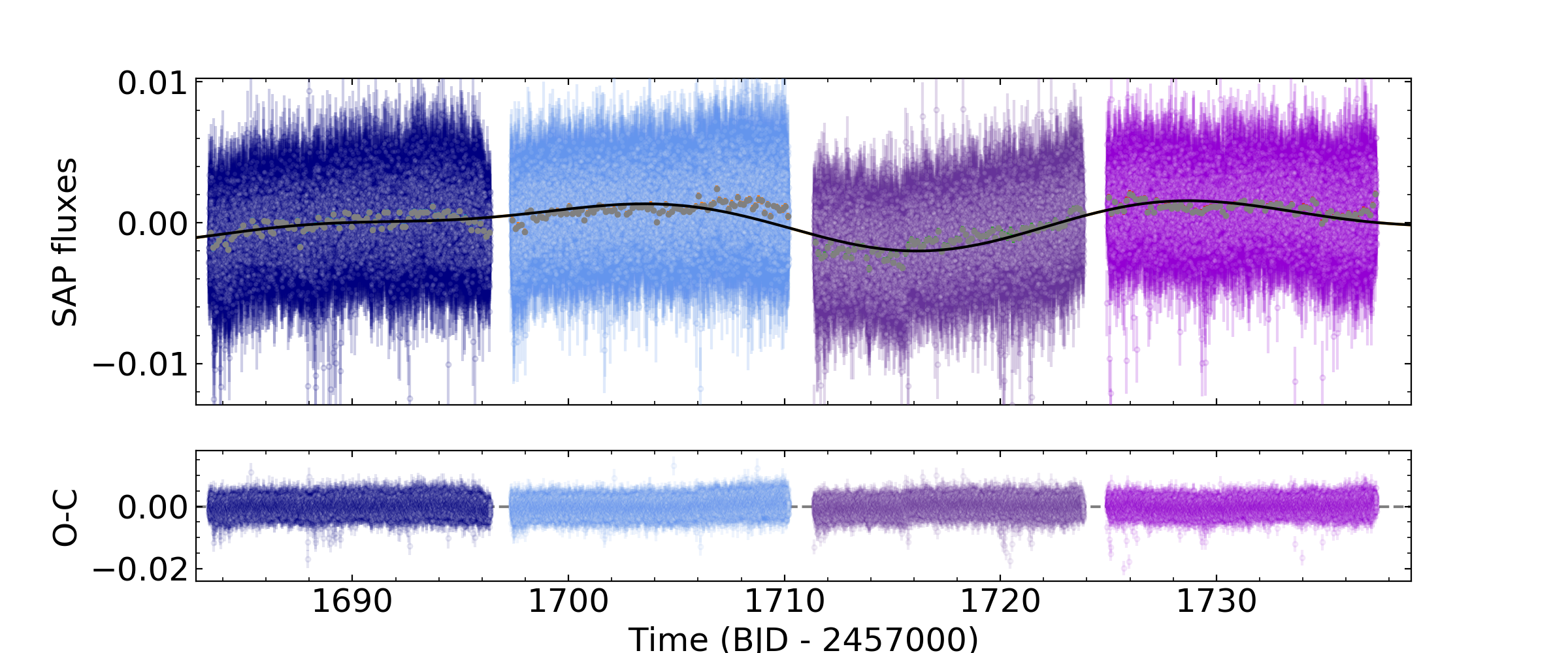}
\includegraphics[width=0.53\textwidth]{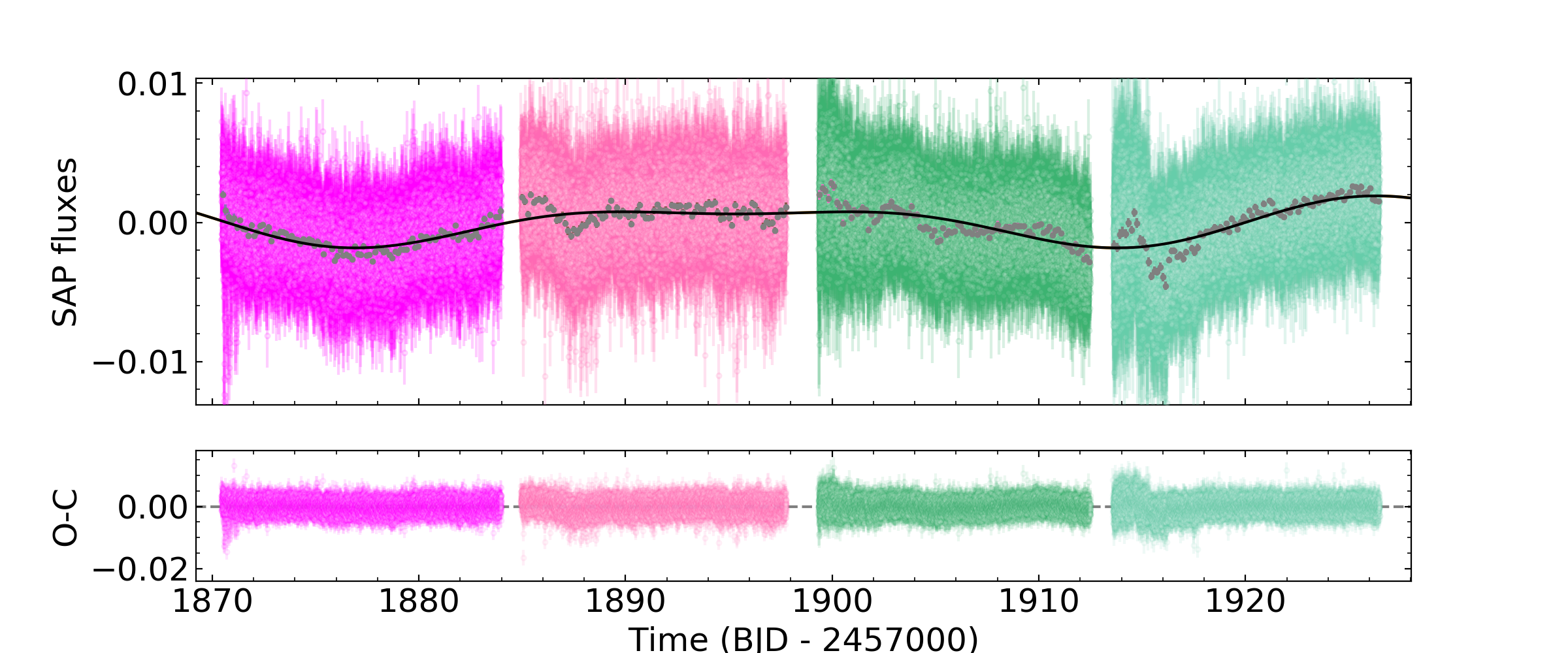}
\caption{\textit{TESS} photometry. (Top panels) TOI-1238 \textit{TESS} SAP photometric fluxes for the four different sectors (each half sector is shown with a different color) normalized to a common reference by fitting sinusoidal functions. The best model fit is plotted as the solid black line, and the binned photometric data are indicated with gray dots. (Bottom panels) Residuals shown as a function of time.}
\label{fig:SAP-flux_two_sin_mod}
\end{figure}

\begin{figure}[]
\centering
\includegraphics[width=0.5\textwidth]{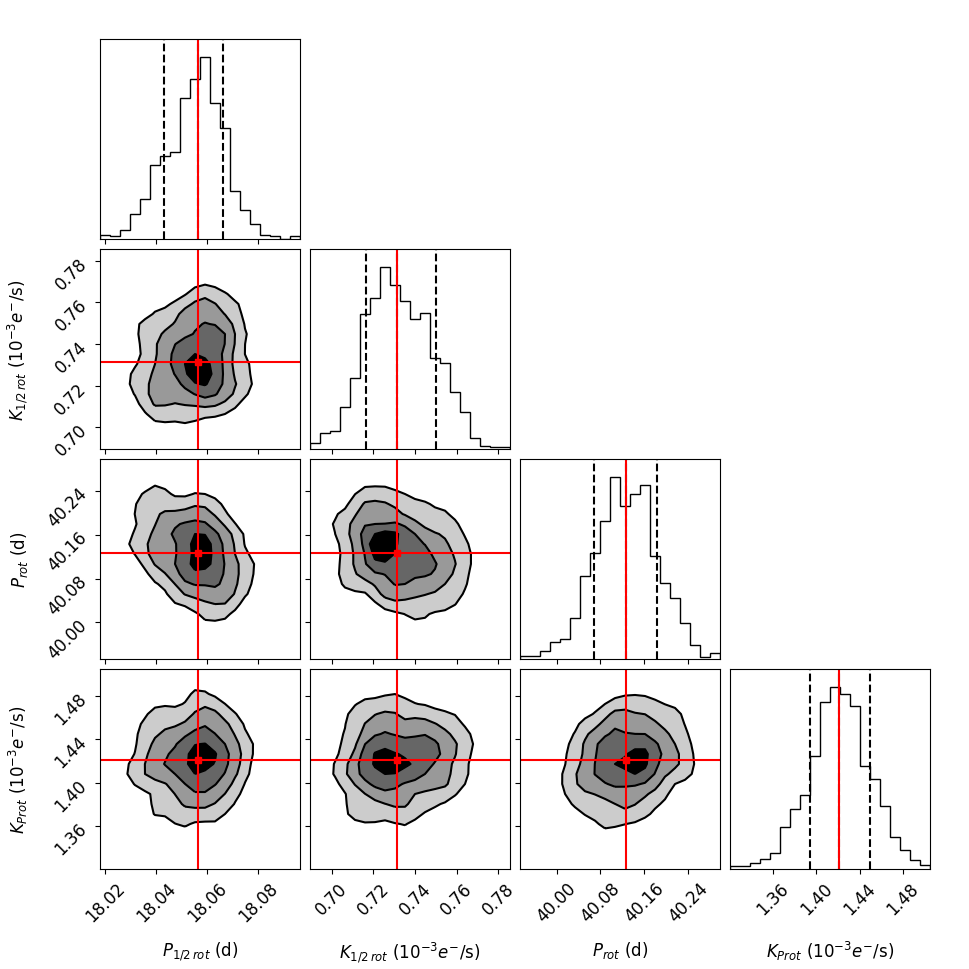}
\caption{Posterior distributions of the sinusoidal fit to the \textit{TESS} SAP fluxes. Vertical dashed lines correspond to the 16\%~and 84\%~quantiles (1 $\sigma$ uncertainty). The red line stands for the mean values of the distributions.}

\label{fig:SAP-corner-plot}
\end{figure}

\subsection{Stellar variability and rotation period}
\label{sec:analysis}

M dwarfs can exhibit a large range of stellar activity levels from inactivity  \citep{2020Sci...368.1477J} to active states with intensities orders of magnitude greater than what is commonly observed in the Sun. The most active stars show inhomogeneities on their surface, such as dark starspots corotating with the star \citep{1997A&A...327.1114L,2005ApJ...621..398O}. It is well known that stellar activity can cause line asymmetries that hinder the very precise measurement of the line center, and consequently induce an apparent RV shift, which may mimic a Keplerian signal or hide the presence of a real planet orbiting the star \citep{2015ApJ...812...42B}. Therefore, before exploring the CARMENES RVs in search for or confirming planets, we analyzed all photometric and spectroscopic index time series of established activity indices available to us in order to identify the characteristic frequencies of the stellar variations and, if possible, the rotation period of TOI-1238. The spectrophotometric spectral type of TOI-1238 is K7--M0. Following previous works \citep[e.g.,][]{2014ApJS..211...24M, 2016A&A...595A..12S, 2018A&A...612A..89S, 2019A&A...624A..27G}, early-M dwarfs typically exhibit measurable rotation periods ranging from 20\,d through $\sim$100\,d, and magnetic activity cycles of several hundred to thousand days.

\subsubsection{\textit{TESS} light curves}
\label{subsec:TESS SAP flux}

We analyzed the SAP fluxes (uncorrected for instrumental features; \citealt{twicken:PA2010SPIE,2020ksci.rept....6M}) of all four \textit{TESS} sectors. We assumed that each sector has a different flux offset in order to bring all sectors to a common reference. In addition, we also assumed intra-sector flux offsets to account for possible drifts in the photometry of the intervals before and after the \textit{TESS} data downlink time. Therefore, in practice we studied the \textit{TESS} LCs divided into eight chunks. The top panel of Fig.~\ref{fig:SAP_and_PDCSAP} shows that the SAP fluxes do not suffer from strong instrumental effects and some low-amplitude variability becomes evident from the data. We modeled the \textit{TESS} SAP fluxes using two sinusoidal functions with four free parameters: two different amplitudes, one period at the rotation period of the star ($P_{\rm rot}$) and other period at approximately half of the stellar rotation period ($P_{\rm 1/2\,rot}$). We consider that this approach can take into account the stellar variability produced by spots located at different latitudes of the stellar surface and also accounts for any possible differential rotation from cycle to cycle. As for the main parameter, we allowed $P_{\rm rot}$ to vary between 20 and 60\,d with a uniform distribution; all other parameters (LC amplitudes and offsets) were explored from initial uniform distributions with a wide range of possible values. The fit was performed using the {\tt juliet} python package \citep{2019MNRAS.490.2262E}, which {uses a nested sampling algorithm in the framework of Bayesian analysis. The MultiNest library \citep{2009MNRAS.398.1601F} was employed via the PyMultiNest wrapper \citep{2014A&A...564A.125B} to explore the parameter space and} to efficiently compute the Bayesian model log-evidence, $\ln{\mathcal{Z}}$.

{The simulations converge on rotation periods of $P_{\rm rot}$ = $40\,\pm\,8$\,d and $P_{\rm 1/2\,rot}$ = $18\pm2$\,d. The model with the rotation period and its half value has a higher statistical significance than the one-sinusoid function model ($\Delta$\,log-evidence $\approx$ 800)}. The flux amplitude of the variability associated with $P_{\rm 1/2\,rot}$ is about the half of that of the main periodicity. Figure~\ref{fig:SAP-flux_two_sin_mod} illustrates the \textit{TESS} SAP fluxes together with the best model, and Fig.~\ref{fig:SAP-corner-plot} shows the resulting corner plot of the fit (the flux offsets were excluded for the clarity of the figure). We remark that all priors were uniform distributions but the distributions of the posteriors are quite Gaussian-like. From Fig.~\ref{fig:SAP-flux_two_sin_mod}, the best model nicely reproduces the flux variability from sector to sector, with a few exceptions: the beginning of the second parts of sectors 21 and 22 (BJD $\sim$ 2,458,888 and $\sim$ 2,458,916). This is likely due to instrumental effects that remain uncorrected in the SAP fluxes. We can make an association ascribing the 40\,d periodicity to the rotation period of TOI-1238, although as discussed in the next sections, this value is likely affected by differential rotation.

\begin{figure}[]
\centering
\includegraphics[width=\columnwidth]{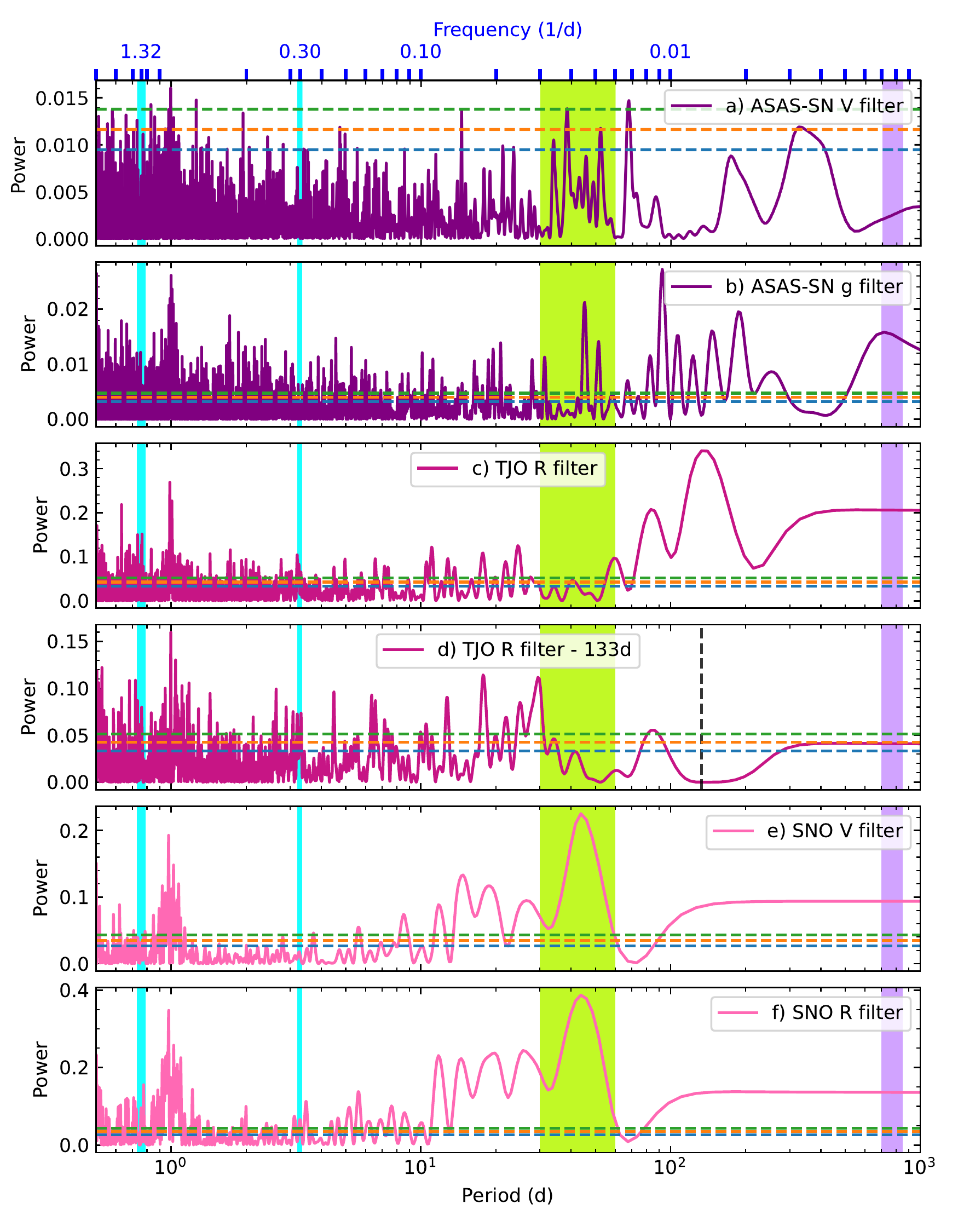}
\caption{GLS periodograms of the ASAS-SN $V$ and $g$, TJO $R$, and SNO $V$ and $R$ LCs {shown in the time space} between 0.5 and 1000\,d. In all panels, the horizontal dashed lines indicate FAP levels of 10\% (blue), 1\% (orange), and 0.1\% (green). The orbital periods of the two transiting planets are marked with vertical blue lines. The highest peak of the CARMENES RV GLS periodogram is shown with a violet line. The greenish band indicates the region where most of the spectroscopic activity indicators have their highest GLS peaks. In the fourth panel, the vertical dashed line indicates the periodicity of the signal removed from the data (see text).}
\label{fig:GLS_phot_multiplot}
\end{figure}

\begin{figure}[!t]
\centering
\includegraphics[width=\columnwidth]{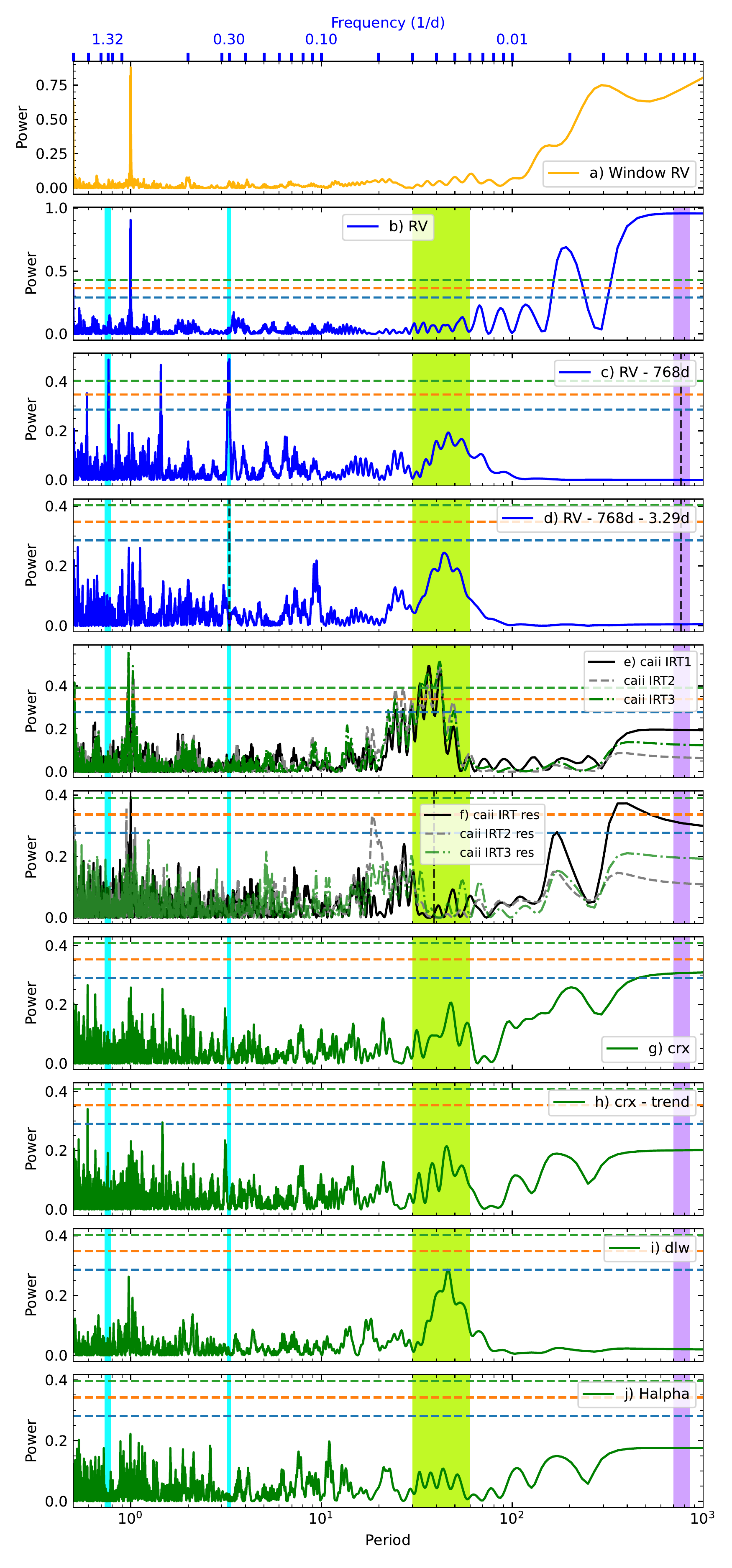}
\caption{Window function of the CARMENES data (top panel) and the GLS periodograms of the CARMENES RVs and various spectroscopic indices (panels 2--10) {shown in the time space} between 0.5 and 1000\,d. The horizontal and vertical lines and bands have the same meaning as in Fig.~\ref{fig:GLS_phot_multiplot}. }
\label{fig:GLS_activity_multiplot}
\end{figure}

\begin{small}
        \begin{table}[]
                \centering
                \caption{{Dispersion ($rms$) of the ground-based photometry of TOI-1238.}}
                \label{tab:phot_amplitudes}
                \begin{tabular}{l c}
                        \hline
                        \hline
                        \noalign{\smallskip}
                        Data set & $rms$ (mmag)  \\
                        \noalign{\smallskip}    
                        \hline  
                        \noalign{\smallskip}
                        ASAS-SN $V$ & 40 \\
                        ASAS-SN $g'$  & 40 \\
                        TJO $R$  & 8 \\
                        SNO $V$  & 4 \\
                        SNO $R$ & 4 \\  
                        TJO $R$ + SNO $R$ & 2.9$\pm$0.2 $^{*}$\\
                        \noalign{\smallskip}
                        \hline
                \end{tabular}
                \tablefoot{$^{*}$ Amplitude of the sinusoidal fit shown in Fig. \ref{fig:TJO_SNO_phot_phase}.}
        \end{table}
\end{small}

\subsubsection{ASAS-SN, TJO, and SNO light curves \label{asas-tjo-sno}}

We computed the generalized Lomb-Scargle (GLS) periodograms \citep{2009A&A...496..577Z} for all ASAS-SN, TJO, and SNO LCs. The periodograms {as a function of time and frequency are shown in Figs.~\ref{fig:GLS_phot_multiplot} and~\ref{fig:GLS_phot_freq_multiplot}, respectively.} We included three different false alarm probability (FAP) levels and the location of the two transiting planets at 0.764 and 3.294\,d as a reference. We also indicate the location of the highest peak of the CARMENES RV GLS periodogram for completeness (see Sect.~\ref{subsec:CARMENES RV data}). In the time line, the ASAS-SN $V$-band photometry is the oldest data set, then ASAS-SN $g$-band photometry was taken with about 1 year overlap with the $V$-band photometry. The TJO $R$-band photometric time series followed with a small overlap of 1 month with the ASAS-SN $g$-band data. Finally, the SNO photometry consists of the most recent data and they overlap for a period of $\sim$1 month with the TJO observations. For all LCs {we evaluated the GLS periodograms from the Nyquist frequency down to 0.001 d$^{-1}$. For plotting purposes, we show the GLS periodograms in the time interval between 0.5\,d (2\,d$^{-1}$) and 1000\,d (0.001\,d$^{-1}$)}. The peak at 1\,d, although visible in all periodograms of Fig.~\ref{fig:GLS_phot_multiplot}, is ascribed to the effects of the observing window (typical of ground-based observations) and will not be considered further.

The ASAS-SN data of the same filter ($V$ and $g$ bands) but taken with different cameras were combined by applying a relative offset so that any possible systematic effect is removed. The highest peak of the ASAS-SN $V$-band GLS periodogram (top panel of Fig.~\ref{fig:GLS_phot_multiplot}) occurs at {$\sim$70\,d} with a second significant peak at $\sim$40\,d. The ASAS-SN $g$-band data (second panel of Fig.~\ref{fig:GLS_phot_multiplot}) has the strongest peak at around 90\,d and a second strong peak at around 45\,d. Given the ratio 2:1 between the two peaks for each filter, it is likely that one of the peaks is the first harmonic of the other. The GLS periodogram of the TJO $R$-band photometry is rather puzzling (third panel of Fig.~\ref{fig:GLS_phot_multiplot}): the highest, significant peak is located at $\sim$133\,d, which we ascribed to a sampling {effect.} In order to see whether this strong peak is veiling other peaks at shorter periodicities (thus in better agreement with the \textit{TESS} and ASAS-SN data), we subtracted the 133 d signal from the original data and performed a new GLS analysis on the residuals. The resulting GLS periodogram is shown in the fourth panel of Fig.~\ref{fig:GLS_phot_multiplot}: there is one significant peak at around 30\,d and another peak at around 19\,d, which are near the periods found in our previous analysis of the \textit{TESS} LC (the TJO data were acquired immediately after \textit{TESS} sector 22). The GLS periodograms of the SNO $V$- and $R$-band LCs are very much alike (two bottom panels of Fig.~\ref{fig:GLS_phot_multiplot}), indicating that a period of $\sim$43\,d and a forest of smaller peaks between 10 and 30\,d are present in both data sets. {The $rms$ values of all ground-based photometric time series are summarized in Table~\ref{tab:phot_amplitudes}}.

For completeness, we combined the TJO and SNO $R$-band LCs and searched for the characteristic period of TOI-1238 by fitting a sinusoidal function, {following the same procedure as for the \textit{TESS} data}. The priors on the stellar rotation period are uniform in the interval 30--150\,d. The best model has a period of 40.3$\pm$0.4\,d, which agrees with all values previously derived from the various photometric time series (\textit{TESS}, ASAS-SN, and SNO $V$ band). The amplitude of the $R$-band stellar variability is 3.0\,mmag, which is small and suggests that TOI-1238 is a quiet star. Figure~\ref{fig:TJO_SNO_phot_phase} displays the TJO and SNO $R$-band photometry folded in phase with the periodicity of 40.3\,d.

Our conclusion is that from all ASAS-SN, TJO, and SNO photometric data sets covering a total of 6.8\,yr of regular monitoring, it is not possible to extract one single, characteristic rotation period for TOI-1238. Instead, we constrained the rotation period of this star to be in the interval between 30 and 45\,d. This is likely explained by the presence of differential rotation. In the following subsection, we demonstrate that the spectroscopic activity indicators also point to a rotation period within this range.

\subsubsection{CARMENES activity indicators and radial velocities \label{spec-indicators}}

We computed the GLS periodograms of some stellar activity indicators included in the CARMENES {\tt serval} pipeline. All ``spectroscopic'' GLS periodograms are shown in panels 5--10 of {Figs.~\ref{fig:GLS_activity_multiplot} and~\ref{fig:GLS_activity_multiplot_freq}}. The top panel of {Fig.~\ref{fig:GLS_activity_multiplot}} displays the window function of the CARMENES observations, where the most significant peaks are located at 1\,d and $\sim$1\,yr.

There are three Ca\,{\sc ii} IRT indices, one per atomic line; their corresponding GLS periodograms are depicted together in the fifth panel of Fig.~\ref{fig:GLS_activity_multiplot}. The dLW index periodogram is shown in the ninth panel of the same figure. The periodograms of the Ca\,{\sc ii} IRT and dLW all show peaks at $\sim$40--45\,d reaching FAP levels above the 0.1\,\%~(Ca\,{\sc ii} IRT) and the 10\,\%~(dLW) significance levels. In both cases, the peak is quite broad with a pedestal extending from $\sim$30 through $\sim$60\,d for the dLW index ({marked} with a greenish band in all panels of Figs.~\ref{fig:GLS_phot_multiplot} and~\ref{fig:GLS_activity_multiplot}). The 40 d feature of the Ca\,{\sc ii} IRT indices is asymmetric. {In order to explore the presence of a signal at half of the stellar rotation period, we removed the 40 d signal (by using a sinusoid function) from the original time series} (shown in the sixth panel of Fig.~\ref{fig:GLS_activity_multiplot}). Albeit with less significance, these residuals have a peak at $\sim$19\,d (about half of the main peak), a pattern that was already seen in the photometric LCs and which hints at the presence of active regions on different locations of TOI-1238's surface. 

The GLS periodogram of the H${\alpha}$ index (bottom panel of Fig.~\ref{fig:GLS_activity_multiplot}) shows no significant peak, which is consistent with TOI-1238 being an H$\alpha$ inactive star. This is further corroborated by the fact that the H$\alpha$ line is observed in absorption in TOI-1238  \citep{2018A&A...614A..76J}. The CRX times series shows a linear trend in the first 2.5 months of observations (Fig.~\ref{fig:toi1238_activity_time}), which is reflected as a long-term periodicity in the GLS periodogram of the seventh panel of Fig.~\ref{fig:GLS_activity_multiplot}. After removing this trend from the data, the resulting GLS periodogram of the CRX residuals (eighth panel of Fig.~\ref{fig:GLS_activity_multiplot}) has no significant peaks above any of the three defined FAP levels. We also investigated correlations between the measured CARMENES RVs and all the individual activity indices provided by the {\tt serval} pipeline {(including He\,{\sc i}\,D$_3$, He\,{\sc i} $\lambda$10833\,\AA, Pa$_\beta$, the Na\,{\sc i}\,D doublet, and the TiO and VO bands)} using Pearson's $r$ and $p$ coefficients to detect correlations and to access the significance of the correlation. We found that there are no strong, or even moderate correlations between the RVs and the activity indices. The lack of correlation or anticorrelation is also confirmed when considering the data of the first season only.

In conclusion, the CARMENES dLW and Ca\,{\sc ii} IRT indices suggest that the characteristic frequency for the stellar variations is around $\sim$40--45\,d, which agrees with the results obtained from the analysis of the coeval TJO and SNO LCs (Sect.~\ref{asas-tjo-sno}). Interestingly, there is no significant peak at this periodicity in the GLS periodogram of the RVs illustrated in Fig.~\ref{fig:toi1238_activity_time}, but the RVs show a strong decreasing trend. We proceeded to remove this trend by fitting a sinusoidal function with a long period of $\sim$800\,d (see {Sect. \ref{sec:radvelmodel}).} The GLS periodogram of the RV residuals is illustrated in the third panel of Fig.~\ref{fig:GLS_activity_multiplot}, where the orbital periods of the two transiting planets become the most prominent peaks. Clearly, the Keplerian signals are present in our CARMENES RV data, thus allowing us to characterize the TOI-1238 planetary system (see next sections). The stellar activity signal expected at $\sim$30--45\,d, although present, it is not significant. Only when the planetary imprint is removed from the RV data, the signal at $\sim$45\,d gains power (fourth panel of Fig.~\ref{fig:GLS_activity_multiplot}) and has a broad structure with a width similar to that of the other spectroscopic indices. Because this periodicity coincides with that observed from the photometric data, we ascribe it to the rotation period of TOI-1238. {To summarize, taking into account all photometric and spectroscopic indices, we conclude that the rotation period of TOI-1238 is $P_{\rm rot}$=40$\pm$5\,d.}

\begin{figure}[]
\centering
\includegraphics[width=0.49\textwidth]{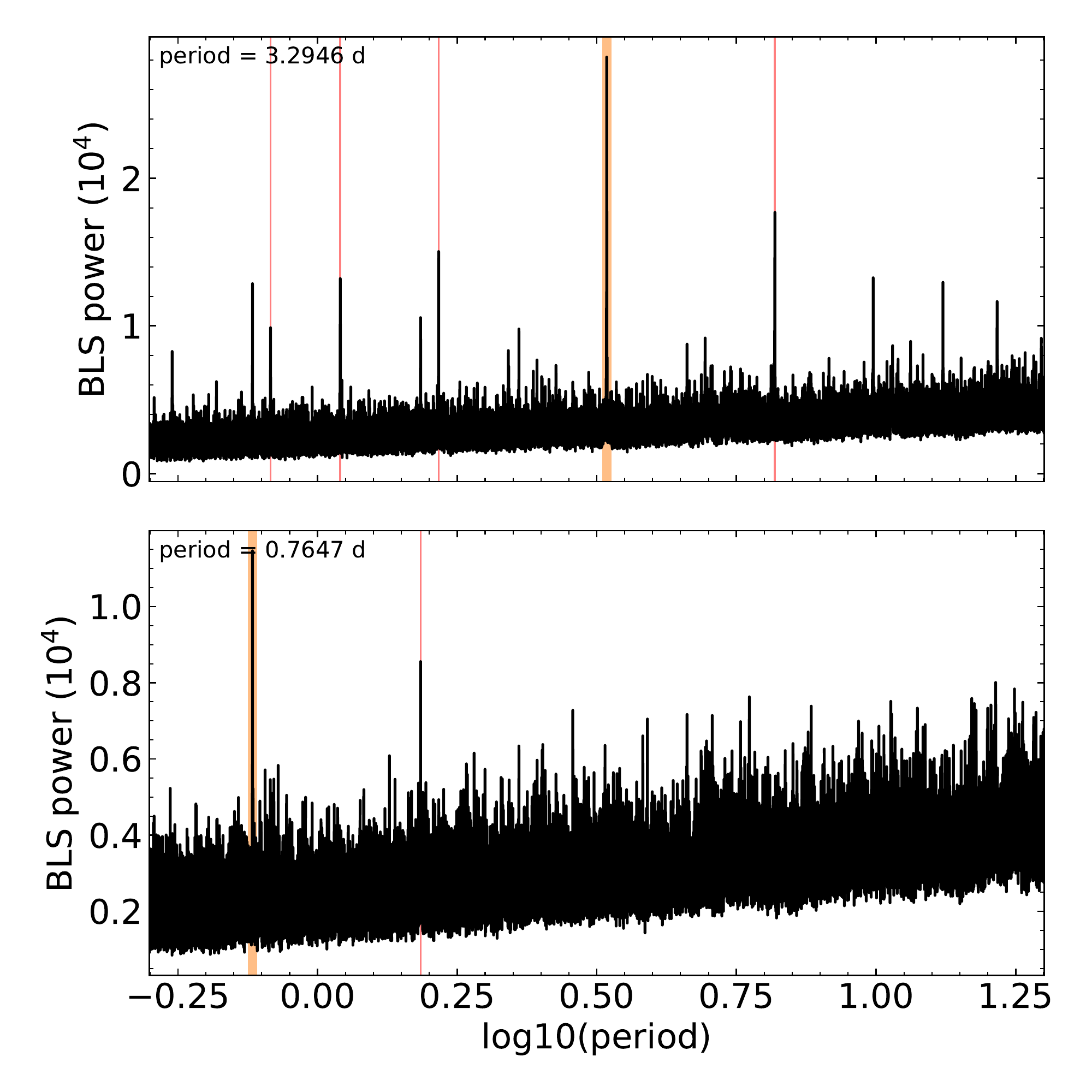}
\caption{BLS periodograms of the four \textit{TESS} sectors. {\sl Top panel:} Detection of the 3.29 d transiting planet (strongest peak marked with an orange vertical line). Various harmonics are also identified with vertical thin red lines. {\sl Bottom panel:} Detection of the 0.76 d transiting planet (strongest peak) after masking out the previous signal. The only harmonic visible in the plot is marked with a thin red line. }
\label{fig:BLS_4LCs}
\end{figure}

\section{TOI-1238 planetary system}
\label{section4}

\subsection{\textit{TESS} transit analysis}
\label{subsec:TESS_light_curve}

We applied the box least squares (BLS) periodogram \citep{2002A&A...391..369K, 2016A&C....17....1H} to the \textit{TESS} time series data to verify that the announced planets are there and to search for additional transits that may have been missed by the TESS pipeline. We employed the BLS algorithm programmed in the {\tt astropy.timeseries python} package and made the process iterative: once a transiting planet candidate is identified, it is masked out and the algorithm is run again to search for additional transiting planets, from the strongest to the weakest signals. The BLS periodograms of the PDCSAP fluxes of the four sectors computed for periodicities between 0.5 and 100\,d are depicted in Fig.~\ref{fig:BLS_4LCs}, although in the diagram we only show up to 20\,d for clarity. The first transiting planet candidate to be detected in the \textit{TESS} data has a BLS peak at 3.29\,d, and the second candidate appears at 0.76\,d. No other significant BLS peaks were identified after these two candidates were properly masked out from the LCs. The python package also delivered first estimates of the periastron passage ($t_0$) and the transit depth ($t_{\rm depth}$) for the two candidates, all of which agree with the \textit{TESS} alerts on TOI-1238. From now on, we name the 0.76 d and the 3.29 d transiting planets TOI-1238\,b and TOI-1238\,c, respectively.

\begin{figure}[]
\centering
\includegraphics[width=0.49\textwidth]{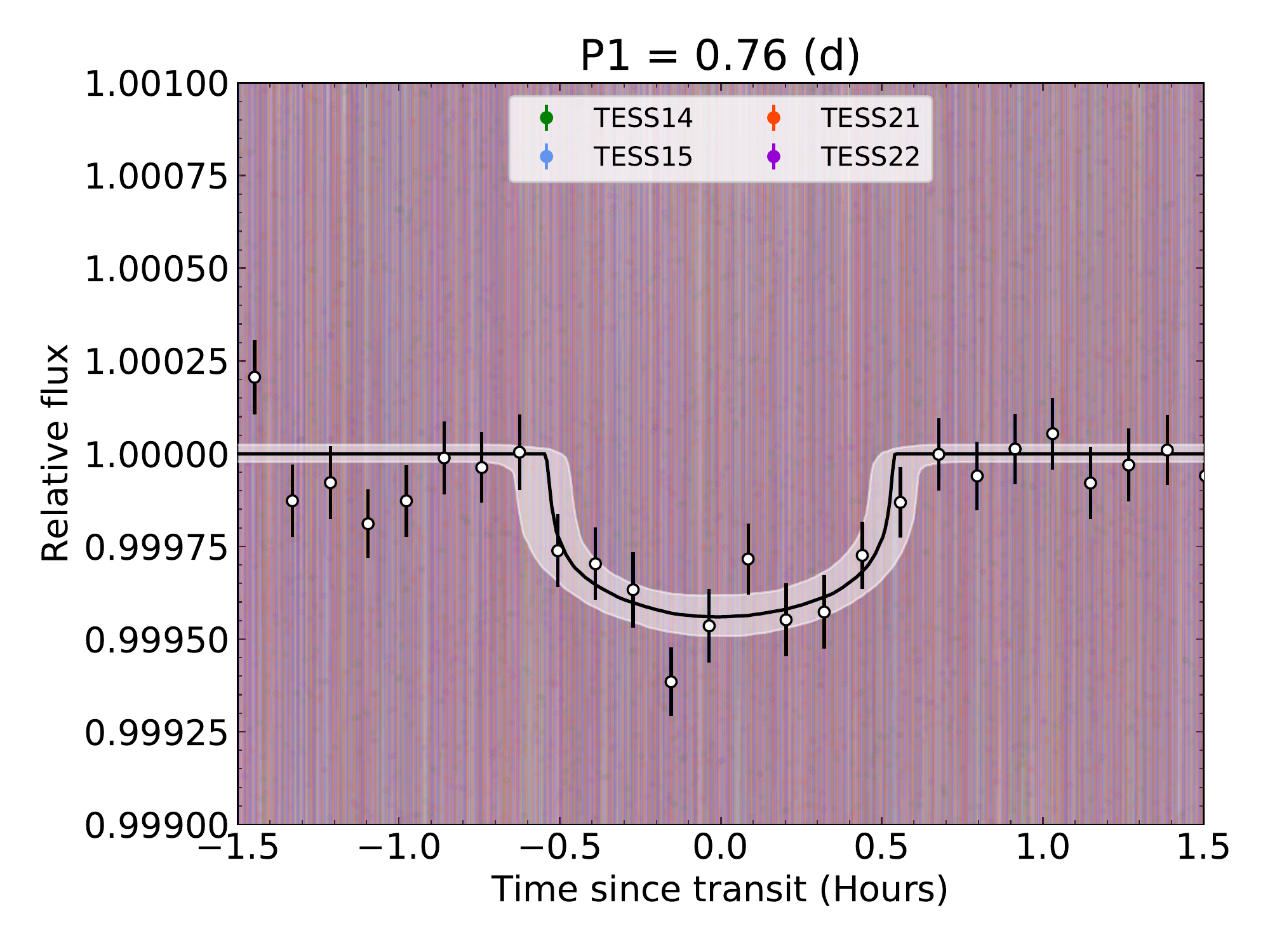}
\includegraphics[width=0.49\textwidth]{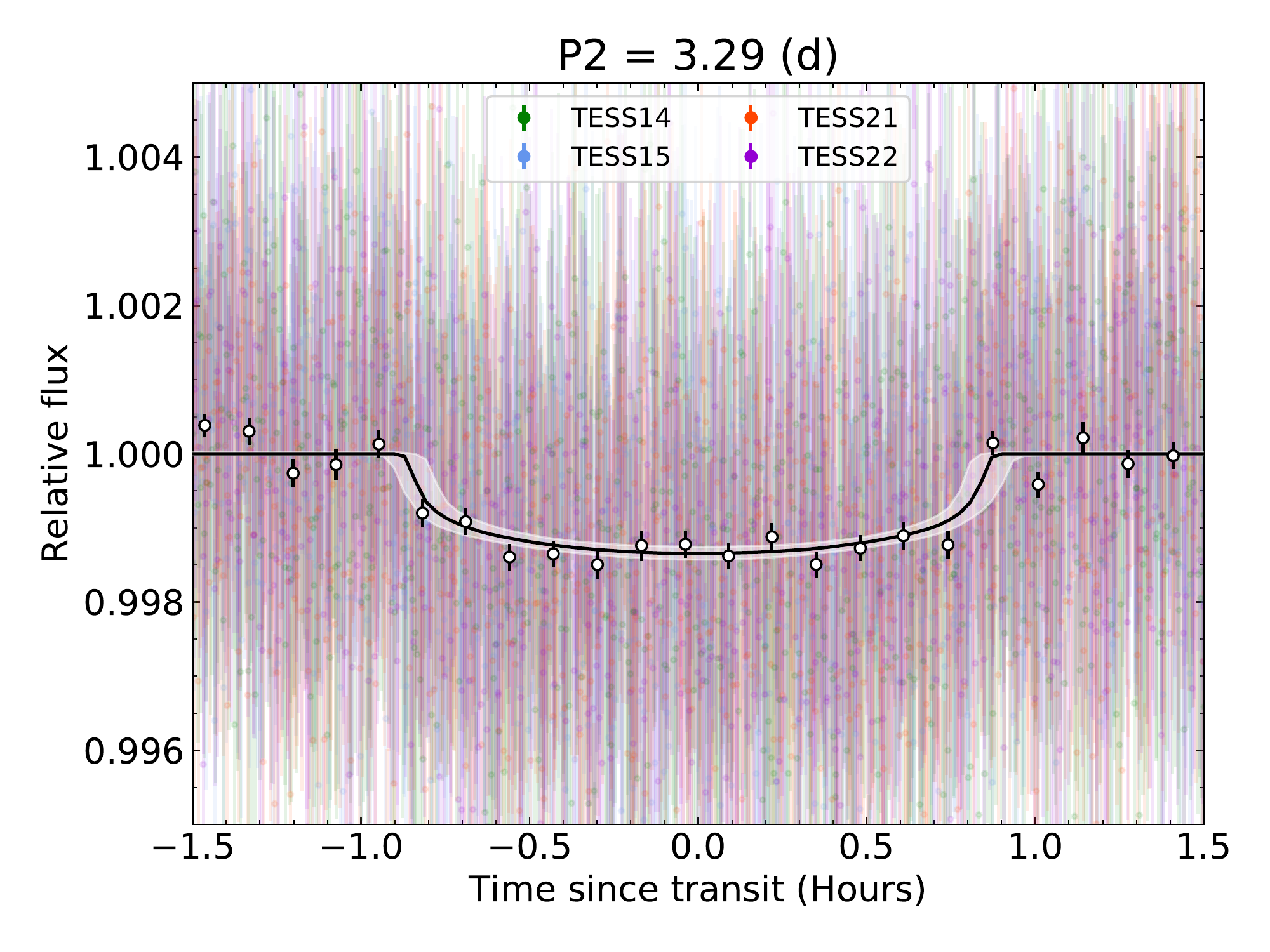}
\caption{\textit{TESS} LCs folded in phase with the orbital periods of TOI-1238\,b (top panel) and TOI-1238\,c (bottom panel). Binned data are plotted as white circles. The best transit model for each planet is shown with a black line (associated uncertainty is illustrated with a whitish color). }
\label{fig:lc_vs_phase_LC-fit_ALL_4LCs}
\end{figure}

\begin{table}[]
\renewcommand{\arraystretch}{1.5}
\begin{tiny}
\centering
\caption{Transit parameters of TOI-1238\,b and TOI-1238\,c from the \textit{TESS} LC fit only.}
\label{tab:toi1238bc_params_from_LC-fit}
\begin{tabular}{l c c c}

\hline
\hline
\noalign{\smallskip}

Parameter &             TOI-1238\,b & TOI-1238\,c \\
\noalign{\smallskip}    
\hline  
\noalign{\smallskip}
\multicolumn{3}{c}{\textit{Fitted transit parameters}}\\
\noalign{\smallskip}
$P$  (d)                                                                                        &$0.764596^{+0.000016}_{-0.000013}$ &  $3.294733^{+0.000040}_{-0.000047}$\\
$t_0$ (BJD-2,457,000)    &$1684.1022^{+0.0027}_{-0.0034} $ & $1707.3521^{+0.0020}_{-0.0018}$\\
$r_1$                                                                                                   &$0.45^{+0.17}_{-0.16}$ & $0.55^{+0.06}_{-0.10}$\\
$r_2$                                                                                           &$0.04^{+0.003}_{-0.002}$ & $0.07^{+0.003}_{-0.003}$\\

\noalign{\smallskip}    
\hline  
\noalign{\smallskip}
\multicolumn{3}{c}{\textit{Derived transit parameters}} \\
\noalign{\smallskip}
$R_{p}/R_{\star}$                                               & $0.0190^{+0.0014}_{-0.0013}$ & $0.0333^{+0.0015}_{-0.0016}$\\
$R_{\rm p}$ (R$_{\oplus}$)   & $1.20^{+0.09}_{-0.08}$ & $2.11^{+0.09}_{-0.10}$\\
$a/R_{\star}$                                                                   & $5.084824^{+0.000079}_{-0.000068}$ & $13.464833^{+0.00016}_{-0.00017}$\\
$a$ (au)                                                                                        &$0.01370587^{+0.00000021}_{-0.00000018}$ & $0.03629373 ^{+0.00000043}_{-0.00000045}$\\
$b=(a/R_{\star})\cos{i}$                                & $0.31^{+0.21}_{-0.20}$ & $0.43^{+0.07}_{-0.13}$\\
$t_{\rm 14}$ (h) $^{(*)}$                                       & $1.12^{+0.05}_{-0.10}$ & $1.76^{+0.09}_{-0.06}$\\
$t_{\rm depth}$ (ppm)                                   &  $359^{+53}_{-46}$ & $1114^{+101}_{-104}$\\
$i$ (deg)                                                                                       &        $86.5^{+2.2}_{-2.3}$ & $88.2^{+0.5}_{-0.3}$\\

\noalign{\smallskip}
\hline
\end{tabular}
\tablefoot{$^{(*)}$ $t_{\rm 14}$ is the transit duration defined as the time between the first and fourth contact.}
\end{tiny}
\end{table}

An inspection of the bottom panels of Fig.~\ref{fig:SAP_and_PDCSAP} shows some structure in the \textit{TESS} LCs, particularly in Sector 22. This may have an impact on the determination of the planetary parameters, particularly for shallow transits. We therefore flattened the \textit{TESS} LCs using Gaussian process (GP) {regressors} with the 
{\tt celerite} M\'atern kernel \citep{2013PASP..125..306F} of the form
\begin{equation}
\lim_{\epsilon \to 0} k_{i,j}(\tau) = \sigma^2 \left( 1 + \frac{\sqrt{3} \tau}{\rho} \right) \exp  \left( \frac{-\sqrt{3}\tau}{\rho} \right),
\end{equation}
where $\tau= \abs{t_{i}-t_{j}}$ is the time-lag, $\sigma$ is the amplitude of the GP, and $\rho$ is the time/length-scale of the GP. All planetary transits were masked out. The fit was done using the {\tt juliet} package, which also required the following parameters: a dilution factor to account for possible contaminating sources in the photometric aperture, and a jitter to be added to the nominal flux error bars. We adopted a dilution factor of 1.0, that is, no contamination in the \textit{TESS} photometry that may mimic a possible planetary transit {(see Sect.~\ref{subsec:TESS photometric time serie} and Fig.~\ref{fig:apertures})}. The priors on the GP $\sigma$ and $\rho$ parameters were based on log-uniform distributions with wide open intervals. We modeled the LCs of all four sectors finding that the models were essentially flat for sectors 14, 15, and 21 (Fig.~\ref{fig:SAP_and_PDCSAP}). We used the GP model to ``detrend'' the data of sector 22. This flattened sector was added to the other three sectors for the characterization of the transiting planets.

In a next step, we modeled the planetary transits from the flattened photometric data to measure the orbital periods ($P$) and relative planet-to-star sizes ($R_p/R_*$), the time-of-transit center of the planet ($t_0$), the inclination of the planetary orbital planes ($i$), and the star–planet distances ($a/R_*)$. The {\tt juliet} package uses the {\tt batman} package \citep{2015PASP..127.1161K} to this end. The stellar limb-darkening coefficients (quadratic law) were parameterized following \cite{2013MNRAS.435.2152K}. We also used the $r_1$ and $r_2$ parameterization \citep{2019MNRAS.490.2262E} instead of determining the relative radii and the impact parameters ($b$) of the planets directly:  $r_1$ and $r_2$ can vary between 0 and 1 and are defined to explore all the physically meaningful ranges for $R_p/R_*$ and $b$. We defined a prior on the stellar density, $\rho_{\star}$ (in kg\,m$^{-3}$) instead of the scaled semimajor axis of the planets, $a$. In this way, a single value of $\rho$ is defined for the system. We assumed that the two transiting planets candidates are in circular orbits. The median of the posterior distributions for each fitted parameter is given in the upper part of Table~\ref{tab:toi1238bc_params_from_LC-fit}, while the bottom part presents the derived transit parameters. Quoted error bars correspond to 1$\sigma$ uncertainties. The outer transiting planet candidate {is twice the size of} the inner planet candidate and is located only 2.6 times farther away from their parent star. Within the error bars, the two objects are in coplanar orbits. The \textit{TESS} LCs folded in phase with orbital periods are illustrated in Fig.~\ref{fig:lc_vs_phase_LC-fit_ALL_4LCs}.

\begin{figure*}[]
\centering
\begin{minipage}{0.48\linewidth}
\includegraphics[angle=0,scale=0.38]{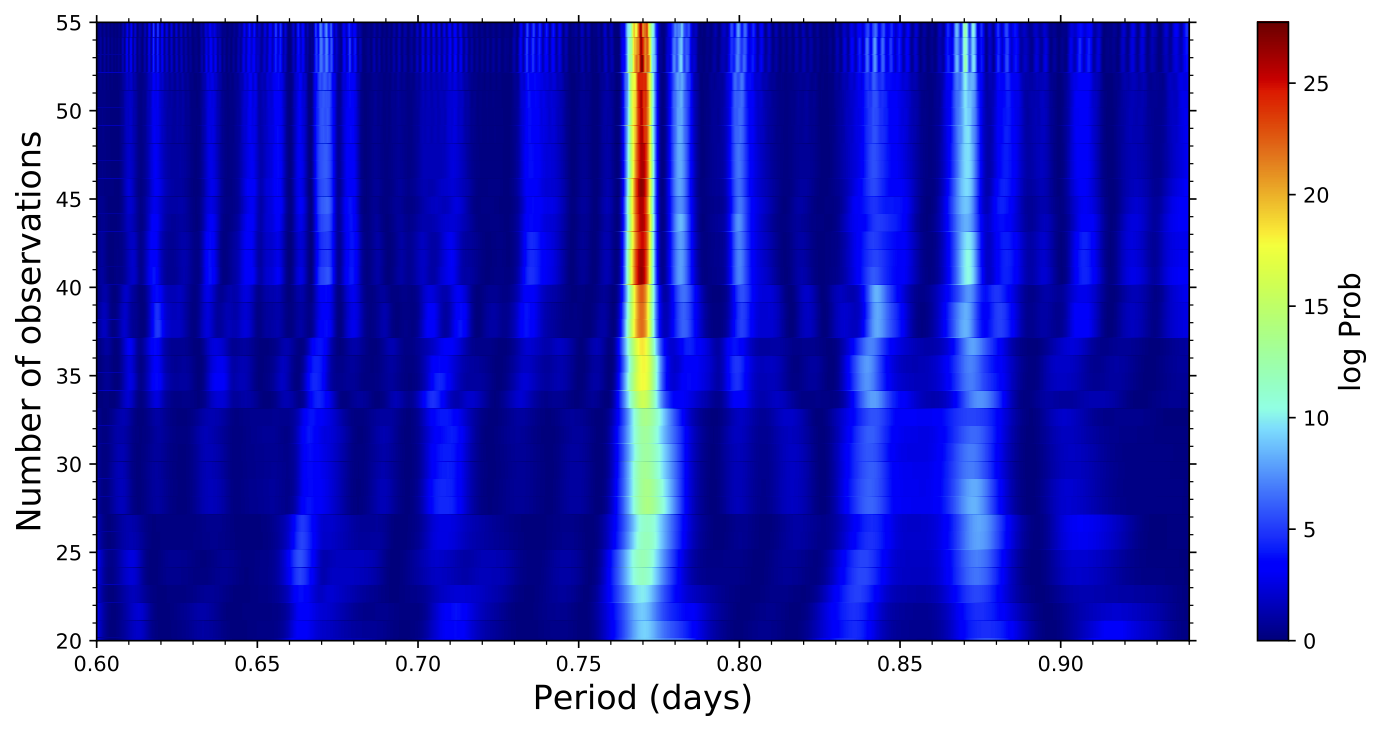}
\end{minipage}
\begin{minipage}{0.48\linewidth}
\includegraphics[angle=0,scale=0.38]{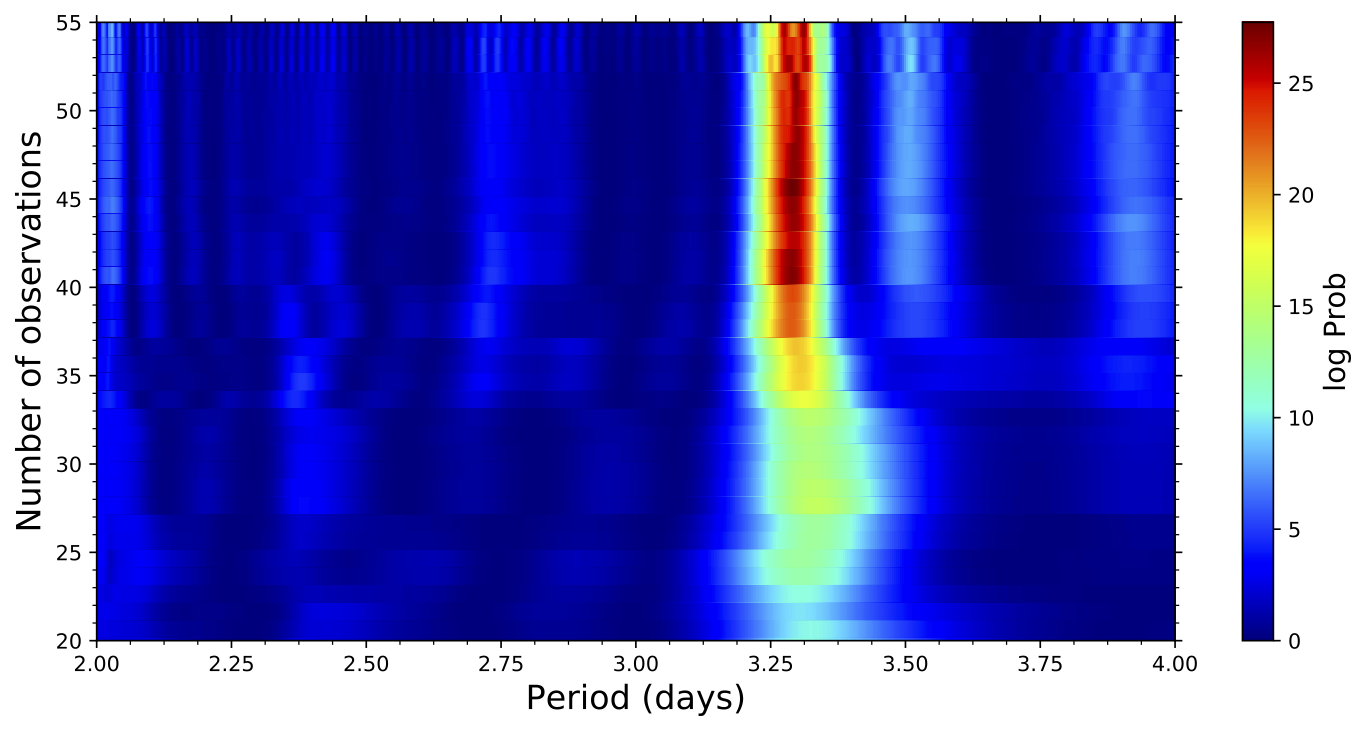}
\end{minipage}
\caption{Evolution of the s-BGLS periodogram of the CARMENES RV data around 0.764\,d (left) and 3.294\,d (right). We used the RV data after subtracting the long-term signal (see text).}
\label{fig:toi1238_color_s-BGLS}
\end{figure*}

\subsection{CARMENES radial velocity analysis}
\label{subsec:CARMENES RV data}

\subsubsection{Pre-whitening}
\label{subsubsec:CARMENES pre-whitening}

The CARMENES RVs show a long-term linear trend whose origin is discussed in Sect.~\ref{external-planet}. After removing this trend, the GLS periodogram of Fig.~\ref{fig:GLS_activity_multiplot} shows strong peaks at the expected planetary orbital periods. This confirms that the \textit{TESS} transit-like features are planetary in nature.

To identify the presence of possible aliasing phenomena in the CARMENES RV data caused by the gaps in the time sampling of the observations \citep[e.g.,][]{2010ApJ...722..937D,2020A&A...636A.119S}, we took into account the spectral window displayed in the top panel of Fig.~\ref{fig:GLS_activity_multiplot}. The strong peaks of the window function may introduce alias peaks in the RV periodogram at frequencies according to the expression $f_{\rm alias}$ = $f_{\rm true} \pm m f_{\rm window}$, where $m$ is an integer, $f_{\rm true}$ is the frequency identified in the RV periodogram and $f_{\rm window}$ the frequency from the window function \citep{1975Ap&SS..36..137D}. Typical aliases affecting ground-based observations are those associated with the year, synodic month, sidereal day, and solar day. In our spectroscopic window function, the highest peaks occur at $\sim$1\,d and at periodicities beyond 300\,d or 1\,yr. We checked that the latter does not introduce any peak in the RV periodogram that can be misinterpreted. However, the orbital periods of TOI-1238\,b and c are, coincidentally, the $\sim$1 d alias of each other. This implies that, without the assistance of any other data free of this aliasing effect, using the RV time series alone will not deliver reliable masses for any of the two transiting planets. As an example, to produce the GLS periodogram of the fourth panel of Fig.~\ref{fig:GLS_activity_multiplot}, where the planetary signal is removed from the RV data using a simple sinusoidal function, we only subtracted the peak at 3.29\,d and the 0.76\,d was automatically subtracted, too. Fortunately, the combined \textit{TESS}--CARMENES data analysis of Sect.~\ref{sec:planet orbiting toi-1238} will overcome this issue.

We verified that the RV signals at 0.76\,d and 3.29\,d are stable and coherent over the entire observational time baseline by producing the stacked Bayesian generalized Lomb-Scargle (s-BGLS) periodogram \citep{2015A&A...573A.101M} shown in Fig.~\ref{fig:toi1238_color_s-BGLS}. For this, we employed the CARMENES RVs free of the long-term trend. The significance or probability of both planetary signals increases with time until a stability is reached at a certain number of observations. Then, the signals become narrower, which is a behavior expected for signals with a Keplerian origin. This provides further support for the exoplanetary nature of TOI-1238\,b and c.


\subsubsection{Nature of the long-term RV signal \label{external-planet}}

TOI-1238 is a K7--M0 star with a projected rotational velocity of $\leq$\,2\,km\,s$^{-1}$, {which is} consistent with our inferred rotation period of 40$\pm$5\,d, typical for field stars of this type. The decreasing, long-term trend of the CARMENES RVs displayed in the top panel of Fig.~\ref{fig:toi1238_activity_time} may have two possible origins: a companion moving around the central star at a long orbital period or a magnetic activity cycle intrinsic to the star. Our spectroscopic data yield a lower limit on the period of such variation, $P \ge 600$\,d {(twice the CARMENES time coverage)}, and a minimum RV amplitude of 70\,m\,s$^{-1}$ {(half the observed RV peak-to-peak variation)}.

The GLS periodogram of the combined ASAS-SN $V$ and $g$ LCs (covering more than 5\,yr of observations) is not conclusive (the high dispersion of the photometry is an issue). We also combined all TJO and SNO photometry by applying suitable offsets to the individual LCs and thus covering about 300\,d of continuous observations, but this time baseline is insufficient to sample one full cycle length of the long-term RV signal. The CARMENES CRX and H$\alpha$ indices and all Ca\,{\sc ii} IRT indices after subtraction of the signal at the stellar rotation period appear to show a shallow slope with time, but it is not possible to relate these variations to the long-term trend seen in the RV time series. Also, the \textit{GALEX} strong NUV emission suggests significant chromospheric activity, thus leaving open the possibility that the $P \ge 600$\,d RV feature could be due to a strong magnetic activity cycle of TOI-1238.

However, the large RV amplitude of the long-term variation ($\approx$140\,m\,s$^{-1}$ from peak to peak) suggests that the most likely scenario is that of the companion. With the exception of young stars (less than several hundred Myr), which show RV variations of tens to hundreds of m\,s$^{-1}$ due to rather strong and rapid stellar activity (\citealt{2020A&A...640A..48L, 2021arXiv211109193S}), no main sequence star with slow rotation period appears to have such high RV variability caused by atmospheric changes. TOI-1238 is likely an old star based on its rotation period and Galactic kinematics. The age of TOI-1238 inferred from gyrochronology relations is likely $>$1\,Gyr (see Fig.~13 of \citealt{2007ApJ...669.1167B}). Furthermore, H$\alpha$ is not seen in emission, and the photometric amplitude of the stellar variations in the \textit{TESS}, TJO, and SNO data is on the order of a few to several mmag.

We also inspected the CARMENES NIR data. These RVs, despite their large error bars and dispersion, are useful for studying the large-amplitude, long-term trend. The downward slope of the RV time series is present in the NIR as well. A joint RV analysis of the CARMENES VIS and NIR data to fit this feature only (following the prescription described in Sect.~\ref{sec:radvelmodel}) revealed that both RV curves share the same amplitude and period. The constant amplitude with wavelength is expected for a Keplerian signal, thus supporting the companion scenario.

\begin{figure}
\centering
\includegraphics[width=0.48\textwidth]{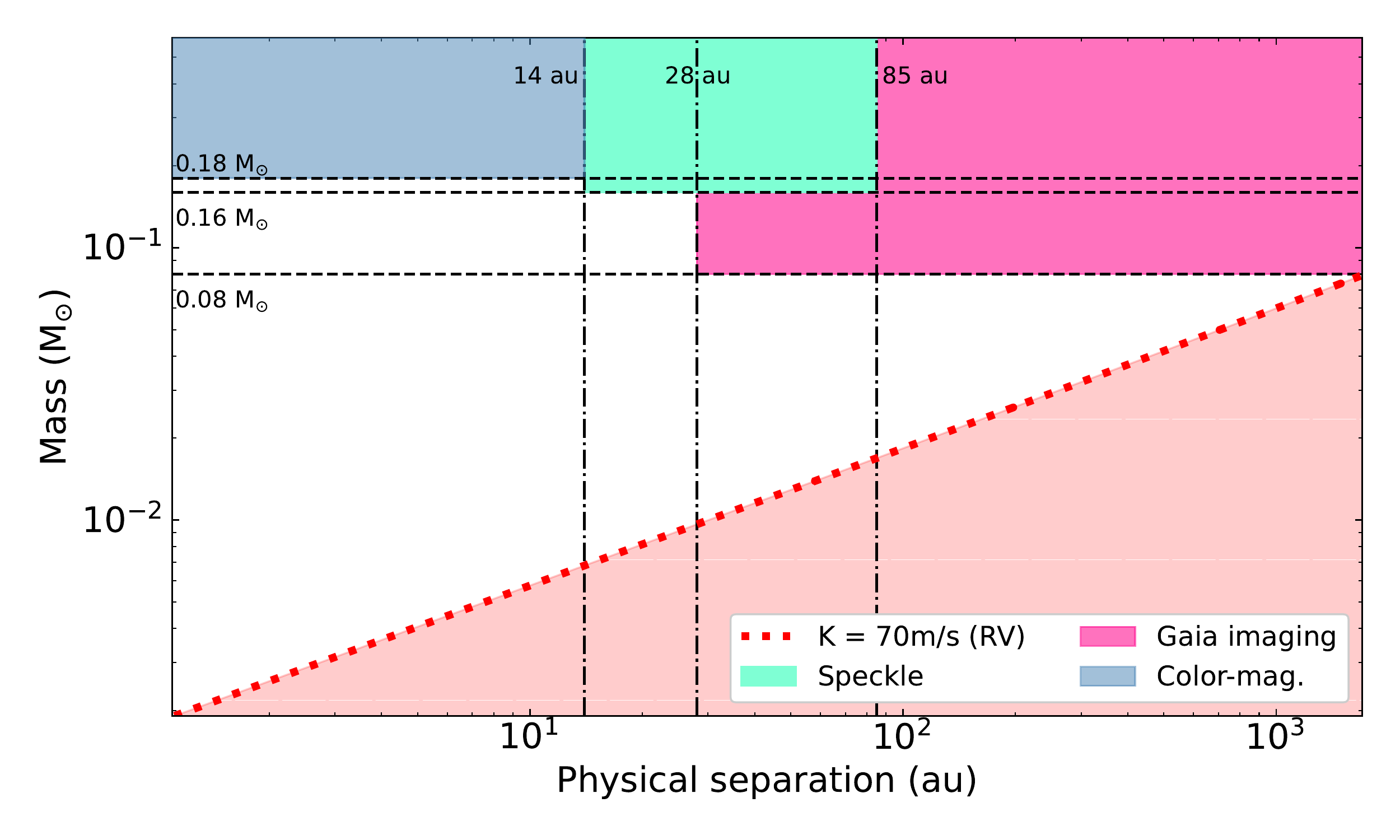}
\caption{Constraints on the mass and semimajor axis of the external companion to TOI-1238 based on photometric, astrometric, and {spectroscopic} observables. Colored areas represent regions excluded by the various observations (see legend and text). Possible masses and orbital separations of this companion are contained within the white area. The limits of the ordinate axis are given by the possible minimum mass of the companion (in a circular orbit) and the stellar mass, while the limits of the abscissa axis are determined by the minimum and maximum orbital separations.}
\label{fig:mass_limit}
\end{figure}

Based on the spectroscopic constraints on the orbital period and RV amplitude of the putative outer companion of TOI-1238 and using Kepler's laws, we imposed the following minimum values on the companion's mass and orbital semimajor axis: $M \geq 2 \sqrt{1-e^2}$ M$_{\rm Jup}$ ($e$ is orbital eccentricity) and $a \geq 1.1$\,au. In what follows, we used all available photometric and astrometric observations to further constrain the companion's properties, particularly its maximum mass. Our results are graphically summarized in Fig.~\ref{fig:mass_limit}, where the constraints derived from the various observables are illustrated in the form of exclusion areas (colored regions) in the mass versus semimajor axis diagram of the figure. The axis of mass goes from the minimum possible mass of the companion in a circular orbit, 2\,M$_{\rm Jup}$, up to the stellar mass of TOI-1238 (0.59\,M$_{\odot}$). The axis of orbital separation starts at 1.1\,au and extends to 1700\,au, which is the value of the cross between the restriction imposed by Kepler's laws to satisfy $K \geq 70$ m\,s$^{-1}$ (plotted as orange triangle in Fig.~\ref{fig:mass_limit}) and another restriction described next.

From the Gemini high spatial resolution image (Fig.~\ref{fig:Gemini-N_image}), TOI-1238 has no stellar companion with {$\Delta m \approx$ 6\,mag} (contrast) at projected angular separations in the interval 0\farcs2--1\farcs2 (i.e., 14--85\,au at the distance of the system). To explore closer separations, we used the precise {\sl Gaia} photometry shown in Fig.~\ref{fig:main_sequence}, where it becomes evident that TOI-1238 is not overluminous with respect to the main sequence of early-M type stars: we thus set {$\Delta m$ = 0.05\,mag} as the maximum possible flux/luminosity contribution of the putative companion to the observed photometry of TOI-1238. Following the mass--magnitude relations of \citet{2013ApJS..208....9P} and \citet{2020A&A...642A.115C} valid for main sequence stars with solar metallicity, both the Gemini flux contrast and the {\sl Gaia} maximum flux contribution translate into masses of $\sim$0.16\,M$_\odot$ and 0.18\,M$_\odot$ (or spectral types in the interval M4--M5). {That is}, companions more massive than $\sim$0.18\,M$_\odot$ at $a < 14$\,au and companions more massive than $\sim$0.16\,M$_\odot$ at $a$ = 14--85\,au can be discarded. These exclusion regions are marked with blue and green colors in Fig.~\ref{fig:mass_limit}. 

{\sl Gaia} imaging and astrometry can also be used to exclude resolved companions to TOI-1238. \textit{Gaia} EDR3 data have an effective spatial resolution of 0\farcs4 and a limiting $G$-band magnitude of 21 (or $\approx$28\,au and $M(G)$ = 16.7 at the distance of the system) in the sky area of TOI-1238 \citep{2021A&A...649A...1G}. We explored all {\sl Gaia} detections in a radius of $2'$ around our star and none has the same proper motion and trigonometric parallax as TOI-1238, thus indicating that there are no proper motion companions identified down to the {\sl Gaia} detectability limit, corresponding to a mass of $\approx$0.077\,M$_{\odot}$ (close to the definition of the substellar mass limit for solar metallicity). This constraint is plotted as a pink exclusion region in Fig.~\ref{fig:mass_limit}. 

{\sl Gaia} EDR3 contains data spanning 34 months (1.4\,yr) of observations out of a nominal five-year duration mission. TOI-1238 has a {\sl Gaia} astrometric excess noise of 98\,$\mu$as, which does not stand out for being very high compared to that of other {\sl Gaia} objects of similar brightness in the same region of the sky, but it is not small either. Actually, this astrometric excess noise is detected with a $\sigma = 13$, which is significant in terms of {\sl Gaia} measurements according to \citet{2021A&A...649A...2L}. Because the time coverage of {\sl Gaia} EDR3 and our minimum limit on the orbital period of the putative external companion are alike, a measurable astrometric excess noise may hint at a relatively short orbital period on the order of a few to several years. We will need to wait for the completion of the {\sl Gaia} mission before we can conclude on the nature of the companion to TOI-1238 using precise astrometric measurements. Also, additional RV measurements will contribute to find a final solution for the mass and other orbital parameters of the putative external companion to TOI-1238. This mass is expected to be contained within the white area of Fig.~\ref{fig:mass_limit}: the outer companion of TOI-1238 could be a very low-mass star with $M \le 0.18$\,M$_\odot$, a brown dwarf or a massive planet.

\subsubsection{Radial velocity models}
 \label{sec:radvelmodel}
 
Despite the aliasing issue described in Sect.~\ref{subsubsec:CARMENES pre-whitening} that prevented us from determining the masses of the transiting planets from the RV data only, we modeled the CARMENES RV time series in order to constrain the most critical ingredients for a subsequent, more sophisticated, joint photometric and spectroscopic analysis of the TOI-1238 planetary system. This step was necessary to save computing time. The models were produced with the {\tt{juliet}} code that calls the {\tt radvel} \citep{2018PASP..130d4504F} package to model Keplerian RV signals. The stellar activity signals were modeled by means of a GP with an exp-sin-squared kernel provided by {\tt george} \citep{2015ITPAM..38..252A} python library, which is suited for learning periodic functions:
\begin{equation}
\label{eq:exp-sin-sqr_kernel}
k(\tau) = \sigma_{\rm GP}^2 \exp \left( -\alpha_{\rm GP} \tau^2 - \Gamma \sin^2 \left( \frac{\pi \, \tau}{P_{\rm rot}} \right)  \right)
,\end{equation}
where $\sigma_{GP}$ is the amplitude of the GP component given in m\,s$^{-1}$, $\Gamma$ is the amplitude of the GP sine-squared component and is dimensionless, $\alpha$ is the inverse length-scale of the GP exponential component given in $\rm d^{-2}$, $P_{\rm rot}$ the period of the GP quasi-periodic component given in days, and $\tau$ is the time lag.

\begin{table}[!t]
\begin{small}
\centering
\caption{Comparison of different {\tt juliet} models for TOI-1238 CARMENES RV data.}
\label{tab:toi1238_rv_model_comparison}
\begin{tabular}{l l l c c}

\hline
\hline
\noalign{\smallskip}
Model  & Description & $\ln \mathcal{Z}$ \\
\noalign{\smallskip}    
\hline  
\noalign{\smallskip}
BM                          & RV offset and jitter                  & -581.6 \\ 

BM$+$2pl$+$LT                       & LT long-term trend         & -210.5 \\ 

BM$+$2pl$+$GP                       & GP $P_{\rm rot} \sim 30-1500$\,d          & -187.5 \\ 

BM$+$3pl                            & External companion at $>$600\,d        & -184.9 \\ 

BM$+$3pl$+$GP                   & GP $P_{\rm rot} \sim 30-60$\,d       & -184.2 \\ 

\noalign{\smallskip}    
\hline  
\noalign{\smallskip}
\end{tabular}
\end{small}
\end{table}

We based the selection of the best model on the rules defined by \cite{2008ConPh..49...71T} based on the Bayesian model log-evidence, $\ln{\mathcal{Z}}$: if $\Delta \ln{\mathcal{Z}} \le 3$ the two models are indistinguishable and neither is preferred, while if $\rm \Delta \ln{\mathcal{Z}} > 3$ the model with the larger Bayesian log-evidence is favored. We performed four different approaches that are summarized in Table~\ref{tab:toi1238_rv_model_comparison}. All include one base model (BM) consisting of a RV offset and a RV jitter. All planetary orbits are assumed to be circular (with one exception indicated below). The other ingredients are as follows. The first is two Keplerian signals at 0.764\,d (TOI-1238\,b) and 3.294\,d (TOI-1238\,c) plus a linear trend to account for the RV long-term trend (BM$+$2pl$+$LT model). The second is two Keplerian signals at 0.764\,d (TOI-1238\,b) and 3.294\,d (TOI-1238\,c) plus the periodic GP with an open $P_{\rm rot}$ prior value to account for the RV long-term trend (BM$+$2pl$+$GP model).

The third is three Keplerian signals (TOI-1238\,b, TOI-1238c, and the external companion), {that is}, we assumed that the RV long-term trend is due to an external companion (BM$+$3pl model). This model is justified by the fact that the stellar rotation signal in the RV data appears to be lower in amplitude than the contribution of the two transiting planets (see Sect.~\ref{spec-indicators}). This model was explored with both a null eccentricity and a free eccentricity for planets TOI-1238\,b and c. The computations indicate that an eccentricity near zero is preferred. 

The final one is the same as above, with the addition of a GP model with $P_{\rm rot}$ in the interval 30--60\,d to simulate the stellar activity due to stellar rotation (BM$+$3pl$+$GP model).

The resulting log-evidence for each model is provided in Table~\ref{tab:toi1238_rv_model_comparison}. Both the BM$+$3pl and BM$+$3pl$+$GP models are equally valid. We interpret this as follows: the former model is successful because the impact and amplitude of the stellar variability at timescales of 30--45\,d in the RV time series are smaller than those of the planetary components. However, the BM$+$3pl$+$GP is physically more plausible because it includes all components observed in the data (two confirmed planets, a long-term signal, and stellar rotation). The GP model does not improve the fit in RVs significantly, and the increase in the number of free parameters is mathematically penalized in the log-evidence parameter. In any case, after the comparison of the models with the two and three Keplerians, the log-evidence criteria suggest that the RV long-term feature is better reproduced by a Keplerian function than by a periodic GP function or a single linear trend, which favors the presence of a third companion in the TOI-1238 system with relatively short orbital periods on the order of years.

\subsection{Masses of the transiting planets}
\label{sec:planet orbiting toi-1238}

\begin{figure*}[]
\centering
\includegraphics[width=\textwidth]{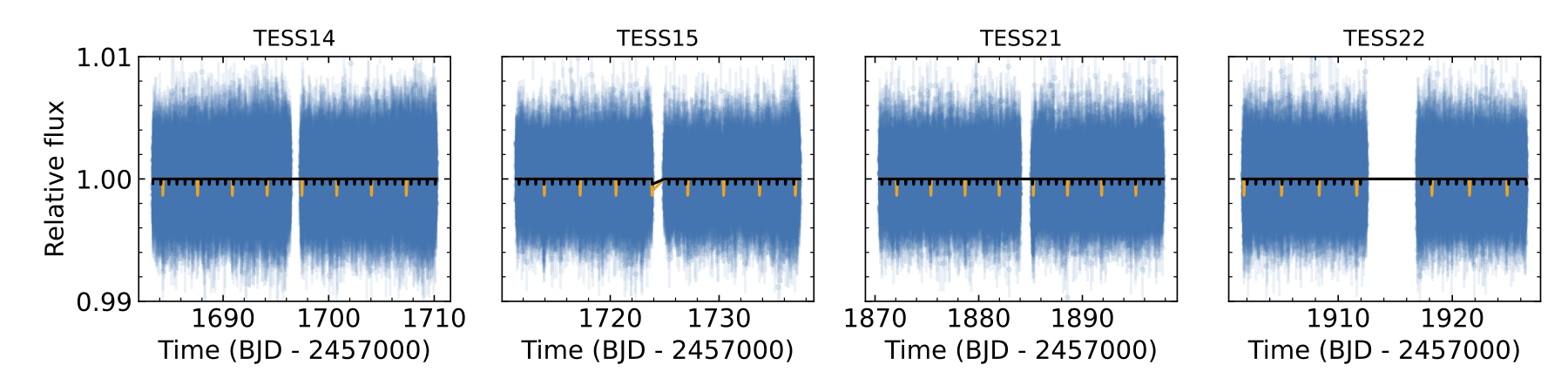}
\caption{Location of the transits of TOI-1238\,b (weakest, black line) and TOI-1238\,c (strongest, orange line) across all four \textit{TESS} sectors.}
\label{fig:lc_vs_time}
\end{figure*}

\begin{figure*}[]
\centering
\includegraphics[width=\textwidth]{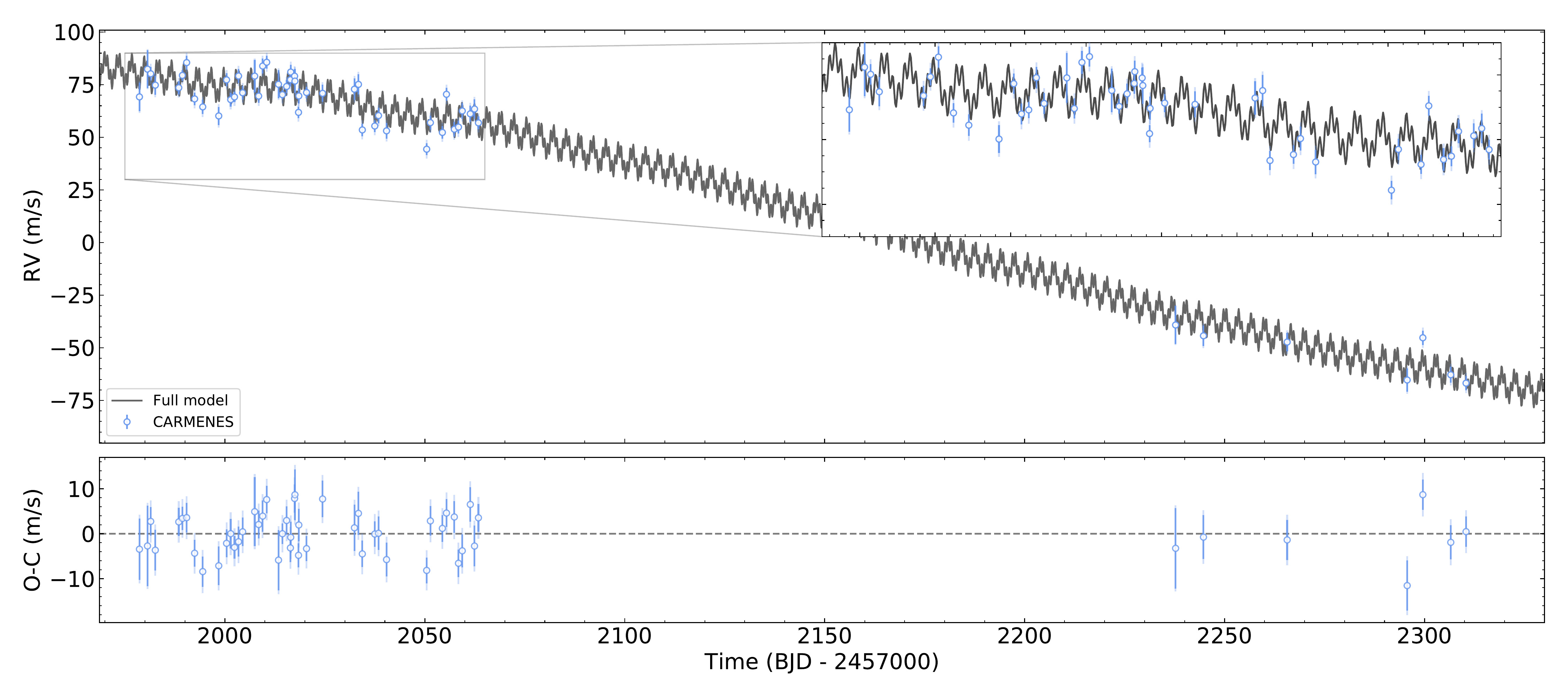}
\includegraphics[width=0.33\textwidth]{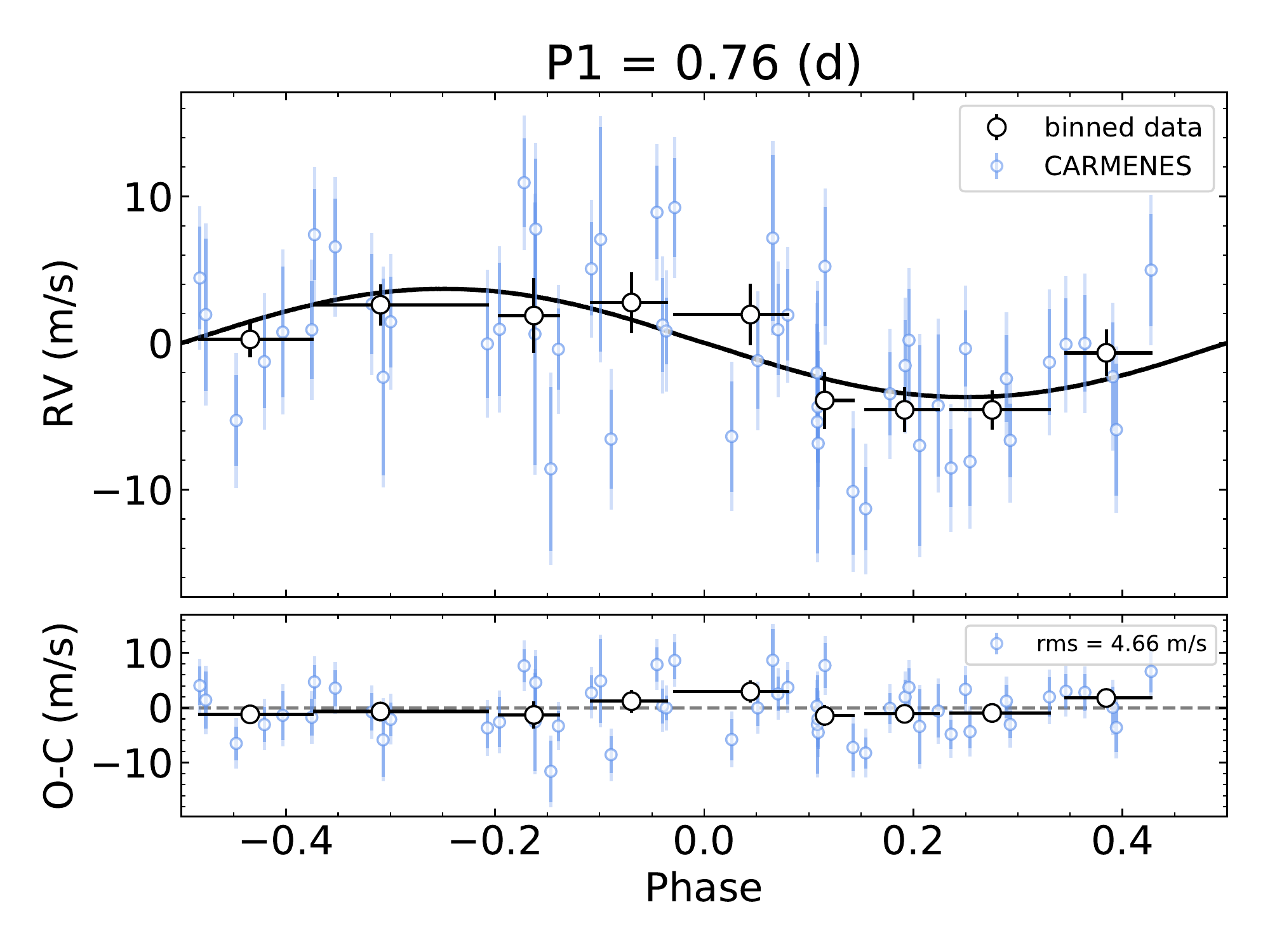}
\includegraphics[width=0.33\textwidth]{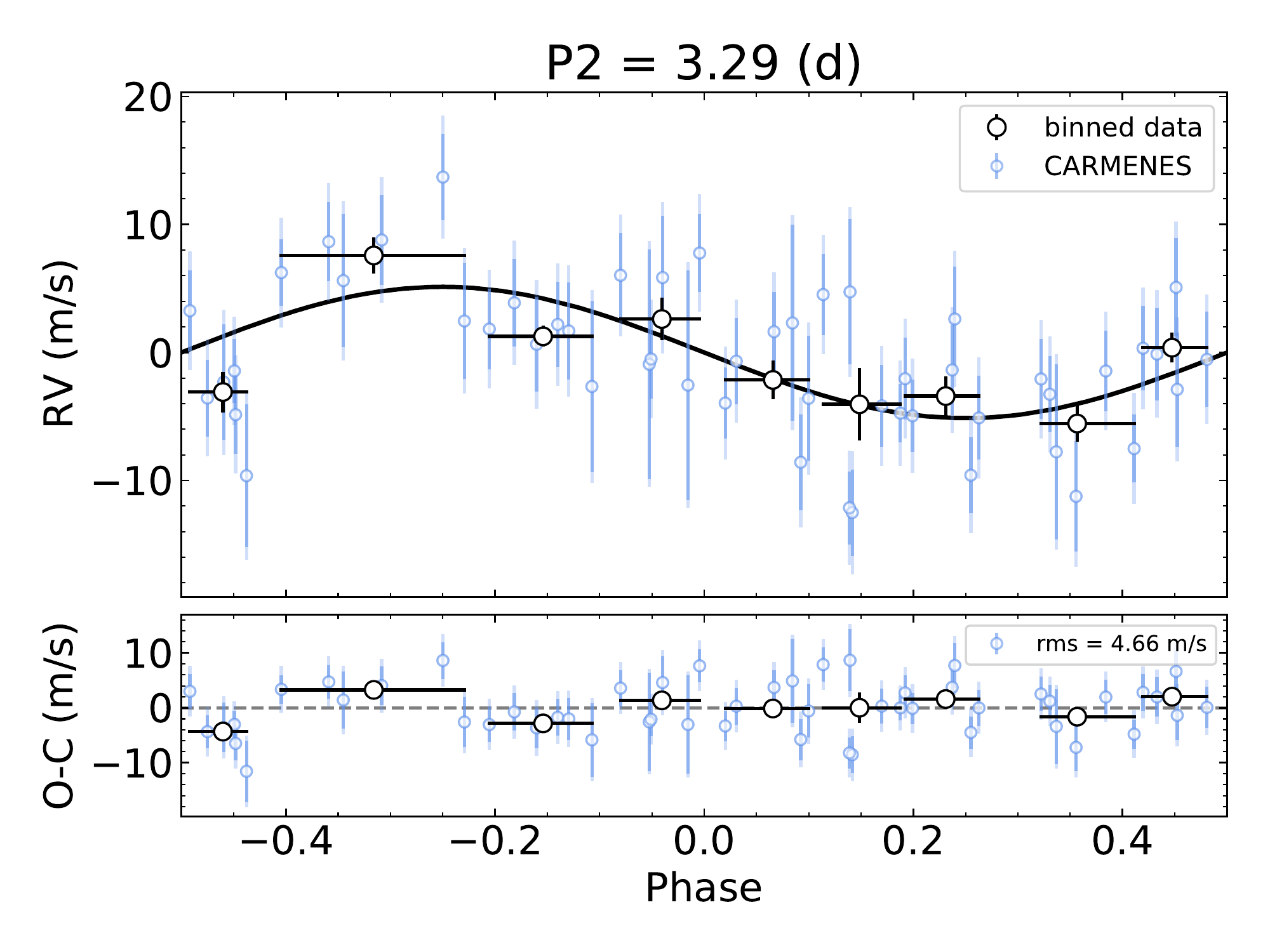}
\caption{TOI-1238 CARMENES RVs (blue dots) and the best model (black line) from the combined photometric and spectroscopic fit. The top panels shows the entire RV time series, where the short-period planets and the long-term feature are clearly visible. The inset displays an enlargement of the first part of the data for a zoomed-in  view of the two transiting planets' contribution to the modeled RV curve. The middle panel illustrates the RV residuals of the fit. The bottom panels present the CARMENES RVs folded in phase with the orbital period of the inner Keplerian components (from left to right: TOI-1238\,b and TOI-1238\,c). Binned data are plotted as open black symbols. All error bars include the quoted CARMENES uncertainties and the RV jitter as obtained from the model added in quadrature.
}
\label{fig:toi1238_rv_time-phase_plots}
\end{figure*}

\begin{table}[!tb]
\renewcommand{\arraystretch}{1.3}
\begin{scriptsize}
\centering
\caption{Final adopted parameters for the TOI-1238 system.}
\label{tab:toi1238bc_params_from_joit-fit}
\begin{tabular}{l c c c}

\hline
\hline
\noalign{\smallskip}

Parameter &              &  &  \\
\hline
\noalign{\smallskip}

\noalign{\smallskip}
\multicolumn{4}{c}{\textit{Photometric parameters} }\\
\noalign{\smallskip}
$\gamma_{\rm TESS, S14}$ ($10^{-6}$\,ppm)                       &    \multicolumn{3}{c}{$-1.7^{+19.1}_{-18.9}$} \\
$\gamma_{\rm TESS, S15}$ ($10^{-6}$\,ppm)                       &    \multicolumn{3}{c}{$-13.8^{+15.2}_{-16.4}$} \\
$\gamma_{\rm TESS, S21}$ ($10^{-6}$\,ppm)                       &    \multicolumn{3}{c}{$-24.3^{+14.8}_{-14.7}$} \\
$\gamma_{\rm TESS, S22}$ ($10^{-6}$\,ppm)                       &    \multicolumn{3}{c}{$-48.3^{+18.7}_{-19.3}$} \\

$\sigma_{\rm TESS, S14}$ (ppm)                                                  &    \multicolumn{3}{c}{$0.0002^{+0.0040}_{-0.0002}$} \\
$\sigma_{\rm TESS, S15}$ (ppm)                                                  &    \multicolumn{3}{c}{$0.0002^{+0.0040}_{-0.0002}$} \\
$\sigma_{\rm TESS, S21}$ (ppm)                                                  &    \multicolumn{3}{c}{$0.0001^{+0.0031}_{-0.0001}$} \\
$\sigma_{\rm TESS, S22}$ (ppm)                                                  &    \multicolumn{3}{c}{$0.0003^{+0.0051}_{-0.0003}$} \\

$q1_{\rm TESS}$                                                                         &    \multicolumn{3}{c}{$0.56^{+0.26}_{-0.27}$} \\

$q2_{\rm TESS}$                                                                         &    \multicolumn{3}{c}{$0.36^{+0.29}_{-0.22}$} \\

\noalign{\smallskip}
\multicolumn{4}{c}{\textit{RV parameters}} \\
\noalign{\smallskip}
$\gamma$ (m/s)                  &    \multicolumn{3}{c}{$-56.9^{+0.8}_{-0.9}$} \\
$\sigma$ (m/s)                          &    \multicolumn{3}{c}{$3.4^{+0.8}_{-0.9}$} \\

\noalign{\smallskip}
\multicolumn{4}{c}{\textit{GP hyperparameters}} \\
\noalign{\smallskip}
$\sigma_{\rm GP,RV}$ (m/s)                                                       &  \multicolumn{3}{c}{$24.6^{+16.3}_{-18.3}$} \\
$\alpha_{\rm GP,RV}$ ($10^{-6}$\,$\rm d^{-2}$)                                  &   \multicolumn{3}{c}{$10.1^{+101.7}_{-9.4}$} \\
$\Gamma_{\rm GP,RV}$                                                    &   \multicolumn{3}{c}{$0.006^{+0.142}_{-0.006}$} \\
$P_{\rm rot, GP,RV}$ (d)                                                                &   \multicolumn{3}{c}{$47.81^{+11.51}_{-8.77}$} \\

\noalign{\smallskip}    
\noalign{\smallskip}
\multicolumn{4}{c}{\textit{Stellar parameters}}\\

$\rho_{\rm \star}$ (kg $\rm m^{-3}$)            &                                        \multicolumn{3}{c}{$4510.6^{+436.5}_{-447.53}$} \\

\noalign{\smallskip}
\hline
\noalign{\smallskip}

Parameter         &             TOI-1238\,b & TOI-1238\,c & Ext. companion \\
\noalign{\smallskip}
\hline
\noalign{\smallskip}

\noalign{\smallskip}
\multicolumn{4}{c}{\textit{Fitted planet parameters}} \\
\noalign{\smallskip}
$P$  (d)                                                                &$0.764597^{+0.000013}_{-0.000011}$ &  $3.294736^{+0.000034}_{-0.000036}$ &  $\geq$600\\
$t_0$ $^{(1)}$                          &$1684.102^{+0.002}_{-0.003} $ & $1707.352^{+0.002}_{-0.001}$ &   \\
$e$                                                                     &  $\le 0.25$ & $\le 0.15$ & \\
$K$ (m/s)                                                       &   $3.74^{+1.03}_{-0.99}$ & $5.10^{+1.02}_{-1.06}$ &  $\geq$70 \\
$r_{1}$                                                         &   $0.45^{+0.14}_{-0.15}$ & $0.51^{+0.07}_{-0.11}$ &  \\
$r_{2}$                                                         &   $0.04^{+0.002}_{-0.002}$ & $0.07^{+0.002}_{-0.003}$ &  \\

\noalign{\smallskip}
\multicolumn{4}{c}{\textit{Derived planet parameters} } \\
\noalign{\smallskip}
$R_{p}/R_{\star}$                               &  $0.019^{+0.001}_{-0.001}$ & $0.033^{+0.001}_{-0.001}$ &  \\
$R_{\rm p}$ (R$_{ \oplus}$)             & $1.21^{+0.11}_{-0.10}$ & $2.11^{+0.14}_{-0.14}$ &  \\
$a/R_{\star}$                                                   &  $5.19^{+0.16}_{-0.17}$ & $13.73^{+0.43}_{-0.47}$ &  \\
$a$ (au)                                                                                                        &$0.0139^{+0.0008}_{-0.0008}$ & $0.037^{+0.002}_{-0.002}$ & $\geq$1.1\\
$b = (a/R_{\star}) \cos i$              &  $0.32^{+0.17}_{-0.19}$ & $0.39^{+0.10}_{-0.13}$ &  \\
$i$ (deg)                                                                       &  $86.51^{+2.11}_{-1.98}$ & $88.38^{+0.57}_{-0.47}$ &  \\

$t_{\rm 14}$ (h)                                                & $1.09^{+0.05}_{-0.08}$ & $1.75^{+0.06}_{-0.06}$ &  \\
$t_{\rm depth}$ (ppm)                   & $366.34^{+44.64}_{-40.73}$ & $1113.42^{+83.63}_{-86.58}$ &  \\

$M_{\rm p} \sin i$ (M$_{ \oplus}$)          & $3.75^{+1.14}_{-1.06}$ & $8.32^{+1.90}_{-1.88}$ &  $\geq$ 2$\sqrt{1-e^2}$ $\rm M_{\rm Jup}$ \\
$M_{\rm p}$ (M$_{ \oplus}$)             & $3.76^{+1.15}_{-1.07}$ & $8.32^{+1.90}_{-1.88}$ &  \\
$\rho_{\rm p}$ (g $\rm cm^{-3}$)                & $11.7^{+4.2}_{-3.4}$ & $4.9^{+2.5}_{-1.8}$ &  \\
$T_{\rm eq}$ (K)$^{(2)}$                                &  {965--1300}\,K &  {590--800}\,K &  \\
$S$ (S$_{\oplus}$)                 & $442^{+39}_{-35}$ & $63^{+6}_{-5}$ & \\

\noalign{\smallskip}
\hline
\end{tabular}
\tablefoot{$^{(1)}$ $t_0$ (BJD $-$ 2,457,000). $^{(2)}$ For Bond albedo in the interval 0.65--0.0. }
\end{scriptsize}
\end{table}

To determine the true masses of TOI-1238\,b and c, we performed a combined photometric and spectroscopic analysis using the \textit{TESS} and CARMENES VIS data. {Our final model consists of two transiting planets at 0.76 and 3.29\,d and a long-term $\ge 400$\,d signal, all three of which are modeled by non-eccentric Keplerian orbits, plus the stellar rotation component at 20--80\,d simulated by the exp-sine-squared GP kernel of {Eq.}~\ref{eq:exp-sin-sqr_kernel}.} On the one hand, we used the results obtained from the transit-only analysis (Sect.~\ref{subsec:TESS_light_curve}) to define normal priors on the orbital period and time of periastron passage of the transiting planets with Gaussian distributions centered at the median values of the posteriors. This is fully justified because these parameters are mainly constrained by the LCs with the RV data adding little information \citep{2020A&A...642A.236K}. On the other hand, the RV amplitudes, $K$, are fit by adopting a prior with a uniform distribution for each Keplerian signal. We also fit a jitter for the \textit{TESS} and CARMENES data. All priors are summarized in Table~\ref{tab:toi1238_priors_details}.

As illustrated by the median posterior parameters of our combined fit presented in Table~\ref{tab:toi1238bc_params_from_joit-fit}, the derived GP period is $47.8^{+11.5}_{-8.7}$\,d, which agrees within the quoted uncertainties with the stellar rotation period determined from independent LC data and from the CARMENES spectroscopic indices. The combined fit also confirmed that TOI-1238 displays a weak activity at optical wavelengths and timescales typical of the stellar rotation period. Figure~\ref{fig:lc_vs_time} shows the four \textit{TESS} sectors and the location of all transits of TOI-1238\,b and c captured by the observations. On three different occasions (BJD = 2,458,713.9 -- sector 15, BJD = 2,458,895.1 -- sector 21, and BJD = 2,458,905.0 -- sector 22), the transits of TOI-1238\,b and c overlap. The LCs folded in phase with the orbital periods of the transiting planets per sector are shown in Fig.~\ref{fig:lc_vs_phase}. For completeness, the corner plot depicting all the posterior distributions of the planetary parameters as obtained from the joint fit is shown in Fig.~\ref{fig:toi1238_cornerplot}.

The resulting RV model is depicted in the top panels of Fig.~\ref{fig:toi1238_rv_time-phase_plots}, whereas the RV curves folded in phase for the two transiting planets are shown in the bottom panels of the figure. The $rms$ of the RV residuals ({that is}, observed RVs minus the best fit) is 4.66\,m\,s$^{-1}$, {which is slightly higher} than the mean value of the CARMENES VIS RV errors (see Sect.~\ref{carmenes_spectroscopy}). This suggests that no other component is detectable at the quality of our data. With RV amplitudes of 3.74$^{+1.03}_{-0.99}$\,m\,s$^{-1}$ (TOI-1238\,b) and 5.10$^{+1.02}_{-1.06}$ m\,s$^{-1}$ (TOI-1238\,c), these transiting planets have true masses of 3.76$^{+1.15}_{-1.07}$\,M$_\oplus$ and 8.32$^{+1.90}_{-1.88}$\,M$_\oplus$, respectively, with a significance of 3.7--5\,$\sigma$. They are moving around their parent star in orbits that are 28 and 11 times smaller than the Sun--Mercury orbital size. TOI-1238\,b and TOI-1238\,c are a small super-Earth and a mini-Neptune, each one located on either side of the radius gap for planetary systems around M-type stars \citep{2020AJ....159..211C}. As for the possible third companion, we can only impose a minimum mass of $M \geq 2 \sqrt{1-e^2}$ M$_{\rm Jup}$ as explained above.

We also explored whether the two inner planets have eccentric orbits by leaving this parameter free in {a new combined photometric and spectroscopic analysis} while maintaining a zero eccentricity for the outer companion. The results revealed that TOI-1238\,b and c are very likely in circular orbits and we derived an upper limit on the orbital eccentricities of $e = 0.25$ and 0.15 at the 1\,$\sigma$ level for planets b and c, respectively. 

In summary, TOI\,1238 is the host of a super-Earth planet (TOI-1238\,b) with a very short orbital period (0.76\,d), a mini-Neptune planet (TOI-1238\,c) also at a close-in orbit (3.29\,d), and a likely more massive and distant companion with an orbital period $\geq$600\,d. We also derived the bulk densities of the transiting planets ($\rho = 11.7^{+4.2}_{-3.4}$ and 4.9$^{+2.5}_{-1.8}$ g\,cm$^{-3}$ for planets b and c, respectively), which are very distinct despite their proximity to the parent star, thus suggesting dissimilar atmospheric conditions. 

\subsection{Aliasing \label{sec:aliasing}}
The 1 d aliasing issue between the 0.764 d and 3.294 d planets in the RV data may cause spectral leakage making the RV amplitude of one signal to be affected by the other. The work by \cite{2010ApJ...722..937D} established that the aliases of an RV signal also includes phase information that depends on the true underlying signal. This information can be used to disentangle the alias signal of the short orbital period planet from the true signal of the second planet and vice versa if the phases are not equal by coincidence. The success in disengaging the two alias signals also depends on the time of the day at which the measurements were performed, {that is,} the time sampling. In our case we obtained a few RV measurements on different occasions during the same nights, which helped break this degeneracy. The additional phase information obtained when performing the combined photometric and spectroscopic analysis delineated in Sect.~\ref{sec:planet orbiting toi-1238} contributed to distinguish the two planetary signals. Another argument indicating that the phases of TOI-1238\,b and c differ enough and that they are well constrained to distinguish the two signals comes from the relation between both RV amplitudes ($K$). If no strong correlation between $K_b$ and $K_c$ is found, the two signals can be safely untangled. 

To confirm that the degeneracy is broken in this way, we performed the following test: we removed the planetary signals from the RV curve using the results of Sect.~\ref{sec:planet orbiting toi-1238} and injected {two} fake signals {with the same planetary orbital periods and phases but different RV amplitudes. We used the same observing CARMENES epochs to produce the fake RV time series. This provides a useful test for checking potential correlations between $K_b$ and $K_c$}. The exact same method (including the priors) described in Sect.~\ref{sec:planet orbiting toi-1238} was then applied to the combined \textit{TESS} photometry and fake RVs. We managed to recover the injected amplitudes within the error bars in all tests, thus supporting the combined analysis for deriving reliable planetary masses of the TOI-1238 system. Furthermore, the box displaying the RV amplitude of TOI-1238\,b versus the RV amplitude of TOI-1238\,c in the corner plot of Fig.~\ref{fig:toi1238_cornerplot} shows a nicely uncorrelated relation resulting from the photometric and spectroscopic joint analysis of the real data. {The Pearson's $r$ and Spearman's $\rho$ coefficients obtained for the planet amplitude relation have a value of $-0.4,$ and the $p$ coefficient is 0.0 for both correlations.} Thanks to the precise phase and period information from the joint analysis described in Sect.~\ref{sec:planet orbiting toi-1238} and an adequate time sampling of the RV observations, we managed to break down the alias issue for planets TOI-1238 b and c. This resulted in {a robust and precise} solution for the RV amplitudes of the two transiting planets, and therefore, the masses of TOI-1238\,b and c are uniquely measured.

\section{Discussion}
\label{discussion}

\subsection{Planetary system stability}
We investigated whether the planetary system is dynamically stable using the angular momentum deficit (AMD) stability criterion \citep{2017A&A...605A..72L}. The AMD can be interpreted as a measure of the excitation of the orbits that limits the close encounters among the planets and ensures long-term stability. The main ingredients are the semimajor axes, the masses and the orbital eccentricities. The result of the analysis rejected high eccentric planetary orbits yielding stable AMD solutions only for those posterior distributions with low eccentricity values: $e\leq0.2$ for the TOI-1238\,b and c and $e\leq0.1$ for the outer companion, in agreement with our findings.

\begin{figure}[]
\centering
\includegraphics[width=0.5\textwidth]{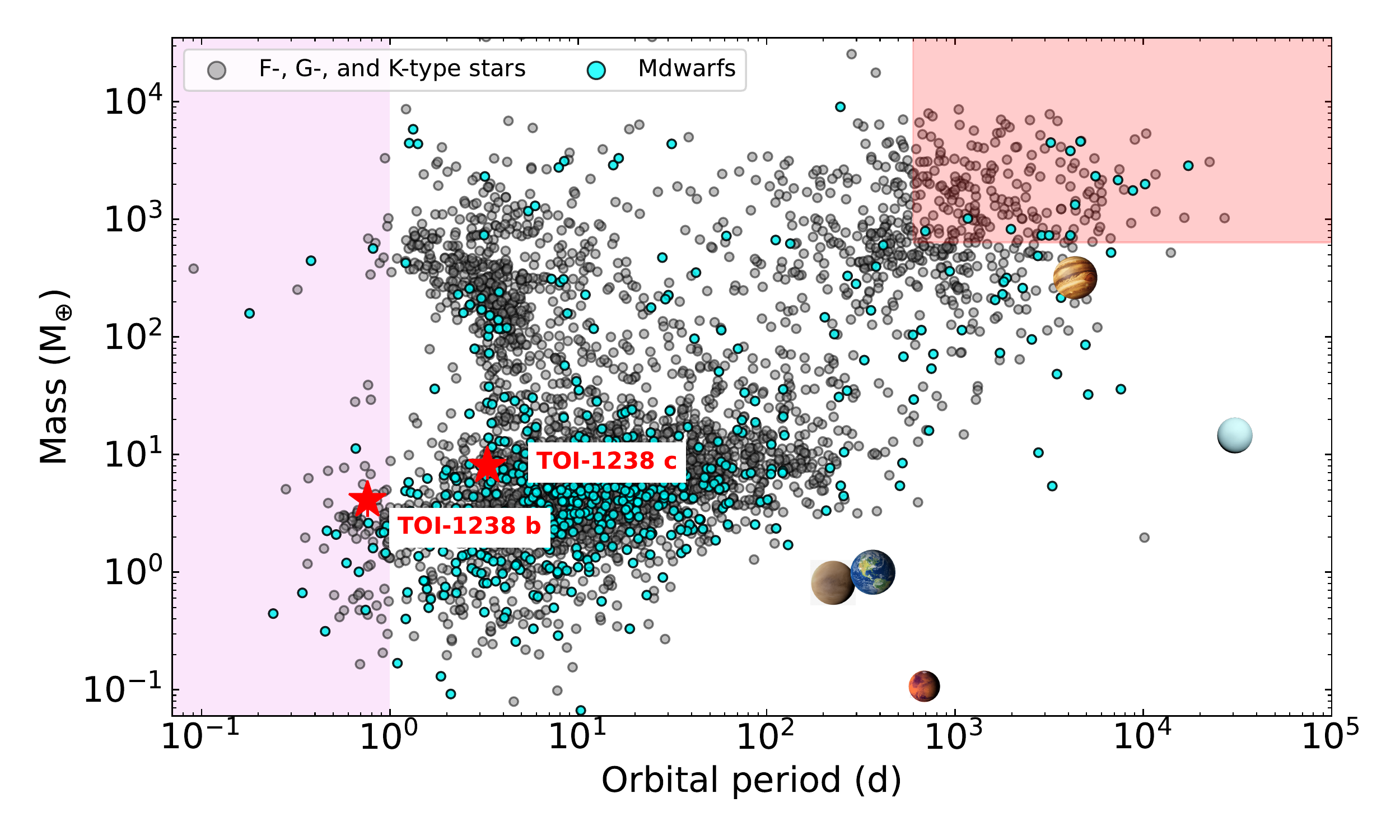}
\caption{Mass versus orbital period diagram for all discovered planets around F-, G-, and K-type (gray dots) and M-type (blue dots) stars from the NASA Exoplanet Archive. The red star symbols represent TOI-1238\,b and c. The shaded area on the left corresponds to the location of USPs. The red shaded area on the right represents the possible locations of the outer companion to TOI-1238 (circular orbits). Venus, the Earth, Mars, Jupiter, and Uranus are also shown for comparison purposes.
}
\label{fig:toi1238_planets_Mass_vs_period}
\end{figure}

\begin{figure*}[]
\centering
\includegraphics[width=0.49\textwidth]{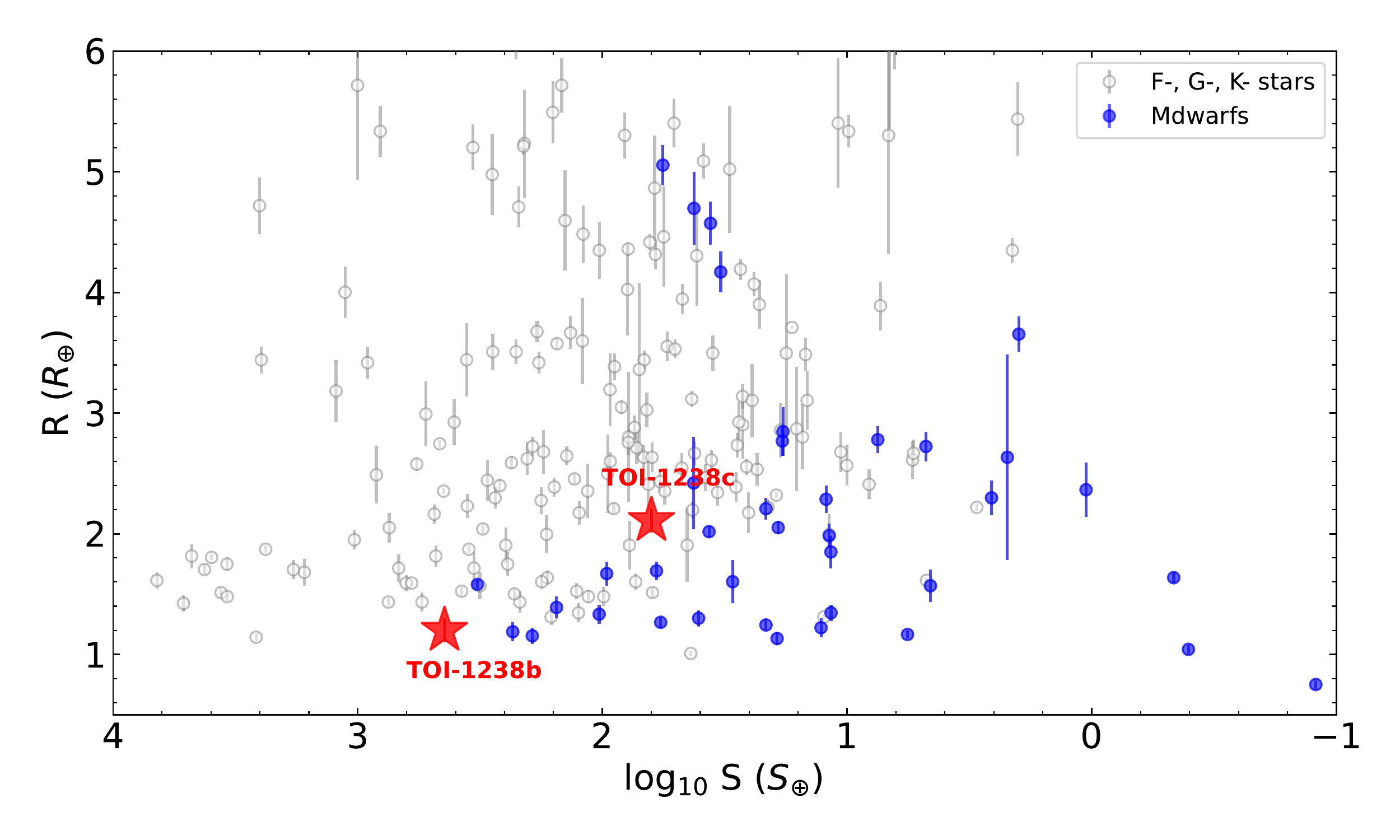}
\includegraphics[width=0.49\textwidth]{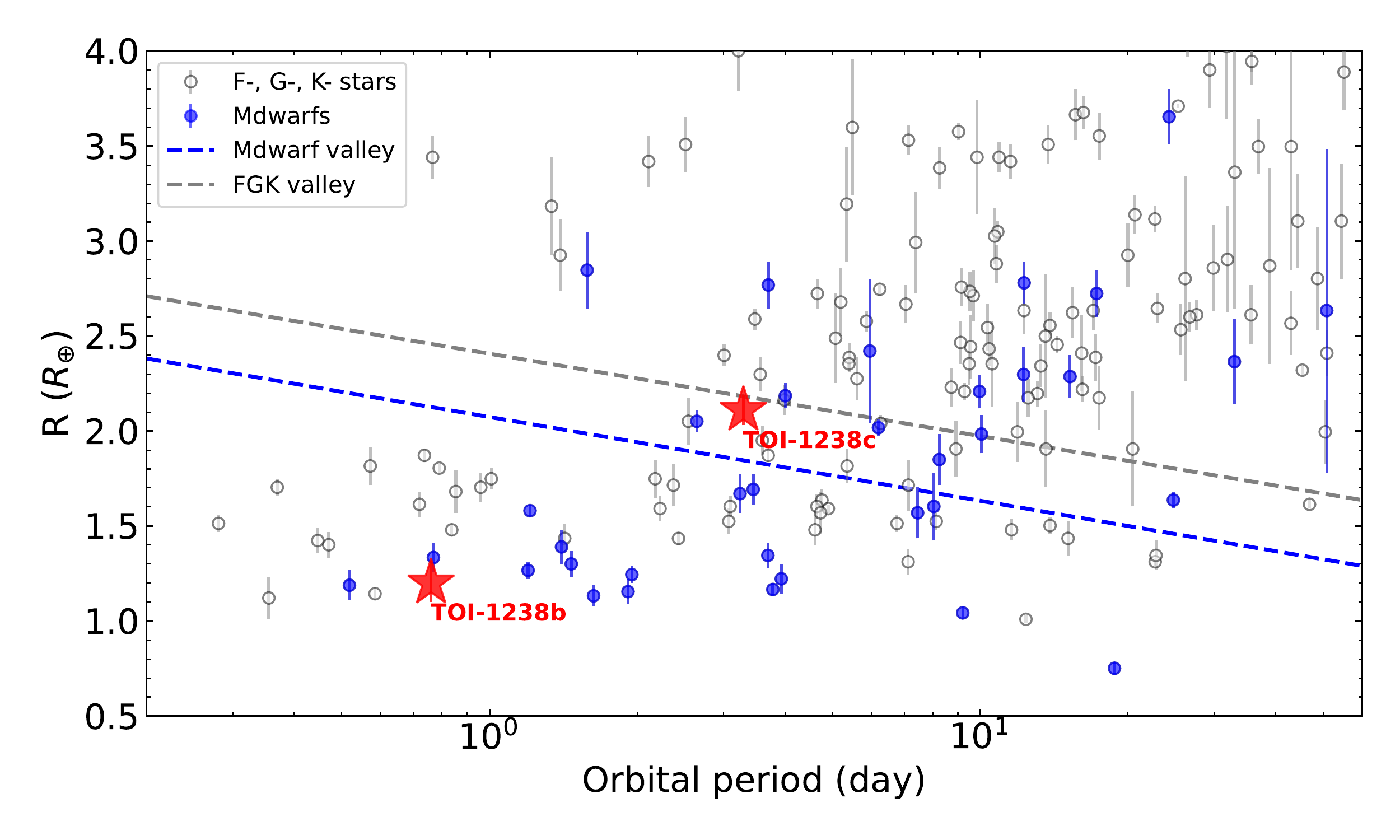}
\caption{Radius of all discovered planets around F-, G-, and K-type stars (gray dots) and M-type stars (blue dots) with planetary mass and radius error bars of 30\,\%~or smaller, plotted as a function of planetary insolation ({\it left panel}) and orbital period ({\it right panel}). These data were extracted from the NASA Exoplanetary Archive. The two transiting planets of the TOI-1238 system are marked with a red star symbol. In the right panel, the dashed gray line indicates the radius valley for planets around solar-type stars, whereas the dashed blue line shows the location of the radius valley of planetary systems with M-type parent stars.} 
\label{fig:toi1238_cmap_S_and_P}
\end{figure*}

\begin{figure*}[]
\centering
\includegraphics[width=0.49\textwidth]{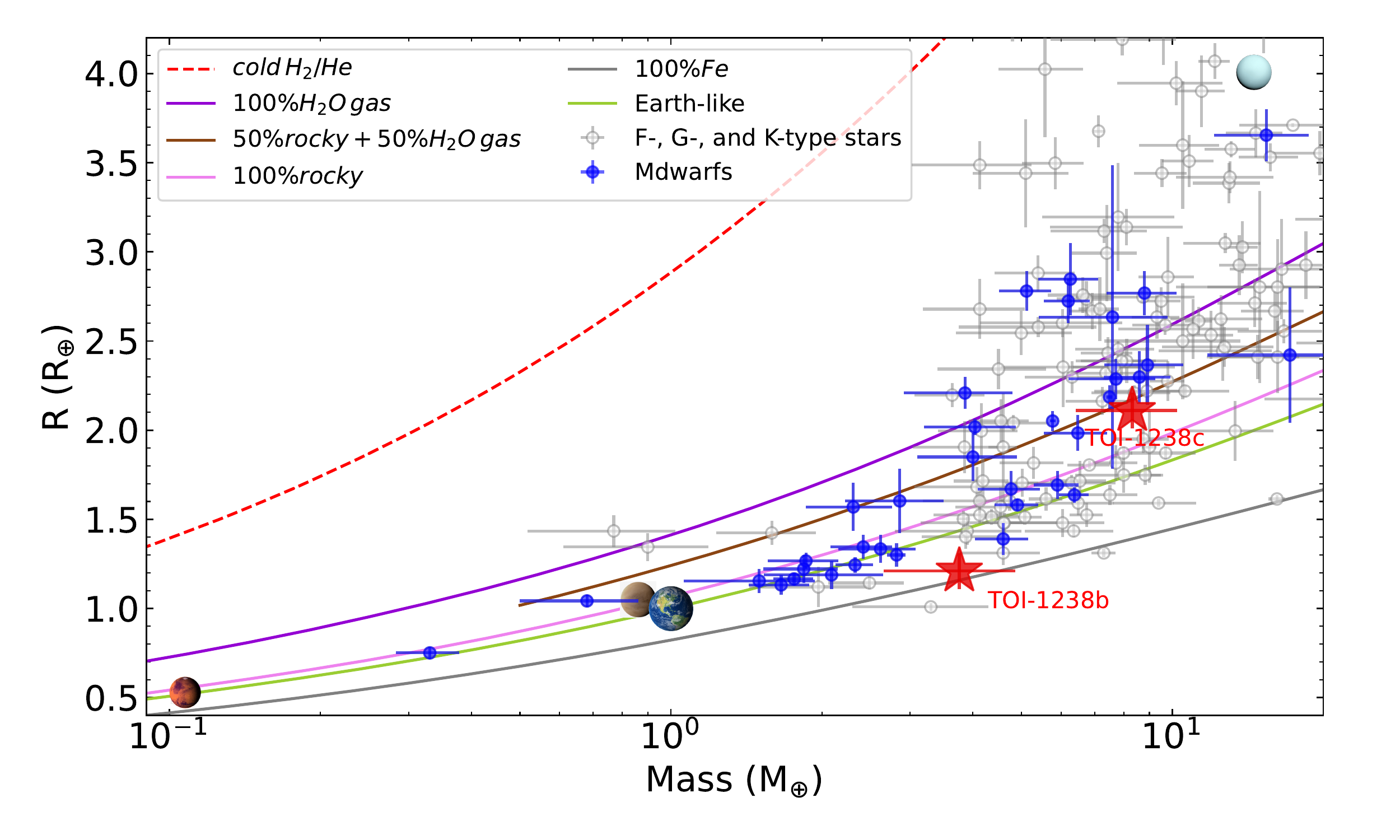}
\includegraphics[width=0.49\textwidth]{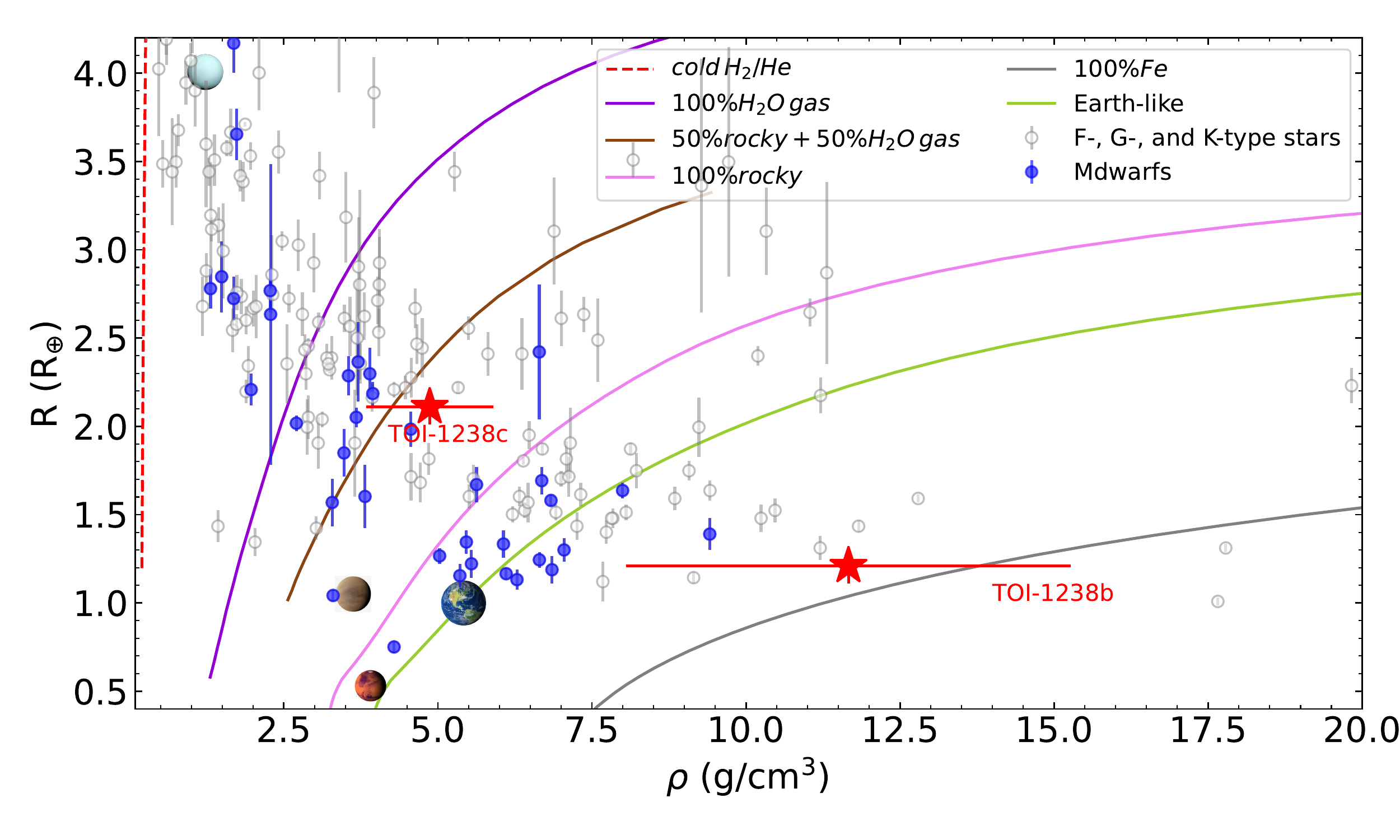}
\caption{Planetary radius shown as a function of mass ({\it left panel}) and density ({\it right panel}) for all planets with measured error bars smaller than 30\,\%. Gray dots stand for F-, G-, and K-type primary stars, and blue dots represent planets orbiting M-type stars. The two transiting planets of TOI-1238, with radius and mass uncertainties of $\sim$5\% and $\sim$25\%, respectively, are shown with red stars. The various planetary composition models of \cite{2016ApJ...819..127Z} are shown with solid colored lines.}
\label{fig:toi1238_planets_around_dM}
\end{figure*}

\subsection{Outermost companion to TOI-1238}
The putative, external, massive companion of TOI-1238 has an estimated radius between $\sim$1.2 and 1.7\,R$\rm _{Jup}$ for masses in the interval 2\,M$_{\rm Jup}$ and 0.18\,M$_\odot$ \citep{2003A&A...402..701B, 2017ApJ...834...17C, 2020A&A...642A.115C}. Despite its large size, the probability for this possible companion to cross the disk of its parent star is very low, on the order of 0.2$\pm$0.2\%, and it is {thus} compatible with it being a non-transiting companion. However, in the case where all companions have coplanar orbits, the transit probability of the external companion significantly increases up to $\sim$20\,\%. Under the low probability hypothesis of a supposed transit with an inclination of 90\,deg, the depth of the transit would be 43456\,$\pm$\,1242\,ppm (adopting 2\,R$_{\rm Jup}$ for the external companion). This is an extremely intense phenomenon that would be visible even in poor-quality LCs. 

Although statistically not very abundant, planetary systems with unresolved, cold Jupiter-mass companions orbiting G, K, and M dwarfs are known to exist (e.g., \citealt{2010ApJ...719..890R,2018AJ....156...92Z,2019Sci...365.1441M, 2020AJ....159..267B, 2020A&A...638A..16T, 2021AJ....161...64L, 2021MNRAS.503.2706K}). If the external companion of TOI-1238 is a substellar object (planet or brown dwarf), TOI-1238 would add to the increasing number of these systems. What is more peculiar about the TOI-1238 system is, however, is that it also contains warm super-Earth and mini-Neptune planets in close-in orbits because massive companions are expected to influence their environment due to their large masses. Planetary systems with the architecture similar to that of TOI-1238 must be described by the theories of planet formation and dynamical evolution. Using the Generation III Bern Model of planet formation and evolution \citep{2020arXiv200705561E}, where most planets are formed via the core accretion mechanism and moved to their final location in protoplanetary disks, \citet{2020arXiv200705563S} investigated the mutual interaction between cold giants and warm super-Earths orbiting solar-type stars. Their findings indicate that there is a positive correlation between the occurrences of inner super-Earths and cold Jupiters, which may indicate that these systems are born from intermediate-mass disks and that the efficiency of the planetary inward migration of the giants is somehow overestimated in planetary formation and evolution models. The presence of this massive outer companion is also compatible with formation by disk gravitational instability leading to the collapse of dense regions in the protoplanetary disk \citep[and references therein]{1997Sci...276.1836B, 2016ARA&A..54..271K}.

\subsection{TOI-1238\,b and c}

\cite{2014ApJ...787L..29K} calculated a conservative estimation of the inner habitable zones (HZs) around stars with stellar effective temperatures in the range 2600--7200\,K for planetary masses between 0.1 and 5\,M$_\oplus$. According to these authors, the inner edge of the HZ around TOI-1238 is located at 0.28\,au with an effective stellar flux incidence of $S_{\rm eff}=1.02$\,S$_{\odot}$. TOI-1238\,b and~c lie between the star and the inner boundary of the HZ. The theoretical equilibrium temperature ($T_{\rm eq}$) of these two  planets was derived by using the Stefan–Boltzmann equation, the stellar parameters given in Table~\ref{tab:stellar_properties_TOI-1238}, and two very different values of the planetary albedo ($A = 0.65$ and~0.0 for the high- and non-reflectance cases). The resulting $T_{\rm eq}$ ranges {are 965--1300\,K and 590--800\,K for TOI-1238\,b and c, respectively (Table~\ref{tab:toi1238bc_params_from_joit-fit}). These temperature intervals include the uncertainties in the stellar $T_{\rm eff}$ and semimajor axis of the planets.} 

Planets with orbital periods of less than one day are known as ultra-short-period planets \citep[USPs;][]{2014ApJ...787...47S,2018NewAR..83...37W}. The number of known USPs is low, and about 30\,\%~of them have been found around M-type stars. To place the TOI-1238 planetary system in context with other known systems, Fig.~\ref{fig:toi1238_planets_Mass_vs_period} shows the planetary mass versus the orbital period diagram for all planets discovered by any observing method (from the NASA Exoplanet Archive\footnote{\url{https://exoplanetarchive.ipac.caltech.edu/index.html}}). All of these systems orbit around F-, G-, K-, and M-type stars. TOI-1238\,b is an USP low-mass planet in a hierarchical planetary system where the inner-most planet is the least massive and the outermost likely companion is the most massive. As shown in the left panel of Fig.~\ref{fig:toi1238_cmap_S_and_P}, TOI-1238\,b is one of the few super-Earths orbiting M dwarfs with the highest stellar irradiation index and highest $T_{\rm eq}$ (in the interval 1000--1258\,K; see Table \ref{tab:toi1238bc_params_from_joit-fit}). Therefore, the day side of TOI-1238\,b is presumably very hot. 

If TOI-1238\,b had an atmosphere, it would be subject to intense erosion \citep{2020AAS...23537507U, 2020DDA....5150302M}, making it a potentially interesting candidate for planetary atmospheric studies. The stellar extended UV (XUV) irradiation ionizes H atoms, giving rise to atmospheric inflation in the planet in the first place, and to mass loss afterward \citep[][and references therein]{2011A&A...532A...6S}. In order to test this possibility, we would need to know the XUV irradiation flux from the star. This flux can be calculated using coronal models, but no X-ray data of TOI-1238 are available in the literature. Instead, we can estimate the current X-ray emission of the star by using the relation between X-rays and the stellar rotational period, as displayed in Fig.~2 of \citet{2011ApJ...743...48W}. According to this relation, using {$P_{\rm rot} = 40\pm5$\,d}, we would expect an X-ray luminosity of $L_{\rm X}=3.3 \times 10^{+28}$\,erg\,s$^{-1}$ in the 5--100~\AA\ spectral range. The stellar extreme ultraviolet (EUV; 100--920~\AA) luminosity can be calculated using the scaling law calculated by \citet{2011A&A...532A...6S}, resulting in $L_{\rm EUV}=2.0 \times 10^{+29}$\,erg\,s$^{-1}$ . Using these quantities, and the planetary system parameters from Table~\ref{tab:toi1238bc_params_from_joit-fit}, we can calculate the mass loss rate in the energy limited regime, as in Eq.~7 of \citet{2011A&A...532A...6S}, resulting in 3.2~M$_{\oplus}$\,Gyr$^{-1}$ {for TOI-1238\,b and 0.9~M$_{\oplus}$\,Gyr$^{-1}$ for TOI-1238\,c}. This formula assumes a planet atmosphere dominated by H, which is probably not the case presently for TOI-1238\,b, given its high density. We can instead estimate what was the expected mass loss rate at some point in the past, assuming that TOI-1238 planets were more massive with an H-dominated atmosphere. Typical densities of planets with heavy XUV irradiation can be as low as 0.3\,g\,cm$^{-3}$. Even assuming that the star had the same XUV emission in the past, we would have a mass loss rate of {125}~M$_{\oplus}$\,Gyr$^{-1}$ for TOI-1238\,b, and {15}~M$_{\oplus}$\,Gyr$^{-1}$ for TOI-1238\,c, assuming the density of the planets reported in Table \ref{tab:toi1238bc_params_from_joit-fit}. The stellar XUV emission decreases as their rotation rate slows down. Therefore, we adopted a conservative assumption, yet we inferred planet photo-evaporation in the past was very strong, likely leading to the loss of a significant fraction of the H {atmosphere}. This might explain the observed planet density, inconsistent with a gas-rich atmosphere. {This scenario is also consistent with an early migration of the planets through the protoplanetary disk.}

Following the metrics proposed by \cite{2018PASP..130k4401K} to identify the \textit{TESS} transiting planets most amenable for atmospheric characterization via transmission and/or emission spectroscopy with the \textit{James Webb Space Telescope (JWST)}, we derived TSM = 2.98 and 33.80, and ESM = 4.63 and 5.12 for TOI-1238\,b and c, respectively, where TSM and ESM stand for the metrics for transmission and emission spectroscopy. These factors depend on the stellar brightness, the host star's temperature, and the planetary properties, including the planet equilibrium and day-side temperatures. As a rule of thumb, for a given host star, the greater radius and temperature of the planet, the larger the TSM and ESM factors are, and the higher the probability of detecting the planetary atmosphere is. Both TOI-1238\,b and c have TSM and ESM factors below the threshold metric values defined for their respective planetary categories (Earth-size and small sub-Neptune) by \cite{2018PASP..130k4401K}, thus indicating that the study of their atmospheres, if they had any, is not an easy task even for the \textit{JWST}.

Various studies have revealed that the radius distribution of planets slightly larger than the Earth is bimodal \citep{2017AJ....154..109F, 2018MNRAS.479.4786V, 2020AJ....160...89P, 2020AJ....160...22C}. Super-Earth planets of up to $\sim$1.5\,R$_{\oplus}$ are relatively common, as are mini-Neptunes in the range of $\sim$2–4\,R$_{\oplus}$. But there appears to be a deficit of planets in between these sizes. The location of the radius gap depends on the orbital period and/or the planet insolation and/or the spectral type of the parent stars \citep{2018MNRAS.479.4786V, 2020AJ....160...22C, 2021MNRAS.tmp.2193V}. The right panel of Fig.~\ref{fig:toi1238_cmap_S_and_P} illustrates the planet radius versus orbital period for all known planets from the NASA Archive with parameters measured to better than 30\,\%. We marked the location of the planetary radius valley for solar-type stars and M-type stars following \citet{2021MNRAS.tmp.2193V}. TOI-1238\,b and c have very different bulk densities and are located at the two sides of the planetary radius gap of close-in planets, which is believed to account for the distinction between small rocky planets and bigger planets with volatile rich envelopes \citep{2015ApJ...807...45D, 2017AJ....154..109F}.

Using the same subsample of well characterized planetary systems as in Fig.~\ref{fig:toi1238_cmap_S_and_P}, we produced the mass-radius diagram shown in the left panel of Fig.~\ref{fig:toi1238_planets_around_dM}, where we overplotted various theoretical models planets with multilayer interior composition \citep{2016ApJ...819..127Z}. The comparison of the loci of TOI-1238\,b and~c with these models yields that the inner, hotter planet is mainly made of iron (thus, very likely solid) whereas the composition of planet c is compatible with 50\%  Earth-like rocky core and 50\% $\rm H_{2}O$-gaseous layer. TOI-1238\,c has a planetary density that resembles those of Venus and the Earth. Interestingly, TOI-1238\,b appears to be denser than the Earth making it one of the densest planets known to date, particularly among the subsample of planets orbiting M dwarfs. This property is more apparent in the right panel of Fig.~\ref{fig:toi1238_planets_around_dM}, where planetary radius is shown as a function of density. It is worth noting that the number of very dense close-in planets ($\rho > 10$ g\,cm$^{-3}$) is quite small as compared to the number of planets with Earth-like densities. In their theoretical study, \citet{2020arXiv200705563S} predicted that systems with close-in, high-density super-Earths are more likely to host an outer giant planet. TOI-1238 matches this theoretical prediction, although we caution that these models are based on several poorly constrained assumptions such as efficient planetary core formation via planetesimal accretion and a simplified disk chemistry.

\section{Summary}
\label{sec:summary_discussion}
We have presented the discovery and validation of two transiting planets, TOI-1238\,b and TOI-1238\,c, orbiting a nearby K7--M0 star using photometric \textit{TESS} and spectroscopic CARMENES data. The RVs also revealed the presence of an outer, more massive companion with a long-period orbit. The intrinsic stellar variability was analyzed using the spectroscopic activity indicators provided by CARMENES and the optical photometric monitoring of the ASAS-SN, TJO, SNO, and \textit{TESS} data. Our findings suggest that TOI-1238 is a rather quiet star with a likely rotation period {of 40$\pm$5\,d.}

The joint analysis of the \textit{TESS} and CARMENES data yields orbital periods of $0.764597^{+0.000013}_{-0.000011}$\,d and $3.294736^{+0.000034}_{-0.000036}$\,d, masses of 3.76$^{+1.15}_{-1.07}$\,M$_{\oplus}$ and 8.32$^{+1.90}_{-1.88}$\,M$_{\oplus}$, and radii of $1.21^{+0.11}_{-0.10}$\,R$_{\oplus}$ and $2.11^{+0.14}_{-0.14}$\,R$_{\oplus}$ for the transiting planets TOI-1238\,b and TOI-1238\,c, respectively, both of which likely have nearly circular orbits with eccentricity below 0.2. From our data, we imposed the following constraints for the outer, massive companion: an orbital period longer than 600\,d and an RV amplitude above 70\,m\,s$^{-1}$, which translates into a minimum mass of $M \geq 2 \sqrt{1-e^2}$\,M$_{\rm Jup}$ and orbital separations $\geq$1.1\,au from its parent star.

The TOI-1238 system contains all the fascinating ingredients of low-mass planetary systems, having two close-in low-mass planets at both sides of the planetary radius valley and a likely massive companion on an outer orbit. One of the inner planets is a very dense super-Earth that has very likely suffered strong irradiation and erosion. The characteristics of the TOI-1238 planetary system offer a compelling laboratory to test innovative models for explaining theories of planet formation and evolution.


\begin{acknowledgements}

CARMENES is an instrument for the Centro Astron\'omico Hispano-Alem\'an de Calar Alto (CAHA, Almer\'ia, Spain). CARMENES is funded by the German Max-Planck- Gesellschaft (MPG), the Spanish Consejo Superior de Investigaciones Cient\'ificas (CSIC), the European Union through FEDER/ERF funds, and the members of the CARMENES Consortium (Max-Planck-Institut f\"ur Astronomie, Instituto de Astrof\'isica de Andaluc\'ia, Landessternwarte Ko\"onigstuhl, Institut de Ci\` encies de l'Espai, Insitut f\"ur Astrophysik G\"ottingen, Universidad, Complutense de Madrid, Th\"uringer Landessternwarte Tautenburg, Instituto de Astrof\'isica de Canarias, Hamburger Sternwarte, Centro de Astrobiolog\'ia and Centro Astron\'omico Hispano-Alem\'an), with additional contributions by the Spanish Ministry of Economy, the state of Baden-W\"uttemberg, the German Science Foundation (DFG), the Klaus Tschira Foundation (KTS), and by the Junta de Andaluc\'ia. This work was based on data from the CARMENES data archive at CAB (CSIC-INTA). This research has made use of the NASA Exoplanet Archive, which is operated by the California Institute of Technology, under contract with the National Aeronautics and Space Administration under the Exoplanet Exploration Program. Funding for the \textit{TESS} mission is provided by NASA’s Science Mission Directorate. This paper includes data collected by the TESS mission that are publicly available from the Mikulski Archive for Space Telescopes. We acknowledge the use of public TESS data from pipelines at the \textit{TESS} Science Office and at the \textit{TESS} Science Processing Operations Center. Resources supporting this work were provided by the NASA High-End Computing (HEC) Program through the NASA Advanced Supercomputing (NAS) Division at Ames Research Center for the production of the SPOC data products. Some of the observations in the paper made use of the High-Resolution Imaging instrument ‘Alopeke obtained under Gemini LLP Proposal Number: GN/S-2021A-LP-105. ‘Alopeke was funded by the NASA Exoplanet Exploration Program and built at the NASA Ames Research Center by Steve B. Howell, Nic Scott, Elliott P. Horch, and Emmett Quigley. `Alopeke was mounted on the Gemini North (and/or South) telescope of the international Gemini Observatory, a program of NSF’s OIR Lab, which is managed by the Association of Universities for Research in Astronomy (AURA) under a cooperative agreement with the National Science Foundation. on behalf of the Gemini partnership: the National Science Foundation (United States), National Research Council (Canada), Agencia Nacional de Investigaci\'on y Desarrollo (Chile), Ministerio de Ciencia, Tecnolog\'ia e Innovaci\'on (Argentina), Minist\'erio da Ci\^encia, Tecnologia, Inova\c{c}\~oes e Comunica\c{c}\~oes (Brazil), and Korea Astronomy and Space Science Institute (Republic of Korea). Data were partly collected with the 90\,cm telescope at the Sierra Nevada Observatory (SNO) operated by the Instituto de Astrof\'\i fica de Andaluc\'\i a (IAA-CSIC). We acknowledge the telescope operators from Observatori Astron\'omic del Montsec, Sierra Nevada Observatory, and Centro Astron\'omico Hispano-Alem\'an de Calar Alto (CAHA). E.G.A., M.R.Z.O., J.A.C., J.S.F, and D.M. acknowledge financial support from the Spanish Ministry of Science and Innovation through project PID2019-109522GBC5[1:4]. E.G.A also acknowledges support from AEI Project No$.$ MDM-2017-0737 Unidad de Excelencia ``Mar\'ia de Maeztu''- Centro de Astrobiolog\'ia (CSIC-INTA). V.M.P. acknowledges financial support from NASA through grant NNX17AG24G. S.V.J. acknowledges the support of the DFG priority program SPP 1992 ``Exploring the Diversity of Extrasolar Planets (JE 701/5-1)''. M.J.L.-G., E.R., C.R.-L., and P.J.A. acknowledge financial support from the Agencia Estatal de Investigaci\'on of the Ministerio de Ciencia e Innovaci\'on through projects PID2019-109522GB-C52, PID2019-107061GB-C64, PID2019-110689RB-100 and the Centre of Excellence Severo Ochoa Instituto de Astrof\'\i sica de Andaluc\'\i a (SEV-2017-0709). G.M. received funding from the European Union’s Horizon 2020 research and innovation programme under the Marie Sk\l{}odowska-Curie grant agreement No. 895525. S.S and S.R. acknowledge support by the DFG Research Unit FOR 2544 Blue Planets around Red Stars, project no. RE 2694/4-1.

\end{acknowledgements}

%
%

\bibliographystyle{aa} 
\bibliography{bibliography.bib} 

\begin{thebibliography}{123}
\expandafter\ifx\csname natexlab\endcsname\relax\def\natexlab#1{#1}\fi

\bibitem[{{Allard} {et~al.}(2012){Allard}, {Homeier}, \&
  {Freytag}}]{2012RSPTA.370.2765A}
{Allard}, F., {Homeier}, D., \& {Freytag}, B. 2012, Philosophical Transactions
  of the Royal Society of London Series A, 370, 2765

\bibitem[{{Aller} {et~al.}(2020){Aller}, {Lillo-Box}, {Jones}, {Miranda}, \&
  {Barcel{\'o} Forteza}}]{2020A&A...635A.128A}
{Aller}, A., {Lillo-Box}, J., {Jones}, D., {Miranda}, L.~F., \& {Barcel{\'o}
  Forteza}, S. 2020, \aap, 635, A128

\bibitem[{{Amado} {et~al.}(2021){Amado}, {Bauer}, {Rodr{\'\i}guez L{\'o}pez},
  {Rodr{\'\i}guez}, {Cardona Guill{\'e}n}, {Perger}, {Caballero},
  {L{\'o}pez-Gonz{\'a}lez}, {Mu{\~n}oz Rodr{\'\i}guez}, {Pozuelos},
  {S{\'a}nchez-Rivero}, {Schlecker}, {Quirrenbach}, {Ribas}, {Reiners},
  {Almenara}, {Astudillo-Defru}, {Azzaro}, {B{\'e}jar}, {Bohemann}, {Bonfils},
  {Bouchy}, {Cifuentes}, {Cort{\'e}s-Contreras}, {Delfosse}, {Dreizler},
  {Forveille}, {Hatzes}, {Henning}, {Jeffers}, {Kaminski}, {K{\"u}rster},
  {Lafarga}, {Lodieu}, {Lovis}, {Mayor}, {Montes}, {Morales}, {Morales},
  {Murgas}, {Ortiz}, {Pall{\'e}}, {Pepe}, {Perdelwitz}, {Pollaco}, {Santos},
  {Sch{\"o}fer}, {Schweitzer}, {S{\'e}gransan}, {Shan}, {Stock}, {Tal-Or},
  {Udry}, {Zapatero Osorio}, \& {Zechmeister}}]{2021A&A...650A.188A}
{Amado}, P.~J., {Bauer}, F.~F., {Rodr{\'\i}guez L{\'o}pez}, C., {et~al.} 2021,
  \aap, 650, A188

\bibitem[{{Ambikasaran} {et~al.}(2015){Ambikasaran}, {Foreman-Mackey},
  {Greengard}, {Hogg}, \& {O'Neil}}]{2015ITPAM..38..252A}
{Ambikasaran}, S., {Foreman-Mackey}, D., {Greengard}, L., {Hogg}, D.~W., \&
  {O'Neil}, M. 2015, IEEE Transactions on Pattern Analysis and Machine
  Intelligence, 38, 252

\bibitem[{{Bakos} {et~al.}(2020){Bakos}, {Bayliss}, {Bento}, {Bhatti}, {Brahm},
  {Csubry}, {Espinoza}, {Hartman}, {Henning}, {Jord{\'a}n}, {Mancini}, {Penev},
  {Rabus}, {Sarkis}, {Suc}, {de Val-Borro}, {Zhou}, {Butler}, {Crane},
  {Durkan}, {Shectman}, {Kim}, {L{\'a}z{\'a}r}, {Papp}, {S{\'a}ri}, {Ricker},
  {Vanderspek}, {Latham}, {Seager}, {Winn}, {Jenkins}, {Chacon},
  {F{\H{u}}r{\'e}sz}, {Goeke}, {Li}, {Quinn}, {Quintana}, {Tenenbaum}, {Teske},
  {Vezie}, {Yu}, {Stockdale}, {Evans}, \& {Relles}}]{2020AJ....159..267B}
{Bakos}, G.~{\'A}., {Bayliss}, D., {Bento}, J., {et~al.} 2020, \aj, 159, 267

\bibitem[{{Baraffe} {et~al.}(2003){Baraffe}, {Chabrier}, {Barman}, {Allard}, \&
  {Hauschildt}}]{2003A&A...402..701B}
{Baraffe}, I., {Chabrier}, G., {Barman}, T.~S., {Allard}, F., \& {Hauschildt},
  P.~H. 2003, \aap, 402, 701

\bibitem[{{Barnes} {et~al.}(2015){Barnes}, {Jeffers}, {Jones}, {Pavlenko},
  {Jenkins}, {Haswell}, \& {Lohr}}]{2015ApJ...812...42B}
{Barnes}, J.~R., {Jeffers}, S.~V., {Jones}, H.~R.~A., {et~al.} 2015, \apj, 812,
  42

\bibitem[{{Barnes}(2007)}]{2007ApJ...669.1167B}
{Barnes}, S.~A. 2007, \apj, 669, 1167

\bibitem[{{Bauer} {et~al.}(2020){Bauer}, {Zechmeister}, {Kaminski},
  {Rodr{\'\i}guez L{\'o}pez}, {Caballero}, {Azzaro}, {Stahl}, {Kossakowski},
  {Quirrenbach}, {Becerril Jarque}, {Rodr{\'\i}guez}, {Amado}, {Seifert},
  {Reiners}, {Sch{\"a}fer}, {Ribas}, {B{\'e}jar}, {Cort{\'e}s-Contreras},
  {Dreizler}, {Hatzes}, {Henning}, {Jeffers}, {K{\"u}rster}, {Lafarga},
  {Montes}, {Morales}, {Schmitt}, {Schweitzer}, \&
  {Solano}}]{2020A&A...640A..50B}
{Bauer}, F.~F., {Zechmeister}, M., {Kaminski}, A., {et~al.} 2020, \aap, 640,
  A50

\bibitem[{{Bayo} {et~al.}(2008){Bayo}, {Rodrigo}, {Barrado Y Navascu{\'e}s},
  {Solano}, {Guti{\'e}rrez}, {Morales-Calder{\'o}n}, \&
  {Allard}}]{2008A&A...492..277B}
{Bayo}, A., {Rodrigo}, C., {Barrado Y Navascu{\'e}s}, D., {et~al.} 2008, \aap,
  492, 277

\bibitem[{{Bianchi} {et~al.}(2017){Bianchi}, {Shiao}, \&
  {Thilker}}]{2017ApJS..230...24B}
{Bianchi}, L., {Shiao}, B., \& {Thilker}, D. 2017, \apjs, 230, 24

\bibitem[{{Blanton} {et~al.}(2017){Blanton}, {Bershady}, {Abolfathi},
  {Albareti}, {Allende Prieto}, {Almeida}, {Alonso-Garc{\'\i}a}, {Anders},
  {Anderson}, {Andrews}, {Aquino-Ort{\'\i}z}, {Arag{\'o}n-Salamanca},
  {Argudo-Fern{\'a}ndez}, {Armengaud}, {Aubourg}, {Avila-Reese}, {Badenes},
  {Bailey}, {Barger}, {Barrera-Ballesteros}, {Bartosz}, {Bates}, {Baumgarten},
  {Bautista}, {Beaton}, {Beers}, {Belfiore}, {Bender}, {Berlind}, {Bernardi},
  {Beutler}, {Bird}, {Bizyaev}, {Blanc}, {Blomqvist}, {Bolton}, {Boquien},
  {Borissova}, {van den Bosch}, {Bovy}, {Brandt}, {Brinkmann}, {Brownstein},
  {Bundy}, {Burgasser}, {Burtin}, {Busca}, {Cappellari}, {Delgado Carigi},
  {Carlberg}, {Carnero Rosell}, {Carrera}, {Chanover}, {Cherinka}, {Cheung},
  {G{\'o}mez Maqueo Chew}, {Chiappini}, {Choi}, {Chojnowski}, {Chuang},
  {Chung}, {Cirolini}, {Clerc}, {Cohen}, {Comparat}, {da Costa}, {Cousinou},
  {Covey}, {Crane}, {Croft}, {Cruz-Gonzalez}, {Garrido Cuadra}, {Cunha},
  {Damke}, {Darling}, {Davies}, {Dawson}, {de la Macorra}, {Dell'Agli}, {De
  Lee}, {Delubac}, {Di Mille}, {Diamond-Stanic}, {Cano-D{\'\i}az}, {Donor},
  {Downes}, {Drory}, {du Mas des Bourboux}, {Duckworth}, {Dwelly}, {Dyer},
  {Ebelke}, {Eigenbrot}, {Eisenstein}, {Emsellem}, {Eracleous}, {Escoffier},
  {Evans}, {Fan}, {Fern{\'a}ndez-Alvar}, {Fernandez-Trincado}, {Feuillet},
  {Finoguenov}, {Fleming}, {Font-Ribera}, {Fredrickson}, {Freischlad},
  {Frinchaboy}, {Fuentes}, {Galbany}, {Garcia-Dias},
  {Garc{\'\i}a-Hern{\'a}ndez}, {Gaulme}, {Geisler}, {Gelfand},
  {Gil-Mar{\'\i}n}, {Gillespie}, {Goddard}, {Gonzalez-Perez}, {Grabowski},
  {Green}, {Grier}, {Gunn}, {Guo}, {Guy}, {Hagen}, {Hahn}, {Hall}, {Harding},
  {Hasselquist}, {Hawley}, {Hearty}, {Gonzalez Hern{\'a}ndez}, {Ho}, {Hogg},
  {Holley-Bockelmann}, {Holtzman}, {Holzer}, {Huehnerhoff}, {Hutchinson},
  {Hwang}, {Ibarra-Medel}, {da Silva Ilha}, {Ivans}, {Ivory}, {Jackson},
  {Jensen}, {Johnson}, {Jones}, {J{\"o}nsson}, {Jullo}, {Kamble}, {Kinemuchi},
  {Kirkby}, {Kitaura}, {Klaene}, {Knapp}, {Kneib}, {Kollmeier}, {Lacerna},
  {Lane}, {Lang}, {Law}, {Lazarz}, {Lee}, {Le Goff}, {Liang}, {Li}, {Li},
  {Lian}, {Lima}, {Lin}, {Lin}, {Bertran de Lis}, {Liu}, {de Icaza Lizaola},
  {Long}, {Lucatello}, {Lundgren}, {MacDonald}, {Deconto Machado}, {MacLeod},
  {Mahadevan}, {Geimba Maia}, {Maiolino}, {Majewski}, {Malanushenko},
  {Malanushenko}, {Manchado}, {Mao}, {Maraston}, {Marques-Chaves}, {Masseron},
  {Masters}, {McBride}, {McDermid}, {McGrath}, {McGreer}, {Medina Pe{\~n}a},
  {Melendez}, {Merloni}, {Merrifield}, {Meszaros}, {Meza}, {Minchev},
  {Minniti}, {Miyaji}, {More}, {Mulchaey}, {M{\"u}ller-S{\'a}nchez}, {Muna},
  {Munoz}, {Myers}, {Nair}, {Nandra}, {Correa do Nascimento}, {Negrete},
  {Ness}, {Newman}, {Nichol}, {Nidever}, {Nitschelm}, {Ntelis}, {O'Connell},
  {Oelkers}, {Oravetz}, {Oravetz}, {Pace}, {Padilla}, {Palanque-Delabrouille},
  {Alonso Palicio}, {Pan}, {Parejko}, {Parikh}, {P{\^a}ris}, {Park}, {Patten},
  {Peirani}, {Pellejero-Ibanez}, {Penny}, {Percival}, {Perez-Fournon},
  {Petitjean}, {Pieri}, {Pinsonneault}, {Pisani}, {Poleski}, {Prada},
  {Prakash}, {Queiroz}, {Raddick}, {Raichoor}, {Barboza Rembold}, {Richstein},
  {Riffel}, {Riffel}, {Rix}, {Robin}, {Rockosi}, {Rodr{\'\i}guez-Torres},
  {Roman-Lopes}, {Rom{\'a}n-Z{\'u}{\~n}iga}, {Rosado}, {Ross}, {Rossi}, {Ruan},
  {Ruggeri}, {Rykoff}, {Salazar-Albornoz}, {Salvato}, {S{\'a}nchez}, {Aguado},
  {S{\'a}nchez-Gallego}, {Santana}, {Santiago}, {Sayres}, {Schiavon}, {da Silva
  Schimoia}, {Schlafly}, {Schlegel}, {Schneider}, {Schultheis}, {Schuster},
  {Schwope}, {Seo}, {Shao}, {Shen}, {Shetrone}, {Shull}, {Simon}, {Skinner},
  {Skrutskie}, {Slosar}, {Smith}, {Sobeck}, {Sobreira}, {Somers}, {Souto},
  {Stark}, {Stassun}, {Stauffer}, {Steinmetz}, {Storchi-Bergmann},
  {Streblyanska}, {Stringfellow}, {Su{\'a}rez}, {Sun}, {Suzuki}, {Szigeti},
  {Taghizadeh-Popp}, {Tang}, {Tao}, {Tayar}, {Tembe}, {Teske}, {Thakar},
  {Thomas}, {Thompson}, {Tinker}, {Tissera}, {Tojeiro}, {Hernandez Toledo}, {de
  la Torre}, {Tremonti}, {Troup}, {Valenzuela}, {Martinez Valpuesta},
  {Vargas-Gonz{\'a}lez}, {Vargas-Maga{\~n}a}, {Vazquez}, {Villanova}, {Vivek},
  {Vogt}, {Wake}, {Walterbos}, {Wang}, {Weaver}, {Weijmans}, {Weinberg},
  {Westfall}, {Whelan}, {Wild}, {Wilson}, {Wood-Vasey}, {Wylezalek}, {Xiao},
  {Yan}, {Yang}, {Ybarra}, {Y{\`e}che}, {Zakamska}, {Zamora}, {Zarrouk},
  {Zasowski}, {Zhang}, {Zhao}, {Zheng}, {Zheng}, {Zhou}, {Zhou}, {Zhu},
  {Zoccali}, \& {Zou}}]{2017AJ....154...28B}
{Blanton}, M.~R., {Bershady}, M.~A., {Abolfathi}, B., {et~al.} 2017, \aj, 154,
  28

\bibitem[{{Bluhm} {et~al.}(2020){Bluhm}, {Luque}, {Espinoza}, {Pall{\'e}},
  {Caballero}, {Dreizler}, {Livingston}, {Mathur}, {Quirrenbach}, {Stock}, {Van
  Eylen}, {Nowak}, {L{\'o}pez}, {Csizmadia}, {Zapatero Osorio}, {Sch{\"o}fer},
  {Lillo-Box}, {Oshagh}, {Gonz{\'a}lez-{\'A}lvarez}, {Amado}, {Barrado},
  {B{\'e}jar}, {Cale}, {Chaturvedi}, {Cifuentes}, {Cochran}, {Collins},
  {Collins}, {Cort{\'e}s-Contreras}, {D{\'\i}ez Alonso}, {El Mufti},
  {Ercolino}, {Fridlund}, {Gaidos}, {Garc{\'\i}a}, {Georgieva},
  {Gonz{\'a}lez-Cuesta}, {Guerra}, {Hatzes}, {Henning}, {Herrero}, {Hidalgo},
  {Isopi}, {Jeffers}, {Jenkins}, {Jensen}, {K{\'a}bath}, {Kaminski}, {Kemmer},
  {Korth}, {Kossakowski}, {K{\"u}rster}, {Lafarga}, {Mallia}, {Montes},
  {Morales}, {Morales-Calder{\'o}n}, {Murgas}, {Narita}, {Passegger}, {Pedraz},
  {Persson}, {Plavchan}, {Rauer}, {Redfield}, {Reffert}, {Reiners}, {Ribas},
  {Ricker}, {Rodr{\'\i}guez-L{\'o}pez}, {Santos}, {Seager}, {Schlecker},
  {Schweitzer}, {Shan}, {Soto}, {Subjak}, {Tal-Or}, {Trifonov}, {Vanaverbeke},
  {Vanderspek}, {Wittrock}, {Zechmeister}, \& {Zohrabi}}]{2020A&A...639A.132B}
{Bluhm}, P., {Luque}, R., {Espinoza}, N., {et~al.} 2020, \aap, 639, A132

\bibitem[{{Bluhm} {et~al.}(2021){Bluhm}, {Pall{\'e}}, {Molaverdikhani},
  {Kemmer}, {Hatzes}, {Kossakowski}, {Stock}, {Caballero}, {Lillo-Box},
  {B{\'e}jar}, {Soto}, {Amado}, {Brown}, {Cadieux}, {Cloutier}, {Collins},
  {Collins}, {Cort{\'e}s-Contreras}, {Doyon}, {Dreizler}, {Espinoza}, {Fukui},
  {Gonz{\'a}lez-{\'A}lvarez}, {Henning}, {Horne}, {Jeffers}, {Jenkins},
  {Jensen}, {Kaminski}, {Kielkopf}, {Kusakabe}, {K{\"u}rster},
  {Lafreni{\`e}re}, {Luque}, {Murgas}, {Montes}, {Morales}, {Narita},
  {Passegger}, {Quirrenbach}, {Sch{\"o}fer}, {Reffert}, {Reiners}, {Ribas},
  {Ricker}, {Seager}, {Schweitzer}, {Schwarz}, {Tamura}, {Trifonov},
  {Vanderspek}, {Winn}, {Zechmeister}, \& {Zapatero
  Osorio}}]{2021A&A...650A..78B}
{Bluhm}, P., {Pall{\'e}}, E., {Molaverdikhani}, K., {et~al.} 2021, \aap, 650,
  A78

\bibitem[{{Boss}(1997)}]{1997Sci...276.1836B}
{Boss}, A.~P. 1997, Science, 276, 1836

\bibitem[{{Buchner} {et~al.}(2014){Buchner}, {Georgakakis}, {Nandra}, {Hsu},
  {Rangel}, {Brightman}, {Merloni}, {Salvato}, {Donley}, \&
  {Kocevski}}]{2014A&A...564A.125B}
{Buchner}, J., {Georgakakis}, A., {Nandra}, K., {et~al.} 2014, \aap, 564, A125

\bibitem[{{Caballero} {et~al.}(2016{\natexlab{a}}){Caballero},
  {Cort{\'e}s-Contreras}, {Alonso-Floriano}, {Montes}, {Quirrenbach}, {Amado},
  {Ribas}, {Reiners}, {Abellan}, {B{\'e}jar}, {Brinkm{\"o}ller}, {Czesla},
  {Dorda}, {Gallardo}, {Gonz{\'a}lez-{\'A}lvarez}, {Hidalgo}, {Holgado},
  {Jeffers}, {Kim}, {Klutsch}, {Lamert}, {Llamas}, {L{\'o}pez-Santiago},
  {Mart{\'{\i}}nez-Rodr{\'{\i}}guez}, {Morales}, {Mundt}, {Passegger},
  {Sch{\"o}fer}, {Seifert}, \& {Zechmeister}}]{2016csss.confE.148C}
{Caballero}, J.~A., {Cort{\'e}s-Contreras}, M., {Alonso-Floriano}, F.~J.,
  {et~al.} 2016{\natexlab{a}}, in 19th Cambridge Workshop on Cool Stars,
  Stellar Systems, and the Sun (CS19), 148

\bibitem[{{Caballero} {et~al.}(2016{\natexlab{b}}){Caballero}, {Gu{\`a}rdia},
  {L{\'o}pez del Fresno}, {Zechmeister}, {de Juan}, {Alonso-Floriano}, {Amado},
  {Colom{\'e}}, {Cort{\'e}s-Contreras}, {Garc{\'{\i}}a-Piquer}, {Gesa}, {de
  Guindos}, {Hagen}, {Helmling}, {Hern{\'a}ndez Casta{\~n}o}, {K{\"u}rster},
  {L{\'o}pez-Santiago}, {Montes}, {Morales Mu{\~n}oz}, {Pavlov}, {Quirrenbach},
  {Reiners}, {Ribas}, {Seifert}, \& {Solano}}]{2016SPIE.9910E..0EC}
{Caballero}, J.~A., {Gu{\`a}rdia}, J., {L{\'o}pez del Fresno}, M., {et~al.}
  2016{\natexlab{b}}, in \procspie, Vol. 9910, Observatory Operations:
  Strategies, Processes, and Systems VI, 99100E

\bibitem[{{Chen} \& {Kipping}(2017)}]{2017ApJ...834...17C}
{Chen}, J. \& {Kipping}, D. 2017, \apj, 834, 17

\bibitem[{{Cifuentes} {et~al.}(2020){Cifuentes}, {Caballero},
  {Cort{\'e}s-Contreras}, {Montes}, {Abell{\'a}n}, {Dorda}, {Holgado},
  {Zapatero Osorio}, {Morales}, {Amado}, {Passegger}, {Quirrenbach}, {Reiners},
  {Ribas}, {Sanz-Forcada}, {Schweitzer}, {Seifert}, \&
  {Solano}}]{2020A&A...642A.115C}
{Cifuentes}, C., {Caballero}, J.~A., {Cort{\'e}s-Contreras}, M., {et~al.} 2020,
  \aap, 642, A115

\bibitem[{{Cloutier} \& {Menou}(2020)}]{2020AJ....159..211C}
{Cloutier}, R. \& {Menou}, K. 2020, \aj, 159, 211

\bibitem[{{Cloutier} {et~al.}(2020){Cloutier}, {Rodriguez}, {Irwin},
  {Charbonneau}, {Stassun}, {Mortier}, {Latham}, {Isaacson}, {Howard}, {Udry},
  {Wilson}, {Watson}, {Pinamonti}, {Lienhard}, {Giacobbe}, {Guerra}, {Collins},
  {Beiryla}, {Esquerdo}, {Matthews}, {Matson}, {Howell}, {Furlan},
  {Crossfield}, {Winters}, {Nava}, {Ment}, {Lopez}, {Ricker}, {Vanderspek},
  {Seager}, {Jenkins}, {Ting}, {Tenenbaum}, {Sozzetti}, {Sha}, {S{\'e}gransan},
  {Schlieder}, {Sasselov}, {Roy}, {Robertson}, {Rice}, {Poretti}, {Piotto},
  {Phillips}, {Pepper}, {Pepe}, {Molinari}, {Mocnik}, {Micela}, {Mayor},
  {Martinez Fiorenzano}, {Mallia}, {Lubin}, {Lovis}, {L{\'o}pez-Morales},
  {Kosiarek}, {Kielkopf}, {Kane}, {Jensen}, {Isopi}, {Huber}, {Hill},
  {Harutyunyan}, {Gonzales}, {Giacalone}, {Ghedina}, {Ercolino}, {Dumusque},
  {Dressing}, {Damasso}, {Dalba}, {Cosentino}, {Conti}, {Col{\'o}n}, {Collins},
  {Cameron}, {Ciardi}, {Christiansen}, {Chontos}, {Cecconi}, {Caldwell},
  {Burke}, {Buchhave}, {Beichman}, {Behmard}, {Beard}, \& {Akana
  Murphy}}]{2020AJ....160...22C}
{Cloutier}, R., {Rodriguez}, J.~E., {Irwin}, J., {et~al.} 2020, \aj, 160, 22

\bibitem[{{Collins} {et~al.}(2017){Collins}, {Kielkopf}, {Stassun}, \&
  {Hessman}}]{2017AJ....153...77C}
{Collins}, K.~A., {Kielkopf}, J.~F., {Stassun}, K.~G., \& {Hessman}, F.~V.
  2017, \aj, 153, 77

\bibitem[{{Colom{\'e}} {et~al.}(2010){Colom{\'e}}, {Casteels}, {Ribas}, \&
  {Francisco}}]{2010SPIE.7740E..3KC}
{Colom{\'e}}, J., {Casteels}, K., {Ribas}, I., \& {Francisco}, X. 2010, in
  Society of Photo-Optical Instrumentation Engineers (SPIE) Conference Series,
  Vol. 7740, Software and Cyberinfrastructure for Astronomy, ed. N.~M.
  {Radziwill} \& A.~{Bridger}, 77403K

\bibitem[{{Colome} \& {Ribas}(2006)}]{2006IAUSS...6E..11C}
{Colome}, J. \& {Ribas}, I. 2006, IAU Special Session, 6, 11

\bibitem[{{Dawson} \& {Fabrycky}(2010)}]{2010ApJ...722..937D}
{Dawson}, R.~I. \& {Fabrycky}, D.~C. 2010, \apj, 722, 937

\bibitem[{{Deeming}(1975)}]{1975Ap&SS..36..137D}
{Deeming}, T.~J. 1975, \apss, 36, 137

\bibitem[{{Dreizler} {et~al.}(2020){Dreizler}, {Crossfield}, {Kossakowski},
  {Plavchan}, {Jeffers}, {Kemmer}, {Luque}, {Espinoza}, {Pall{\'e}}, {Stassun},
  {Matthews}, {Cale}, {Caballero}, {Schlecker}, {Lillo-Box}, {Zechmeister},
  {Lalitha}, {Reiners}, {Soubkiou}, {Bitsch}, {Zapatero Osorio}, {Chaturvedi},
  {Hatzes}, {Ricker}, {Vanderspek}, {Latham}, {Seager}, {Winn}, {Jenkins},
  {Aceituno}, {Amado}, {Barkaoui}, {Barbieri}, {Batalha}, {Bauer}, {Benneke},
  {Benkhaldoun}, {Beichman}, {Berberian}, {Burt}, {Butler}, {Caldwell},
  {Chintada}, {Chontos}, {Christiansen}, {Ciardi}, {Cifuentes}, {Collins},
  {Collins}, {Combs}, {Cort{\'e}s-Contreras}, {Crane}, {Daylan}, {Dragomir},
  {Esparza-Borges}, {Evans}, {Feng}, {Flowers}, {Fukui}, {Fulton}, {Furlan},
  {Gaidos}, {Geneser}, {Giacalone}, {Gillon}, {Gonzales}, {Gorjian}, {Hellier},
  {Hidalgo}, {Howard}, {Howell}, {Huber}, {Isaacson}, {Jehin}, {Jensen},
  {Kaminski}, {Kane}, {Kawauchi}, {Kielkopf}, {Klahr}, {Kosiarek}, {Kreidberg},
  {K{\"u}rster}, {Lafarga}, {Livingston}, {Louie}, {Mann}, {Madrigal-Aguado},
  {Matson}, {Mocnik}, {Morales}, {Muirhead}, {Murgas}, {Nandakumar}, {Narita},
  {Nowak}, {Oshagh}, {Parviainen}, {Passegger}, {Pollacco}, {Pozuelos},
  {Quirrenbach}, {Reefe}, {Ribas}, {Robertson}, {Rodr{\'\i}guez-L{\'o}pez},
  {Rose}, {Roy}, {Schweitzer}, {Schlieder}, {Shectman}, {Tanner},
  {{\c{S}}enavc{\i}}, {Teske}, {Twicken}, {Villasenor}, {Wang}, {Weiss},
  {Wittrock}, {Y{\i}lmaz}, \& {Zohrabi}}]{2020A&A...644A.127D}
{Dreizler}, S., {Crossfield}, I.~J.~M., {Kossakowski}, D., {et~al.} 2020, \aap,
  644, A127

\bibitem[{{Dressing} \& {Charbonneau}(2013)}]{2013ApJ...767...95D}
{Dressing}, C.~D. \& {Charbonneau}, D. 2013, \apj, 767, 95

\bibitem[{{Dressing} \& {Charbonneau}(2015)}]{2015ApJ...807...45D}
{Dressing}, C.~D. \& {Charbonneau}, D. 2015, \apj, 807, 45

\bibitem[{{Emsenhuber} {et~al.}(2020){Emsenhuber}, {Mordasini}, {Burn},
  {Alibert}, {Benz}, \& {Asphaug}}]{2020arXiv200705561E}
{Emsenhuber}, A., {Mordasini}, C., {Burn}, R., {et~al.} 2020, arXiv e-prints,
  arXiv:2007.05561

\bibitem[{{Espinoza} {et~al.}(2019){Espinoza}, {Kossakowski}, \&
  {Brahm}}]{2019MNRAS.490.2262E}
{Espinoza}, N., {Kossakowski}, D., \& {Brahm}, R. 2019, \mnras, 490, 2262

\bibitem[{{Feroz} {et~al.}(2009){Feroz}, {Hobson}, \&
  {Bridges}}]{2009MNRAS.398.1601F}
{Feroz}, F., {Hobson}, M.~P., \& {Bridges}, M. 2009, \mnras, 398, 1601

\bibitem[{{Foreman-Mackey} {et~al.}(2013){Foreman-Mackey}, {Hogg}, {Lang}, \&
  {Goodman}}]{2013PASP..125..306F}
{Foreman-Mackey}, D., {Hogg}, D.~W., {Lang}, D., \& {Goodman}, J. 2013, \pasp,
  125, 306

\bibitem[{{Fulton} {et~al.}(2018){Fulton}, {Petigura}, {Blunt}, \&
  {Sinukoff}}]{2018PASP..130d4504F}
{Fulton}, B.~J., {Petigura}, E.~A., {Blunt}, S., \& {Sinukoff}, E. 2018, \pasp,
  130, 044504

\bibitem[{{Fulton} {et~al.}(2017){Fulton}, {Petigura}, {Howard}, {Isaacson},
  {Marcy}, {Cargile}, {Hebb}, {Weiss}, {Johnson}, \&
  {Morton}}]{2017AJ....154..109F}
{Fulton}, B.~J., {Petigura}, E.~A., {Howard}, A.~W., {et~al.} 2017, \aj, 154,
  109

\bibitem[{{Gagn{\'e}} {et~al.}(2018){Gagn{\'e}}, {Mamajek}, {Malo}, {Riedel},
  {Rodriguez}, {Lafreni{\`e}re}, {Faherty}, {Roy-Loubier}, {Pueyo}, {Robin}, \&
  {Doyon}}]{2018ApJ...856...23G}
{Gagn{\'e}}, J., {Mamajek}, E.~E., {Malo}, L., {et~al.} 2018, \apj, 856, 23

\bibitem[{{Gaia Collaboration} {et~al.}(2018){Gaia Collaboration}, {Brown},
  {Vallenari}, {Prusti}, {de Bruijne}, {Babusiaux}, {Bailer-Jones}, {Biermann},
  {Evans}, {Eyer}, {Jansen}, {Jordi}, {Klioner}, {Lammers}, {Lindegren},
  {Luri}, {Mignard}, {Panem}, {Pourbaix}, {Randich}, {Sartoretti}, {Siddiqui},
  {Soubiran}, {van Leeuwen}, {Walton}, {Arenou}, {Bastian}, {Cropper},
  {Drimmel}, {Katz}, {Lattanzi}, {Bakker}, {Cacciari}, {Casta{\~n}eda},
  {Chaoul}, {Cheek}, {De Angeli}, {Fabricius}, {Guerra}, {Holl}, {Masana},
  {Messineo}, {Mowlavi}, {Nienartowicz}, {Panuzzo}, {Portell}, {Riello},
  {Seabroke}, {Tanga}, {Th{\'e}venin}, {Gracia-Abril}, {Comoretto},
  {Garcia-Reinaldos}, {Teyssier}, {Altmann}, {Andrae}, {Audard},
  {Bellas-Velidis}, {Benson}, {Berthier}, {Blomme}, {Burgess}, {Busso},
  {Carry}, {Cellino}, {Clementini}, {Clotet}, {Creevey}, {Davidson}, {De
  Ridder}, {Delchambre}, {Dell'Oro}, {Ducourant},
  {Fern{\'a}ndez-Hern{\'a}ndez}, {Fouesneau}, {Fr{\'e}mat}, {Galluccio},
  {Garc{\'\i}a-Torres}, {Gonz{\'a}lez-N{\'u}{\~n}ez}, {Gonz{\'a}lez-Vidal},
  {Gosset}, {Guy}, {Halbwachs}, {Hambly}, {Harrison}, {Hern{\'a}ndez},
  {Hestroffer}, {Hodgkin}, {Hutton}, {Jasniewicz}, {Jean-Antoine-Piccolo},
  {Jordan}, {Korn}, {Krone-Martins}, {Lanzafame}, {Lebzelter}, {L{\"o}ffler},
  {Manteiga}, {Marrese}, {Mart{\'\i}n-Fleitas}, {Moitinho}, {Mora}, {Muinonen},
  {Osinde}, {Pancino}, {Pauwels}, {Petit}, {Recio-Blanco}, {Richards},
  {Rimoldini}, {Robin}, {Sarro}, {Siopis}, {Smith}, {Sozzetti}, {S{\"u}veges},
  {Torra}, {van Reeven}, {Abbas}, {Abreu Aramburu}, {Accart}, {Aerts},
  {Altavilla}, {{\'A}lvarez}, {Alvarez}, {Alves}, {Anderson}, {Andrei},
  {Anglada Varela}, {Antiche}, {Antoja}, {Arcay}, {Astraatmadja}, {Bach},
  {Baker}, {Balaguer-N{\'u}{\~n}ez}, {Balm}, {Barache}, {Barata}, {Barbato},
  {Barblan}, {Barklem}, {Barrado}, {Barros}, {Barstow}, {Bartholom{\'e}
  Mu{\~n}oz}, {Bassilana}, {Becciani}, {Bellazzini}, {Berihuete}, {Bertone},
  {Bianchi}, {Bienaym{\'e}}, {Blanco-Cuaresma}, {Boch}, {Boeche}, {Bombrun},
  {Borrachero}, {Bossini}, {Bouquillon}, {Bourda}, {Bragaglia}, {Bramante},
  {Breddels}, {Bressan}, {Brouillet}, {Br{\"u}semeister}, {Brugaletta},
  {Bucciarelli}, {Burlacu}, {Busonero}, {Butkevich}, {Buzzi}, {Caffau},
  {Cancelliere}, {Cannizzaro}, {Cantat-Gaudin}, {Carballo}, {Carlucci},
  {Carrasco}, {Casamiquela}, {Castellani}, {Castro-Ginard}, {Charlot},
  {Chemin}, {Chiavassa}, {Cocozza}, {Costigan}, {Cowell}, {Crifo}, {Crosta},
  {Crowley}, {Cuypers}, {Dafonte}, {Damerdji}, {Dapergolas}, {David}, {David},
  {de Laverny}, {De Luise}, {De March}, {de Martino}, {de Souza}, {de Torres},
  {Debosscher}, {del Pozo}, {Delbo}, {Delgado}, {Delgado}, {Di Matteo},
  {Diakite}, {Diener}, {Distefano}, {Dolding}, {Drazinos}, {Dur{\'a}n},
  {Edvardsson}, {Enke}, {Eriksson}, {Esquej}, {Eynard Bontemps}, {Fabre},
  {Fabrizio}, {Faigler}, {Falc{\~a}o}, {Farr{\`a}s Casas}, {Federici},
  {Fedorets}, {Fernique}, {Figueras}, {Filippi}, {Findeisen}, {Fonti},
  {Fraile}, {Fraser}, {Fr{\'e}zouls}, {Gai}, {Galleti}, {Garabato},
  {Garc{\'\i}a-Sedano}, {Garofalo}, {Garralda}, {Gavel}, {Gavras}, {Gerssen},
  {Geyer}, {Giacobbe}, {Gilmore}, {Girona}, {Giuffrida}, {Glass}, {Gomes},
  {Granvik}, {Gueguen}, {Guerrier}, {Guiraud}, {Guti{\'e}rrez-S{\'a}nchez},
  {Haigron}, {Hatzidimitriou}, {Hauser}, {Haywood}, {Heiter}, {Helmi}, {Heu},
  {Hilger}, {Hobbs}, {Hofmann}, {Holland}, {Huckle}, {Hypki}, {Icardi},
  {Jan{\ss}en}, {Jevardat de Fombelle}, {Jonker}, {Juh{\'a}sz}, {Julbe},
  {Karampelas}, {Kewley}, {Klar}, {Kochoska}, {Kohley}, {Kolenberg},
  {Kontizas}, {Kontizas}, {Koposov}, {Kordopatis}, {Kostrzewa-Rutkowska},
  {Koubsky}, {Lambert}, {Lanza}, {Lasne}, {Lavigne}, {Le Fustec}, {Le
  Poncin-Lafitte}, {Lebreton}, {Leccia}, {Leclerc}, {Lecoeur-Taibi},
  {Lenhardt}, {Leroux}, {Liao}, {Licata}, {Lindstr{\o}m}, {Lister}, {Livanou},
  {Lobel}, {L{\'o}pez}, {Managau}, {Mann}, {Mantelet}, {Marchal}, {Marchant},
  {Marconi}, {Marinoni}, {Marschalk{\'o}}, {Marshall}, {Martino}, {Marton},
  {Mary}, {Massari}, {Matijevi{\v{c}}}, {Mazeh}, {McMillan}, {Messina},
  {Michalik}, {Millar}, {Molina}, {Molinaro}, {Moln{\'a}r}, {Montegriffo},
  {Mor}, {Morbidelli}, {Morel}, {Morris}, {Mulone}, {Muraveva}, {Musella},
  {Nelemans}, {Nicastro}, {Noval}, {O'Mullane}, {Ord{\'e}novic},
  {Ord{\'o}{\~n}ez-Blanco}, {Osborne}, {Pagani}, {Pagano}, {Pailler},
  {Palacin}, {Palaversa}, {Panahi}, {Pawlak}, {Piersimoni}, {Pineau}, {Plachy},
  {Plum}, {Poggio}, {Poujoulet}, {Pr{\v{s}}a}, {Pulone}, {Racero}, {Ragaini},
  {Rambaux}, {Ramos-Lerate}, {Regibo}, {Reyl{\'e}}, {Riclet}, {Ripepi}, {Riva},
  {Rivard}, {Rixon}, {Roegiers}, {Roelens}, {Romero-G{\'o}mez}, {Rowell},
  {Royer}, {Ruiz-Dern}, {Sadowski}, {Sagrist{\`a} Sell{\'e}s}, {Sahlmann},
  {Salgado}, {Salguero}, {Sanna}, {Santana-Ros}, {Sarasso}, {Savietto},
  {Schultheis}, {Sciacca}, {Segol}, {Segovia}, {S{\'e}gransan}, {Shih},
  {Siltala}, {Silva}, {Smart}, {Smith}, {Solano}, {Solitro}, {Sordo}, {Soria
  Nieto}, {Souchay}, {Spagna}, {Spoto}, {Stampa}, {Steele},
  {Steidelm{\"u}ller}, {Stephenson}, {Stoev}, {Suess}, {Surdej}, {Szabados},
  {Szegedi-Elek}, {Tapiador}, {Taris}, {Tauran}, {Taylor}, {Teixeira},
  {Terrett}, {Teyssandier}, {Thuillot}, {Titarenko}, {Torra Clotet}, {Turon},
  {Ulla}, {Utrilla}, {Uzzi}, {Vaillant}, {Valentini}, {Valette}, {van Elteren},
  {Van Hemelryck}, {van Leeuwen}, {Vaschetto}, {Vecchiato}, {Veljanoski},
  {Viala}, {Vicente}, {Vogt}, {von Essen}, {Voss}, {Votruba}, {Voutsinas},
  {Walmsley}, {Weiler}, {Wertz}, {Wevers}, {Wyrzykowski}, {Yoldas},
  {{\v{Z}}erjal}, {Ziaeepour}, {Zorec}, {Zschocke}, {Zucker}, {Zurbach}, \&
  {Zwitter}}]{2018A&A...616A...1G}
{Gaia Collaboration}, {Brown}, A.~G.~A., {Vallenari}, A., {et~al.} 2018, \aap,
  616, A1

\bibitem[{{Gaia Collaboration} {et~al.}(2021){Gaia Collaboration}, {Brown},
  {Vallenari}, {Prusti}, {de Bruijne}, {Babusiaux}, {Biermann}, {Creevey},
  {Evans}, {Eyer}, {Hutton}, {Jansen}, {Jordi}, {Klioner}, {Lammers},
  {Lindegren}, {Luri}, {Mignard}, {Panem}, {Pourbaix}, {Randich}, {Sartoretti},
  {Soubiran}, {Walton}, {Arenou}, {Bailer-Jones}, {Bastian}, {Cropper},
  {Drimmel}, {Katz}, {Lattanzi}, {van Leeuwen}, {Bakker}, {Cacciari},
  {Casta{\~n}eda}, {De Angeli}, {Ducourant}, {Fabricius}, {Fouesneau},
  {Fr{\'e}mat}, {Guerra}, {Guerrier}, {Guiraud}, {Jean-Antoine Piccolo},
  {Masana}, {Messineo}, {Mowlavi}, {Nicolas}, {Nienartowicz}, {Pailler},
  {Panuzzo}, {Riclet}, {Roux}, {Seabroke}, {Sordo}, {Tanga}, {Th{\'e}venin},
  {Gracia-Abril}, {Portell}, {Teyssier}, {Altmann}, {Andrae}, {Bellas-Velidis},
  {Benson}, {Berthier}, {Blomme}, {Brugaletta}, {Burgess}, {Busso}, {Carry},
  {Cellino}, {Cheek}, {Clementini}, {Damerdji}, {Davidson}, {Delchambre},
  {Dell'Oro}, {Fern{\'a}ndez-Hern{\'a}ndez}, {Galluccio}, {Garc{\'\i}a-Lario},
  {Garcia-Reinaldos}, {Gonz{\'a}lez-N{\'u}{\~n}ez}, {Gosset}, {Haigron},
  {Halbwachs}, {Hambly}, {Harrison}, {Hatzidimitriou}, {Heiter},
  {Hern{\'a}ndez}, {Hestroffer}, {Hodgkin}, {Holl}, {Jan{\ss}en}, {Jevardat de
  Fombelle}, {Jordan}, {Krone-Martins}, {Lanzafame}, {L{\"o}ffler}, {Lorca},
  {Manteiga}, {Marchal}, {Marrese}, {Moitinho}, {Mora}, {Muinonen}, {Osborne},
  {Pancino}, {Pauwels}, {Petit}, {Recio-Blanco}, {Richards}, {Riello},
  {Rimoldini}, {Robin}, {Roegiers}, {Rybizki}, {Sarro}, {Siopis}, {Smith},
  {Sozzetti}, {Ulla}, {Utrilla}, {van Leeuwen}, {van Reeven}, {Abbas}, {Abreu
  Aramburu}, {Accart}, {Aerts}, {Aguado}, {Ajaj}, {Altavilla}, {{\'A}lvarez},
  {{\'A}lvarez Cid-Fuentes}, {Alves}, {Anderson}, {Anglada Varela}, {Antoja},
  {Audard}, {Baines}, {Baker}, {Balaguer-N{\'u}{\~n}ez}, {Balbinot}, {Balog},
  {Barache}, {Barbato}, {Barros}, {Barstow}, {Bartolom{\'e}}, {Bassilana},
  {Bauchet}, {Baudesson-Stella}, {Becciani}, {Bellazzini}, {Bernet}, {Bertone},
  {Bianchi}, {Blanco-Cuaresma}, {Boch}, {Bombrun}, {Bossini}, {Bouquillon},
  {Bragaglia}, {Bramante}, {Breedt}, {Bressan}, {Brouillet}, {Bucciarelli},
  {Burlacu}, {Busonero}, {Butkevich}, {Buzzi}, {Caffau}, {Cancelliere},
  {C{\'a}novas}, {Cantat-Gaudin}, {Carballo}, {Carlucci}, {Carnerero},
  {Carrasco}, {Casamiquela}, {Castellani}, {Castro-Ginard}, {Castro Sampol},
  {Chaoul}, {Charlot}, {Chemin}, {Chiavassa}, {Cioni}, {Comoretto}, {Cooper},
  {Cornez}, {Cowell}, {Crifo}, {Crosta}, {Crowley}, {Dafonte}, {Dapergolas},
  {David}, {David}, {de Laverny}, {De Luise}, {De March}, {De Ridder}, {de
  Souza}, {de Teodoro}, {de Torres}, {del Peloso}, {del Pozo}, {Delbo},
  {Delgado}, {Delgado}, {Delisle}, {Di Matteo}, {Diakite}, {Diener},
  {Distefano}, {Dolding}, {Eappachen}, {Edvardsson}, {Enke}, {Esquej}, {Fabre},
  {Fabrizio}, {Faigler}, {Fedorets}, {Fernique}, {Fienga}, {Figueras},
  {Fouron}, {Fragkoudi}, {Fraile}, {Franke}, {Gai}, {Garabato},
  {Garcia-Gutierrez}, {Garc{\'\i}a-Torres}, {Garofalo}, {Gavras}, {Gerlach},
  {Geyer}, {Giacobbe}, {Gilmore}, {Girona}, {Giuffrida}, {Gomel}, {Gomez},
  {Gonzalez-Santamaria}, {Gonz{\'a}lez-Vidal}, {Granvik},
  {Guti{\'e}rrez-S{\'a}nchez}, {Guy}, {Hauser}, {Haywood}, {Helmi}, {Hidalgo},
  {Hilger}, {H{\l}adczuk}, {Hobbs}, {Holland}, {Huckle}, {Jasniewicz},
  {Jonker}, {Juaristi Campillo}, {Julbe}, {Karbevska}, {Kervella}, {Khanna},
  {Kochoska}, {Kontizas}, {Kordopatis}, {Korn}, {Kostrzewa-Rutkowska},
  {Kruszy{\'n}ska}, {Lambert}, {Lanza}, {Lasne}, {Le Campion}, {Le Fustec},
  {Lebreton}, {Lebzelter}, {Leccia}, {Leclerc}, {Lecoeur-Taibi}, {Liao},
  {Licata}, {Lindstr{\o}m}, {Lister}, {Livanou}, {Lobel}, {Madrero Pardo},
  {Managau}, {Mann}, {Marchant}, {Marconi}, {Marcos Santos}, {Marinoni},
  {Marocco}, {Marshall}, {Martin Polo}, {Mart{\'\i}n-Fleitas}, {Masip},
  {Massari}, {Mastrobuono-Battisti}, {Mazeh}, {McMillan}, {Messina},
  {Michalik}, {Millar}, {Mints}, {Molina}, {Molinaro}, {Moln{\'a}r},
  {Montegriffo}, {Mor}, {Morbidelli}, {Morel}, {Morris}, {Mulone}, {Munoz},
  {Muraveva}, {Murphy}, {Musella}, {Noval}, {Ord{\'e}novic}, {Orr{\`u}},
  {Osinde}, {Pagani}, {Pagano}, {Palaversa}, {Palicio}, {Panahi}, {Pawlak},
  {Pe{\~n}alosa Esteller}, {Penttil{\"a}}, {Piersimoni}, {Pineau}, {Plachy},
  {Plum}, {Poggio}, {Poretti}, {Poujoulet}, {Pr{\v{s}}a}, {Pulone}, {Racero},
  {Ragaini}, {Rainer}, {Raiteri}, {Rambaux}, {Ramos}, {Ramos-Lerate}, {Re
  Fiorentin}, {Regibo}, {Reyl{\'e}}, {Ripepi}, {Riva}, {Rixon}, {Robichon},
  {Robin}, {Roelens}, {Rohrbasser}, {Romero-G{\'o}mez}, {Rowell}, {Royer},
  {Rybicki}, {Sadowski}, {Sagrist{\`a} Sell{\'e}s}, {Sahlmann}, {Salgado},
  {Salguero}, {Samaras}, {Sanchez Gimenez}, {Sanna}, {Santove{\~n}a},
  {Sarasso}, {Schultheis}, {Sciacca}, {Segol}, {Segovia}, {S{\'e}gransan},
  {Semeux}, {Shahaf}, {Siddiqui}, {Siebert}, {Siltala}, {Slezak}, {Smart},
  {Solano}, {Solitro}, {Souami}, {Souchay}, {Spagna}, {Spoto}, {Steele},
  {Steidelm{\"u}ller}, {Stephenson}, {S{\"u}veges}, {Szabados}, {Szegedi-Elek},
  {Taris}, {Tauran}, {Taylor}, {Teixeira}, {Thuillot}, {Tonello}, {Torra},
  {Torra}, {Turon}, {Unger}, {Vaillant}, {van Dillen}, {Vanel}, {Vecchiato},
  {Viala}, {Vicente}, {Voutsinas}, {Weiler}, {Wevers}, {Wyrzykowski}, {Yoldas},
  {Yvard}, {Zhao}, {Zorec}, {Zucker}, {Zurbach}, \&
  {Zwitter}}]{2021A&A...649A...1G}
{Gaia Collaboration}, {Brown}, A.~G.~A., {Vallenari}, A., {et~al.} 2021, \aap,
  649, A1

\bibitem[{{Gaia Collaboration} {et~al.}(2016){Gaia Collaboration}, {Prusti},
  {de Bruijne}, {Brown}, {Vallenari}, {Babusiaux}, {Bailer-Jones}, {Bastian},
  {Biermann}, {Evans}, \& et~al.}]{2016A&A...595A...1G}
{Gaia Collaboration}, {Prusti}, T., {de Bruijne}, J.~H.~J., {et~al.} 2016,
  \aap, 595, A1

\bibitem[{{Gonz{\'a}lez-{\'A}lvarez} {et~al.}(2019){Gonz{\'a}lez-{\'A}lvarez},
  {Micela}, {Maldonado}, {Affer}, {Maggio}, {Lanza}, {Covino}, {Benatti},
  {Bignamini}, {Cosentino}, {Damasso}, {Desidera}, {Gonz{\'a}lez
  Hern{\'a}ndez}, {Mart{\'\i}nez-Fiorenzano}, {Pagano}, {Perger}, {Piotto},
  {Pinamonti}, {Rainer}, {Rebolo}, {Ribas}, {Scandariato}, {Sozzetti},
  {Su{\'a}rez Mascare{\~n}o}, \& {Toledo-Padr{\'o}n}}]{2019A&A...624A..27G}
{Gonz{\'a}lez-{\'A}lvarez}, E., {Micela}, G., {Maldonado}, J., {et~al.} 2019,
  \aap, 624, A27

\bibitem[{{Hartman} \& {Bakos}(2016)}]{2016A&C....17....1H}
{Hartman}, J.~D. \& {Bakos}, G.~{\'A}. 2016, Astronomy and Computing, 17, 1

\bibitem[{{Henden} {et~al.}(2015){Henden}, {Levine}, {Terrell}, \&
  {Welch}}]{2015AAS...22533616H}
{Henden}, A.~A., {Levine}, S., {Terrell}, D., \& {Welch}, D.~L. 2015, in
  American Astronomical Society Meeting Abstracts, Vol. 225, American
  Astronomical Society Meeting Abstracts \#225, 336.16

\bibitem[{{Howard} {et~al.}(2010){Howard}, {Marcy}, {Johnson}, {Fischer},
  {Wright}, {Isaacson}, {Valenti}, {Anderson}, {Lin}, \&
  {Ida}}]{2010Sci...330..653H}
{Howard}, A.~W., {Marcy}, G.~W., {Johnson}, J.~A., {et~al.} 2010, Science, 330,
  653

\bibitem[{{Howell} {et~al.}(2016){Howell}, {Everett}, {Horch}, {Winters},
  {Hirsch}, {Nusdeo}, \& {Scott}}]{2016ApJ...829L...2H}
{Howell}, S.~B., {Everett}, M.~E., {Horch}, E.~P., {et~al.} 2016, \apjl, 829,
  L2

\bibitem[{{Howell} {et~al.}(2011){Howell}, {Everett}, {Sherry}, {Horch}, \&
  {Ciardi}}]{2011AJ....142...19H}
{Howell}, S.~B., {Everett}, M.~E., {Sherry}, W., {Horch}, E., \& {Ciardi},
  D.~R. 2011, \aj, 142, 19

\bibitem[{Ida {et~al.}(2013)Ida, Lin, \& Nagasawa}]{Ida_2013}
Ida, S., Lin, D. N.~C., \& Nagasawa, M. 2013, The Astrophysical Journal, 775,
  42

\bibitem[{{Jeffers} {et~al.}(2020){Jeffers}, {Dreizler}, {Barnes}, {Haswell},
  {Nelson}, {Rodr{\'\i}guez}, {L{\'o}pez-Gonz‧lez}, {Morales}, {Luque},
  {Zechmeister}, {Vogt}, {Jenkins}, {Palle}, {Berdi {\~n}as}, {Coleman},
  {D{\'\i}az}, {Ribas}, {Jones}, {Butler}, {Tinney}, {Bailey}, {Carter},
  {O'Toole}, {Wittenmyer}, {Crane}, {Feng}, {Shectman}, {Teske}, {Reiners},
  {Amado}, \& {Anglada-Escud{\'e}}}]{2020Sci...368.1477J}
{Jeffers}, S.~V., {Dreizler}, S., {Barnes}, J.~R., {et~al.} 2020, Science, 368,
  1477

\bibitem[{{Jeffers} {et~al.}(2018){Jeffers}, {Sch{\"o}fer}, {Lamert},
  {Reiners}, {Montes}, {Caballero}, {Cort{\'e}s-Contreras}, {Marvin},
  {Passegger}, {Zechmeister}, {Quirrenbach}, {Alonso-Floriano}, {Amado},
  {Bauer}, {Casal}, {Diez Alonso}, {Herrero}, {Morales}, {Mundt}, {Ribas}, \&
  {Sarmiento}}]{2018A&A...614A..76J}
{Jeffers}, S.~V., {Sch{\"o}fer}, P., {Lamert}, A., {et~al.} 2018, \aap, 614,
  A76

\bibitem[{{Jenkins}(2002)}]{2002ApJ...575..493J}
{Jenkins}, J.~M. 2002, \apj, 575, 493

\bibitem[{{Jenkins} {et~al.}(2010){Jenkins}, {Chandrasekaran}, {McCauliff},
  {Caldwell}, {Tenenbaum}, {Li}, {Klaus}, {Cote}, \&
  {Middour}}]{2010SPIE.7740E..0DJ}
{Jenkins}, J.~M., {Chandrasekaran}, H., {McCauliff}, S.~D., {et~al.} 2010, in
  Society of Photo-Optical Instrumentation Engineers (SPIE) Conference Series,
  Vol. 7740, Software and Cyberinfrastructure for Astronomy, ed. N.~M.
  {Radziwill} \& A.~{Bridger}, 77400D

\bibitem[{{Jenkins} {et~al.}(2016){Jenkins}, {Twicken}, {McCauliff},
  {Campbell}, {Sanderfer}, {Lung}, {Mansouri-Samani}, {Girouard}, {Tenenbaum},
  {Klaus}, {Smith}, {Caldwell}, {Chacon}, {Henze}, {Heiges}, {Latham},
  {Morgan}, {Swade}, {Rinehart}, \& {Vanderspek}}]{2016SPIE.9913E..3EJ}
{Jenkins}, J.~M., {Twicken}, J.~D., {McCauliff}, S., {et~al.} 2016, in Society
  of Photo-Optical Instrumentation Engineers (SPIE) Conference Series, Vol.
  9913, Software and Cyberinfrastructure for Astronomy IV, ed. G.~{Chiozzi} \&
  J.~C. {Guzman}, 99133E

\bibitem[{{Johnson} \& {Soderblom}(1987)}]{1987AJ.....93..864J}
{Johnson}, D.~R.~H. \& {Soderblom}, D.~R. 1987, \aj, 93, 864

\bibitem[{{Kemmer} {et~al.}(2020){Kemmer}, {Stock}, {Kossakowski}, {Kaminski},
  {Molaverdikhani}, {Schlecker}, {Caballero}, {Amado}, {Astudillo-Defru},
  {Bonfils}, {Ciardi}, {Collins}, {Espinoza}, {Fukui}, {Hirano}, {Jenkins},
  {Latham}, {Matthews}, {Narita}, {Pall{\'e}}, {Parviainen}, {Quirrenbach},
  {Reiners}, {Ribas}, {Ricker}, {Schlieder}, {Seager}, {Vanderspek}, {Winn},
  {Almenara}, {B{\'e}jar}, {Bluhm}, {Bouchy}, {Boyd}, {Christiansen},
  {Cifuentes}, {Cloutier}, {Collins}, {Cort{\'e}s-Contreras}, {Crossfield},
  {Crouzet}, {de Leon}, {Della-Rose}, {Delfosse}, {Dreizler}, {Esparza-Borges},
  {Essack}, {Forveille}, {Figueira}, {Galad{\'\i}-Enr{\'\i}quez}, {Gan},
  {Glidden}, {Gonzales}, {Guerra}, {Harakawa}, {Hatzes}, {Henning}, {Herrero},
  {Hodapp}, {Hori}, {Howell}, {Ikoma}, {Isogai}, {Jeffers}, {K{\"u}rster},
  {Kawauchi}, {Kimura}, {Klagyivik}, {Kotani}, {Kurokawa}, {Kusakabe},
  {Kuzuhara}, {Lafarga}, {Livingston}, {Luque}, {Matson}, {Morales}, {Mori},
  {Muirhead}, {Murgas}, {Nishikawa}, {Nishiumi}, {Omiya}, {Reffert},
  {Rodr{\'\i}guez L{\'o}pez}, {Santos}, {Sch{\"o}fer}, {Schwarz}, {Shiao},
  {Tamura}, {Terada}, {Twicken}, {Ueda}, {Vievard}, {Watanabe}, \&
  {Zechmeister}}]{2020A&A...642A.236K}
{Kemmer}, J., {Stock}, S., {Kossakowski}, D., {et~al.} 2020, \aap, 642, A236

\bibitem[{{Kempton} {et~al.}(2018){Kempton}, {Bean}, {Louie}, {Deming}, {Koll},
  {Mansfield}, {Christiansen}, {L{\'o}pez-Morales}, {Swain}, {Zellem},
  {Ballard}, {Barclay}, {Barstow}, {Batalha}, {Beatty}, {Berta-Thompson},
  {Birkby}, {Buchhave}, {Charbonneau}, {Cowan}, {Crossfield}, {de Val-Borro},
  {Doyon}, {Dragomir}, {Gaidos}, {Heng}, {Hu}, {Kane}, {Kreidberg}, {Mallonn},
  {Morley}, {Narita}, {Nascimbeni}, {Pall{\'e}}, {Quintana}, {Rauscher},
  {Seager}, {Shkolnik}, {Sing}, {Sozzetti}, {Stassun}, {Valenti}, \& {von
  Essen}}]{2018PASP..130k4401K}
{Kempton}, E. M.~R., {Bean}, J.~L., {Louie}, D.~R., {et~al.} 2018, \pasp, 130,
  114401

\bibitem[{{Kim} {et~al.}(2021){Kim}, {Chung}, {Udalski}, {Gould}, {Albrow},
  {Jung}, {Hwang}, {Han}, {Ryu}, {Shin}, {Shvartzvald}, {Yee}, {Zang}, {Cha},
  {Kim}, {Kim}, {Kim}, {Lee}, {Lee}, {Lee}, {Park}, {Pogge}, {KMTNet
  Collaboration}, {Mr{\'o}z}, {Poleski}, {Wrona}, {Iwanek}, {Szyma{\'n}ski},
  {Skowron}, {Soszy{\'n}ski}, {Koz{\l}owski}, {Pietrukowicz}, {Ulaczyk}, \&
  {Rybicki}}]{2021MNRAS.503.2706K}
{Kim}, Y.~H., {Chung}, S.-J., {Udalski}, A., {et~al.} 2021, \mnras, 503, 2706

\bibitem[{{Kipping}(2013)}]{2013MNRAS.435.2152K}
{Kipping}, D.~M. 2013, \mnras, 435, 2152

\bibitem[{{Kochanek} {et~al.}(2017){Kochanek}, {Shappee}, {Stanek}, {Holoien},
  {Thompson}, {Prieto}, {Dong}, {Shields}, {Will}, {Britt}, {Perzanowski}, \&
  {Pojma{\'n}ski}}]{2017PASP..129j4502K}
{Kochanek}, C.~S., {Shappee}, B.~J., {Stanek}, K.~Z., {et~al.} 2017, \pasp,
  129, 104502

\bibitem[{{Kopparapu} {et~al.}(2014){Kopparapu}, {Ramirez}, {SchottelKotte},
  {Kasting}, {Domagal-Goldman}, \& {Eymet}}]{2014ApJ...787L..29K}
{Kopparapu}, R.~K., {Ramirez}, R.~M., {SchottelKotte}, J., {et~al.} 2014,
  \apjl, 787, L29

\bibitem[{{Kov{\'a}cs} {et~al.}(2002){Kov{\'a}cs}, {Zucker}, \&
  {Mazeh}}]{2002A&A...391..369K}
{Kov{\'a}cs}, G., {Zucker}, S., \& {Mazeh}, T. 2002, \aap, 391, 369

\bibitem[{{Kratter} \& {Lodato}(2016)}]{2016ARA&A..54..271K}
{Kratter}, K. \& {Lodato}, G. 2016, \araa, 54, 271

\bibitem[{{Kreidberg}(2015)}]{2015PASP..127.1161K}
{Kreidberg}, L. 2015, \pasp, 127, 1161

\bibitem[{{Laskar} \& {Petit}(2017)}]{2017A&A...605A..72L}
{Laskar}, J. \& {Petit}, A.~C. 2017, \aap, 605, A72

\bibitem[{{Laughlin} {et~al.}(2004){Laughlin}, {Bodenheimer}, \&
  {Adams}}]{2004ApJ...612L..73L}
{Laughlin}, G., {Bodenheimer}, P., \& {Adams}, F.~C. 2004, \apjl, 612, L73

\bibitem[{{Leto} {et~al.}(1997){Leto}, {Pagano}, {Buemi}, \&
  {Rodono}}]{1997A&A...327.1114L}
{Leto}, G., {Pagano}, I., {Buemi}, C.~S., \& {Rodono}, M. 1997, \aap, 327, 1114

\bibitem[{{Li} {et~al.}(2019){Li}, {Tenenbaum}, {Twicken}, {Burke}, {Jenkins},
  {Quintana}, {Rowe}, \& {Seader}}]{Li:DVmodelFit2019}
{Li}, J., {Tenenbaum}, P., {Twicken}, J.~D., {et~al.} 2019, \pasp, 131, 024506

\bibitem[{{Lillo-Box} {et~al.}(2020{\natexlab{a}}){Lillo-Box}, {Figueira},
  {Leleu}, {Acu{\~n}a}, {Faria}, {Hara}, {Santos}, {Correia}, {Robutel},
  {Deleuil}, {Barrado}, {Sousa}, {Bonfils}, {Mousis}, {Almenara},
  {Astudillo-Defru}, {Marcq}, {Udry}, {Lovis}, \& {Pepe}}]{2020A&A...642A.121L}
{Lillo-Box}, J., {Figueira}, P., {Leleu}, A., {et~al.} 2020{\natexlab{a}},
  \aap, 642, A121

\bibitem[{{Lillo-Box} {et~al.}(2020{\natexlab{b}}){Lillo-Box}, {Lopez},
  {Santerne}, {Nielsen}, {Barros}, {Deleuil}, {Acu{\~n}a}, {Mousis}, {Sousa},
  {Adibekyan}, {Armstrong}, {Barrado}, {Bayliss}, {Brown}, {Demangeon},
  {Dumusque}, {Figueira}, {Hojjatpanah}, {Osborn}, {Santos}, \&
  {Udry}}]{2020A&A...640A..48L}
{Lillo-Box}, J., {Lopez}, T.~A., {Santerne}, A., {et~al.} 2020{\natexlab{b}},
  \aap, 640, A48

\bibitem[{{Lindegren} {et~al.}(2021){Lindegren}, {Klioner}, {Hern{\'a}ndez},
  {Bombrun}, {Ramos-Lerate}, {Steidelm{\"u}ller}, {Bastian}, {Biermann}, {de
  Torres}, {Gerlach}, {Geyer}, {Hilger}, {Hobbs}, {Lammers}, {McMillan},
  {Stephenson}, {Casta{\~n}eda}, {Davidson}, {Fabricius}, {Gracia-Abril},
  {Portell}, {Rowell}, {Teyssier}, {Torra}, {Bartolom{\'e}}, {Clotet},
  {Garralda}, {Gonz{\'a}lez-Vidal}, {Torra}, {Abbas}, {Altmann}, {Anglada
  Varela}, {Balaguer-N{\'u}{\~n}ez}, {Balog}, {Barache}, {Becciani}, {Bernet},
  {Bertone}, {Bianchi}, {Bouquillon}, {Brown}, {Bucciarelli}, {Busonero},
  {Butkevich}, {Buzzi}, {Cancelliere}, {Carlucci}, {Charlot}, {Cioni},
  {Crosta}, {Crowley}, {del Peloso}, {del Pozo}, {Drimmel}, {Esquej}, {Fienga},
  {Fraile}, {Gai}, {Garcia-Reinaldos}, {Guerra}, {Hambly}, {Hauser},
  {Jan{\ss}en}, {Jordan}, {Kostrzewa-Rutkowska}, {Lattanzi}, {Liao}, {Licata},
  {Lister}, {L{\"o}ffler}, {Marchant}, {Masip}, {Mignard}, {Mints}, {Molina},
  {Mora}, {Morbidelli}, {Murphy}, {Pagani}, {Panuzzo}, {Pe{\~n}alosa Esteller},
  {Poggio}, {Re Fiorentin}, {Riva}, {Sagrist{\`a} Sell{\'e}s}, {Sanchez
  Gimenez}, {Sarasso}, {Sciacca}, {Siddiqui}, {Smart}, {Souami}, {Spagna},
  {Steele}, {Taris}, {Utrilla}, {van Reeven}, \&
  {Vecchiato}}]{2021A&A...649A...2L}
{Lindegren}, L., {Klioner}, S.~A., {Hern{\'a}ndez}, J., {et~al.} 2021, \aap,
  649, A2

\bibitem[{{Lindor} {et~al.}(2021){Lindor}, {Hartman}, {Bakos}, {Bhatti},
  {Csubry}, {Penev}, {Bieryla}, {Latham}, {Torres}, {Buchhave}, {de Val-Borro},
  {Howard}, {Isaacson}, {Fulton}, {Boisse}, {Santerne}, {H{\'e}brard},
  {Kov{\'a}cs}, {Huang}, {Dembicky}, {Falco}, {Everett}, {Horch},
  {L{\'a}z{\'a}r}, {Papp}, \& {S{\'a}ri}}]{2021AJ....161...64L}
{Lindor}, B.~M., {Hartman}, J.~D., {Bakos}, G.~{\'A}., {et~al.} 2021, \aj, 161,
  64

\bibitem[{{Luhman}(2018)}]{2018AJ....156..271L}
{Luhman}, K.~L. 2018, \aj, 156, 271

\bibitem[{{Luque} {et~al.}(2019){Luque}, {Pall{\'e}}, {Kossakowski},
  {Dreizler}, {Kemmer}, {Espinoza}, {Burt}, {Anglada-Escud{\'e}}, {B{\'e}jar},
  {Caballero}, {Collins}, {Collins}, {Cort{\'e}s-Contreras},
  {D{\'\i}ez-Alonso}, {Feng}, {Hatzes}, {Hellier}, {Henning}, {Jeffers},
  {Kaltenegger}, {K{\"u}rster}, {Madden}, {Molaverdikhani}, {Montes}, {Narita},
  {Nowak}, {Ofir}, {Oshagh}, {Parviainen}, {Quirrenbach}, {Reffert}, {Reiners},
  {Rodr{\'\i}guez-L{\'o}pez}, {Schlecker}, {Stock}, {Trifonov}, {Winn},
  {Zapatero Osorio}, {Zechmeister}, {Amado}, {Anderson}, {Batalha}, {Bauer},
  {Bluhm}, {Burke}, {Butler}, {Caldwell}, {Chen}, {Crane}, {Dragomir},
  {Dressing}, {Dynes}, {Jenkins}, {Kaminski}, {Klahr}, {Kotani}, {Lafarga},
  {Latham}, {Lewin}, {McDermott}, {Monta{\~n}{\'e}s-Rodr{\'\i}guez}, {Morales},
  {Murgas}, {Nagel}, {Pedraz}, {Ribas}, {Ricker}, {Rowden}, {Seager},
  {Shectman}, {Tamura}, {Teske}, {Twicken}, {Vanderspeck}, {Wang}, \&
  {Wohler}}]{2019A&A...628A..39L}
{Luque}, R., {Pall{\'e}}, E., {Kossakowski}, D., {et~al.} 2019, \aap, 628, A39

\bibitem[{{Madhusudhan}(2019)}]{2019ARA&A..57..617M}
{Madhusudhan}, N. 2019, \araa, 57, 617

\bibitem[{{Mainzer} {et~al.}(2011){Mainzer}, {Bauer}, {Grav}, {Masiero},
  {Cutri}, {Dailey}, {Eisenhardt}, {McMillan}, {Wright}, {Walker}, {Jedicke},
  {Spahr}, {Tholen}, {Alles}, {Beck}, {Brandenburg}, {Conrow}, {Evans},
  {Fowler}, {Jarrett}, {Marsh}, {Masci}, {McCallon}, {Wheelock}, {Wittman},
  {Wyatt}, {DeBaun}, {Elliott}, {Elsbury}, {Gautier}, {Gomillion}, {Leisawitz},
  {Maleszewski}, {Micheli}, \& {Wilkins}}]{2011ApJ...731...53M}
{Mainzer}, A., {Bauer}, J., {Grav}, T., {et~al.} 2011, \apj, 731, 53

\bibitem[{{Mart{\'\i}nez-Rodr{\'\i}guez}
  {et~al.}(2019){Mart{\'\i}nez-Rodr{\'\i}guez}, {Caballero}, {Cifuentes},
  {Piro}, \& {Barnes}}]{2019ApJ...887..261M}
{Mart{\'\i}nez-Rodr{\'\i}guez}, H., {Caballero}, J.~A., {Cifuentes}, C.,
  {Piro}, A.~L., \& {Barnes}, R. 2019, \apj, 887, 261

\bibitem[{{McQuillan} {et~al.}(2014){McQuillan}, {Mazeh}, \&
  {Aigrain}}]{2014ApJS..211...24M}
{McQuillan}, A., {Mazeh}, T., \& {Aigrain}, S. 2014, \apjs, 211, 24

\bibitem[{{Millholland} \& {Spalding}(2020)}]{2020DDA....5150302M}
{Millholland}, S. \& {Spalding}, C. 2020, in AAS/Division of Dynamical
  Astronomy Meeting, Vol.~52, 503.02

\bibitem[{{Morales} {et~al.}(2019){Morales}, {Mustill}, {Ribas}, {Davies},
  {Reiners}, {Bauer}, {Kossakowski}, {Herrero}, {Rodr{\'\i}guez},
  {L{\'o}pez-Gonz{\'a}lez}, {Rodr{\'\i}guez-L{\'o}pez}, {B{\'e}jar},
  {Gonz{\'a}lez-Cuesta}, {Luque}, {Pall{\'e}}, {Perger}, {Baroch}, {Johansen},
  {Klahr}, {Mordasini}, {Anglada-Escud{\'e}}, {Caballero},
  {Cort{\'e}s-Contreras}, {Dreizler}, {Lafarga}, {Nagel}, {Passegger},
  {Reffert}, {Rosich}, {Schweitzer}, {Tal-Or}, {Trifonov}, {Zechmeister},
  {Quirrenbach}, {Amado}, {Guenther}, {Hagen}, {Henning}, {Jeffers},
  {Kaminski}, {K{\"u}rster}, {Montes}, {Seifert}, {Abell{\'a}n}, {Abril},
  {Aceituno}, {Aceituno}, {Alonso-Floriano}, {Ammler-von Eiff}, {Antona},
  {Arroyo-Torres}, {Azzaro}, {Barrado}, {Becerril-Jarque}, {Ben{\'\i}tez},
  {Berdi{\~n}as}, {Bergond}, {Brinkm{\"o}ller}, {del Burgo}, {Burn},
  {Calvo-Ortega}, {Cano}, {C{\'a}rdenas}, {Guill{\'e}n}, {Carro}, {Casal},
  {Casanova}, {Casasayas-Barris}, {Chaturvedi}, {Cifuentes}, {Claret},
  {Colom{\'e}}, {Czesla}, {D{\'\i}ez-Alonso}, {Dorda}, {Emsenhuber},
  {Fern{\'a}ndez}, {Fern{\'a}ndez-Mart{\'\i}n}, {Ferro}, {Fuhrmeister},
  {Galad{\'\i}-Enr{\'\i}quez}, {Cava}, {Vargas}, {Garcia-Piquer}, {Gesa},
  {Gonz{\'a}lez-{\'A}lvarez}, {Hern{\'a}ndez}, {Gonz{\'a}lez-Peinado},
  {Gu{\`a}rdia}, {Guijarro}, {de Guindos}, {Hatzes}, {Hauschildt}, {Hedrosa},
  {Hermelo}, {Arabi}, {Otero}, {Hintz}, {Holgado}, {Huber}, {Huke}, {Johnson},
  {de Juan}, {Kehr}, {Kemmer}, {Kim}, {Kl{\"u}ter}, {Klutsch}, {Labarga},
  {Labiche}, {Lalitha}, {Lamp{\'o}n}, {Lara}, {Launhardt}, {L{\'a}zaro},
  {Lizon}, {Llamas}, {Lodieu}, {L{\'o}pez del Fresno}, {Salas},
  {L{\'o}pez-Santiago}, {Madinabeitia}, {Mall}, {Mancini}, {Mand el}, {Marfil},
  {Molina}, {Mart{\'\i}n}, {Mart{\'\i}n-Fern{\'a}ndez}, {Mart{\'\i}n-Ruiz},
  {Mart{\'\i}nez-Rodr{\'\i}guez}, {Marvin}, {Mirabet}, {Moya}, {Naranjo},
  {Nelson}, {Nortmann}, {Nowak}, {Ofir}, {Pascual}, {Pavlov}, {Pedraz},
  {Medialdea}, {P{\'e}rez-Calpena}, {Perryman}, {Rabaza}, {Ballesta}, {Rebolo},
  {Redondo}, {Rix}, {Rodler}, {Trinidad}, {Sabotta}, {Sadegi}, {Salz},
  {S{\'a}nchez-Blanco}, {Carrasco}, {S{\'a}nchez-L{\'o}pez}, {Sanz-Forcada},
  {Sarkis}, {Sarmiento}, {Sch{\"a}fer}, {Schlecker}, {Schmitt}, {Sch{\"o}fer},
  {Solano}, {Sota}, {Stahl}, {Stock}, {Stuber}, {St{\"u}rmer}, {Su{\'a}rez},
  {Tabernero}, {Tulloch}, {Veredas}, {Vico-Linares}, {Vilardell}, {Wagner},
  {Winkler}, {Wolthoff}, {Yan}, \& {Osorio}}]{2019Sci...365.1441M}
{Morales}, J.~C., {Mustill}, A.~J., {Ribas}, I., {et~al.} 2019, Science, 365,
  1441

\bibitem[{{Morris} {et~al.}(2020){Morris}, {Twicken}, {Smith}, {Clarke},
  {Jenkins}, {Bryson}, {Girouard}, \& {Klaus}}]{2020ksci.rept....6M}
{Morris}, R.~L., {Twicken}, J.~D., {Smith}, J.~C., {et~al.} 2020, {Kepler Data
  Processing Handbook: Photometric Analysis}, Kepler Science Document
  KSCI-19081-003

\bibitem[{{Mortier} {et~al.}(2015){Mortier}, {Faria}, {Correia}, {Santerne}, \&
  {Santos}}]{2015A&A...573A.101M}
{Mortier}, A., {Faria}, J.~P., {Correia}, C.~M., {Santerne}, A., \& {Santos},
  N.~C. 2015, \aap, 573, A101

\bibitem[{{Nowak} {et~al.}(2020){Nowak}, {Luque}, {Parviainen}, {Pall{\'e}},
  {Molaverdikhani}, {B{\'e}jar}, {Lillo-Box}, {Rodr{\'\i}guez-L{\'o}pez},
  {Caballero}, {Zechmeister}, {Passegger}, {Cifuentes}, {Schweitzer}, {Narita},
  {Cale}, {Espinoza}, {Murgas}, {Hidalgo}, {Zapatero Osorio}, {Pozuelos},
  {Aceituno}, {Amado}, {Barkaoui}, {Barrado}, {Bauer}, {Benkhaldoun},
  {Caldwell}, {Casasayas Barris}, {Chaturvedi}, {Chen}, {Collins}, {Collins},
  {Cort{\'e}s-Contreras}, {Crossfield}, {de Le{\'o}n}, {D{\'\i}ez Alonso},
  {Dreizler}, {El Mufti}, {Esparza-Borges}, {Essack}, {Fukui}, {Gaidos},
  {Gillon}, {Gonzales}, {Guerra}, {Hatzes}, {Henning}, {Herrero}, {Hesse},
  {Hirano}, {Howell}, {Jeffers}, {Jehin}, {Jenkins}, {Kaminski}, {Kemmer},
  {Kielkopf}, {Kossakowski}, {Kotani}, {K{\"u}rster}, {Lafarga}, {Latham},
  {Law}, {Lissauer}, {Lodieu}, {Madrigal-Aguado}, {Mann}, {Massey}, {Matson},
  {Matthews}, {Monta{\~n}{\'e}s-Rodr{\'\i}guez}, {Montes}, {Morales}, {Mori},
  {Nagel}, {Oshagh}, {Pedraz}, {Plavchan}, {Pollacco}, {Quirrenbach},
  {Reffert}, {Reiners}, {Ribas}, {Ricker}, {Rose}, {Schlecker}, {Schlieder},
  {Seager}, {Stangret}, {Stock}, {Tamura}, {Tanner}, {Teske}, {Trifonov},
  {Twicken}, {Vanderspek}, {Watanabe}, {Wittrock}, {Ziegler}, \&
  {Zohrabi}}]{2020A&A...642A.173N}
{Nowak}, G., {Luque}, R., {Parviainen}, H., {et~al.} 2020, \aap, 642, A173

\bibitem[{{Osten} {et~al.}(2005){Osten}, {Hawley}, {Allred}, {Johns-Krull}, \&
  {Roark}}]{2005ApJ...621..398O}
{Osten}, R.~A., {Hawley}, S.~L., {Allred}, J.~C., {Johns-Krull}, C.~M., \&
  {Roark}, C. 2005, \apj, 621, 398

\bibitem[{{Passegger} {et~al.}(2019){Passegger}, {Schweitzer}, {Shulyak},
  {Nagel}, {Hauschildt}, {Reiners}, {Amado}, {Caballero},
  {Cort{\'e}s-Contreras}, {Dom{\'\i}nguez-Fern{\'a}ndez}, {Quirrenbach},
  {Ribas}, {Azzaro}, {Anglada-Escud{\'e}}, {Bauer}, {B{\'e}jar}, {Dreizler},
  {Guenther}, {Henning}, {Jeffers}, {Kaminski}, {K{\"u}rster}, {Lafarga},
  {Mart{\'\i}n}, {Montes}, {Morales}, {Schmitt}, \&
  {Zechmeister}}]{2019A&A...627A.161P}
{Passegger}, V.~M., {Schweitzer}, A., {Shulyak}, D., {et~al.} 2019, \aap, 627,
  A161

\bibitem[{{Pecaut} \& {Mamajek}(2013)}]{2013ApJS..208....9P}
{Pecaut}, M.~J. \& {Mamajek}, E.~E. 2013, \apjs, 208, 9

\bibitem[{{Petigura}(2020)}]{2020AJ....160...89P}
{Petigura}, E.~A. 2020, \aj, 160, 89

\bibitem[{{Quirrenbach} {et~al.}(2016){Quirrenbach}, {Amado}, {Caballero},
  {Mundt}, {Reiners}, {Ribas}, {Seifert}, {Abril}, {Aceituno},
  {Alonso-Floriano}, {Anwand-Heerwart}, {Azzaro}, {Bauer}, {Barrado},
  {Becerril}, {Bejar}, {Benitez}, {Berdinas}, {Brinkm{\"o}ller}, {Cardenas},
  {Casal}, {Claret}, {Colom{\'e}}, {Cortes-Contreras}, {Czesla}, {Doellinger},
  {Dreizler}, {Feiz}, {Fernandez}, {Ferro}, {Fuhrmeister}, {Galadi},
  {Gallardo}, {G{\'a}lvez-Ortiz}, {Garcia-Piquer}, {Garrido}, {Gesa},
  {G{\'o}mez Galera}, {Gonz{\'a}lez Hern{\'a}ndez}, {Gonzalez Peinado},
  {Gr{\"o}zinger}, {Gu{\`a}rdia}, {Guenther}, {de Guindos}, {Hagen}, {Hatzes},
  {Hauschildt}, {Helmling}, {Henning}, {Hermann}, {Hern{\'a}ndez Arabi},
  {Hern{\'a}ndez Casta{\~n}o}, {Hern{\'a}ndez Hernando}, {Herrero}, {Huber},
  {Huber}, {Huke}, {Jeffers}, {de Juan}, {Kaminski}, {Kehr}, {Kim}, {Klein},
  {Kl{\"u}ter}, {K{\"u}rster}, {Lafarga}, {Lara}, {Lamert}, {Laun},
  {Launhardt}, {Lemke}, {Lenzen}, {Llamas}, {Lopez del Fresno},
  {L{\'o}pez-Puertas}, {L{\'o}pez-Santiago}, {Lopez Salas}, {Magan
  Madinabeitia}, {Mall}, {Mandel}, {Mancini}, {Marin Molina}, {Maroto
  Fern{\'a}ndez}, {Mart{\'{\i}}n}, {Mart{\'{\i}}n-Ruiz}, {Marvin}, {Mathar},
  {Mirabet}, {Montes}, {Morales}, {Morales Mu{\~n}oz}, {Nagel}, {Naranjo},
  {Nowak}, {Palle}, {Panduro}, {Passegger}, {Pavlov}, {Pedraz}, {Perez},
  {P{\'e}rez-Medialdea}, {Perger}, {Pluto}, {Ram{\'o}n}, {Rebolo}, {Redondo},
  {Reffert}, {Reinhart}, {Rhode}, {Rix}, {Rodler}, {Rodr{\'{\i}}guez},
  {Rodr{\'{\i}}guez L{\'o}pez}, {Rohloff}, {Rosich}, {Sanchez Carrasco},
  {Sanz-Forcada}, {Sarkis}, {Sarmiento}, {Sch{\"a}fer}, {Schiller}, {Schmidt},
  {Schmitt}, {Sch{\"o}fer}, {Schweitzer}, {Shulyak}, {Solano}, {Stahl},
  {Storz}, {Tabernero}, {Tala}, {Tal-Or}, {Ulbrich}, {Veredas}, {Vico Linares},
  {Vilardell}, {Wagner}, {Winkler}, {Zapatero Osorio}, {Zechmeister},
  {Ammler-von Eiff}, {Anglada-Escud{\'e}}, {del Burgo}, {Garcia-Vargas},
  {Klutsch}, {Lizon}, {Lopez-Morales}, {Ofir}, {P{\'e}rez-Calpena}, {Perryman},
  {S{\'a}nchez-Blanco}, {Strachan}, {St{\"u}rmer}, {Su{\'a}rez}, {Trifonov},
  {Tulloch}, \& {Xu}}]{2016SPIE.9908E..12Q}
{Quirrenbach}, A., {Amado}, P.~J., {Caballero}, J.~A., {et~al.} 2016, in
  \procspie, Vol. 9908, Ground-based and Airborne Instrumentation for Astronomy
  VI, 990812

\bibitem[{{Quirrenbach} {et~al.}(2018){Quirrenbach}, {Amado}, {Ribas},
  {Reiners}, {Caballero}, {Seifert}, {Aceituno}, {Azzaro}, {Baroch}, \&
  {Barrado}}]{2018SPIE10702E..0WQ}
{Quirrenbach}, A., {Amado}, P.~J., {Ribas}, I., {et~al.} 2018, in Society of
  Photo-Optical Instrumentation Engineers (SPIE) Conference Series, Vol. 10702,
  Ground-based and Airborne Instrumentation for Astronomy VII, 107020W

\bibitem[{{Reiners} {et~al.}(2018{\natexlab{a}}){Reiners}, {Ribas},
  {Zechmeister}, {Caballero}, {Trifonov}, {Dreizler}, {Morales}, {Tal-Or},
  {Lafarga}, {Quirrenbach}, {Amado}, {Kaminski}, {Jeffers}, {Aceituno},
  {B{\'e}jar}, {Gu{\`a}rdia}, {Guenther}, {Hagen}, {Montes}, {Passegger},
  {Seifert}, {Schweitzer}, {Cort{\'e}s-Contreras}, {Abril}, {Alonso-Floriano},
  {Eiff}, {Antona}, {Anglada-Escud{\'e}}, {Anwand-Heerwart}, {Arroyo-Torres},
  {Azzaro}, {Baroch}, {Barrado}, {Bauer}, {Becerril}, {Ben{\'{\i}}tez},
  {Berdi{\~n}as}, {Bergond}, {Bl{\"u}mcke}, {Brinkm{\"o}ller}, {del Burgo},
  {Cano}, {C{\'a}rdenas V{\'a}zquez}, {Casal}, {Cifuentes}, {Claret},
  {Colom{\'e}}, {Czesla}, {D{\'{\i}}ez-Alonso}, {Feiz}, {Fern{\'a}ndez},
  {Ferro}, {Fuhrmeister}, {Galad{\'{\i}}-Enr{\'{\i}}quez}, {Garcia-Piquer},
  {Garc{\'{\i}}a Vargas}, {Gesa}, {G{\'o}mez Galera}, {Gonz{\'a}lez
  Hern{\'a}ndez}, {Gonz{\'a}lez-Peinado}, {Gr{\"o}zinger}, {Grohnert},
  {Guijarro}, {de Guindos}, {Guti{\'e}rrez-Soto}, {Hatzes}, {Hauschildt},
  {Hedrosa}, {Helmling}, {Henning}, {Hermelo}, {Hern{\'a}ndez Arab{\'{\i}}},
  {Hern{\'a}ndez Casta{\~n}o}, {Hern{\'a}ndez Hernando}, {Herrero}, {Huber},
  {Huke}, {Johnson}, {de Juan}, {Kim}, {Klein}, {Kl{\"u}ter}, {Klutsch},
  {K{\"u}rster}, {Labarga}, {Lamert}, {Lamp{\'o}n}, {Lara}, {Laun}, {Lemke},
  {Lenzen}, {Launhardt}, {L{\'o}pez del Fresno}, {L{\'o}pez-Gonz{\'a}lez},
  {L{\'o}pez-Puertas}, {L{\'o}pez Salas}, {L{\'o}pez-Santiago}, {Luque},
  {Mag{\'a}n Madinabeitia}, {Mall}, {Mancini}, {Mandel}, {Marfil},
  {Mar{\'{\i}}n Molina}, {Maroto Fern{\'a}ndez}, {Mart{\'{\i}}n},
  {Mart{\'{\i}}n-Ruiz}, {Marvin}, {Mathar}, {Mirabet}, {Moreno-Raya}, {Moya},
  {Mundt}, {Nagel}, {Naranjo}, {Nortmann}, {Nowak}, {Ofir}, {Oreiro},
  {Pall{\'e}}, {Panduro}, {Pascual}, {Pavlov}, {Pedraz}, {P{\'e}rez-Calpena},
  {P{\'e}rez Medialdea}, {Perger}, {Perryman}, {Pluto}, {Rabaza}, {Ram{\'o}n},
  {Rebolo}, {Redondo}, {Reffert}, {Reinhart}, {Rhode}, {Rix}, {Rodler},
  {Rodr{\'{\i}}guez}, {Rodr{\'{\i}}guez-L{\'o}pez}, {Rodr{\'{\i}}guez
  Trinidad}, {Rohloff}, {Rosich}, {Sadegi}, {S{\'a}nchez-Blanco}, {S{\'a}nchez
  Carrasco}, {S{\'a}nchez-L{\'o}pez}, {Sanz-Forcada}, {Sarkis}, {Sarmiento},
  {Sch{\"a}fer}, {Schmitt}, {Schiller}, {Sch{\"o}fer}, {Solano}, {Stahl},
  {Strachan}, {St{\"u}rmer}, {Su{\'a}rez}, {Tabernero}, {Tala}, {Tulloch},
  {Ulbrich}, {Veredas}, {Vico Linares}, {Vilardell}, {Wagner}, {Winkler},
  {Wolthoff}, {Xu}, {Yan}, \& {Zapatero Osorio}}]{2018A&A...609L...5R}
{Reiners}, A., {Ribas}, I., {Zechmeister}, M., {et~al.} 2018{\natexlab{a}},
  \aap, 609, L5

\bibitem[{{Reiners} {et~al.}(2018{\natexlab{b}}){Reiners}, {Zechmeister},
  {Caballero}, {Ribas}, {Morales}, {Jeffers}, {Sch{\"o}fer}, {Tal-Or},
  {Quirrenbach}, {Amado}, {Kaminski}, {Seifert}, {Abril}, {Aceituno},
  {Alonso-Floriano}, {Ammler-von Eiff}, {Antona}, {Anglada-Escud{\'e}},
  {Anwand-Heerwart}, {Arroyo-Torres}, {Azzaro}, {Baroch}, {Barrado}, {Bauer},
  {Becerril}, {B{\'e}jar}, {Ben{\'{\i}}tez}, {Berdi{\~n}as}, {Bergond},
  {Bl{\"u}mcke}, {Brinkm{\"o}ller}, {del Burgo}, {Cano}, {C{\'a}rdenas
  V{\'a}zquez}, {Casal}, {Cifuentes}, {Claret}, {Colom{\'e}},
  {Cort{\'e}s-Contreras}, {Czesla}, {D{\'{\i}}ez-Alonso}, {Dreizler}, {Feiz},
  {Fern{\'a}ndez}, {Ferro}, {Fuhrmeister}, {Galad{\'{\i}}-Enr{\'{\i}}quez},
  {Garcia-Piquer}, {Garc{\'{\i}}a Vargas}, {Gesa}, {G{\'o}mez Galera},
  {Gonz{\'a}lez Hern{\'a}ndez}, {Gonz{\'a}lez-Peinado}, {Gr{\"o}zinger},
  {Grohnert}, {Gu{\`a}rdia}, {Guenther}, {Guijarro}, {de Guindos},
  {Guti{\'e}rrez-Soto}, {Hagen}, {Hatzes}, {Hauschildt}, {Hedrosa}, {Helmling},
  {Henning}, {Hermelo}, {Hern{\'a}ndez Arab{\'{\i}}}, {Hern{\'a}ndez
  Casta{\~n}o}, {Hern{\'a}ndez Hernando}, {Herrero}, {Huber}, {Huke},
  {Johnson}, {de Juan}, {Kim}, {Klein}, {Kl{\"u}ter}, {Klutsch}, {K{\"u}rster},
  {Lafarga}, {Lamert}, {Lamp{\'o}n}, {Lara}, {Laun}, {Lemke}, {Lenzen},
  {Launhardt}, {L{\'o}pez del Fresno}, {L{\'o}pez-Gonz{\'a}lez},
  {L{\'o}pez-Puertas}, {L{\'o}pez Salas}, {L{\'o}pez-Santiago}, {Luque},
  {Mag{\'a}n Madinabeitia}, {Mall}, {Mancini}, {Mandel}, {Marfil},
  {Mar{\'{\i}}n Molina}, {Maroto Fern{\'a}ndez}, {Mart{\'{\i}}n},
  {Mart{\'{\i}}n-Ruiz}, {Marvin}, {Mathar}, {Mirabet}, {Montes}, {Moreno-Raya},
  {Moya}, {Mundt}, {Nagel}, {Naranjo}, {Nortmann}, {Nowak}, {Ofir}, {Oreiro},
  {Pall{\'e}}, {Panduro}, {Pascual}, {Passegger}, {Pavlov}, {Pedraz},
  {P{\'e}rez-Calpena}, {P{\'e}rez Medialdea}, {Perger}, {Perryman}, {Pluto},
  {Rabaza}, {Ram{\'o}n}, {Rebolo}, {Redondo}, {Reffert}, {Reinhart}, {Rhode},
  {Rix}, {Rodler}, {Rodr{\'{\i}}guez}, {Rodr{\'{\i}}guez-L{\'o}pez},
  {Rodr{\'{\i}}guez Trinidad}, {Rohloff}, {Rosich}, {Sadegi},
  {S{\'a}nchez-Blanco}, {S{\'a}nchez Carrasco}, {S{\'a}nchez-L{\'o}pez},
  {Sanz-Forcada}, {Sarkis}, {Sarmiento}, {Sch{\"a}fer}, {Schmitt}, {Schiller},
  {Schweitzer}, {Solano}, {Stahl}, {Strachan}, {St{\"u}rmer}, {Su{\'a}rez},
  {Tabernero}, {Tala}, {Trifonov}, {Tulloch}, {Ulbrich}, {Veredas}, {Vico
  Linares}, {Vilardell}, {Wagner}, {Winkler}, {Wolthoff}, {Xu}, {Yan}, \&
  {Zapatero Osorio}}]{2018A&A...612A..49R}
{Reiners}, A., {Zechmeister}, M., {Caballero}, J.~A., {et~al.}
  2018{\natexlab{b}}, \aap, 612, A49

\bibitem[{{Reyl{\'e}} {et~al.}(2021){Reyl{\'e}}, {Jardine}, {Fouqu{\'e}},
  {Caballero}, {Smart}, \& {Sozzetti}}]{2021A&A...650A.201R}
{Reyl{\'e}}, C., {Jardine}, K., {Fouqu{\'e}}, P., {et~al.} 2021, \aap, 650,
  A201

\bibitem[{{Ricker} {et~al.}(2015){Ricker}, {Winn}, {Vanderspek}, {Latham},
  {Bakos}, {Bean}, {Berta-Thompson}, {Brown}, {Buchhave}, {Butler}, {Butler},
  {Chaplin}, {Charbonneau}, {Christensen-Dalsgaard}, {Clampin}, {Deming},
  {Doty}, {De Lee}, {Dressing}, {Dunham}, {Endl}, {Fressin}, {Ge}, {Henning},
  {Holman}, {Howard}, {Ida}, {Jenkins}, {Jernigan}, {Johnson}, {Kaltenegger},
  {Kawai}, {Kjeldsen}, {Laughlin}, {Levine}, {Lin}, {Lissauer}, {MacQueen},
  {Marcy}, {McCullough}, {Morton}, {Narita}, {Paegert}, {Palle}, {Pepe},
  {Pepper}, {Quirrenbach}, {Rinehart}, {Sasselov}, {Sato}, {Seager},
  {Sozzetti}, {Stassun}, {Sullivan}, {Szentgyorgyi}, {Torres}, {Udry}, \&
  {Villasenor}}]{2015JATIS...1a4003R}
{Ricker}, G.~R., {Winn}, J.~N., {Vanderspek}, R., {et~al.} 2015, Journal of
  Astronomical Telescopes, Instruments, and Systems, 1, 014003

\bibitem[{{Rivera} {et~al.}(2010){Rivera}, {Laughlin}, {Butler}, {Vogt},
  {Haghighipour}, \& {Meschiari}}]{2010ApJ...719..890R}
{Rivera}, E.~J., {Laughlin}, G., {Butler}, R.~P., {et~al.} 2010, \apj, 719, 890

\bibitem[{{Sanchis-Ojeda} {et~al.}(2014){Sanchis-Ojeda}, {Rappaport}, {Winn},
  {Kotson}, {Levine}, \& {El Mellah}}]{2014ApJ...787...47S}
{Sanchis-Ojeda}, R., {Rappaport}, S., {Winn}, J.~N., {et~al.} 2014, \apj, 787,
  47

\bibitem[{{Sanz-Forcada} {et~al.}(2011){Sanz-Forcada}, {Micela}, {Ribas},
  {Pollock}, {Eiroa}, {Velasco}, {Solano}, \&
  {Garc{\'{\i}}a-{\'A}lvarez}}]{2011A&A...532A...6S}
{Sanz-Forcada}, J., {Micela}, G., {Ribas}, I., {et~al.} 2011, \aap, 532, A6

\bibitem[{{Schlecker} {et~al.}(2020){Schlecker}, {Mordasini}, {Emsenhuber},
  {Klahr}, {Henning}, {Burn}, {Alibert}, \& {Benz}}]{2020arXiv200705563S}
{Schlecker}, M., {Mordasini}, C., {Emsenhuber}, A., {et~al.} 2020, arXiv
  e-prints, arXiv:2007.05563

\bibitem[{{Sch{\"o}fer} {et~al.}(2019){Sch{\"o}fer}, {Jeffers}, {Reiners},
  {Shulyak}, {Fuhrmeister}, {Johnson}, {Zechmeister}, {Ribas}, {Quirrenbach},
  {Amado}, {Caballero}, {Anglada-Escud{\'e}}, {Bauer}, {B{\'e}jar},
  {Cort{\'e}s-Contreras}, {Dreizler}, {Guenther}, {Kaminski}, {K{\"u}rster},
  {Lafarga}, {Montes}, {Morales}, {Pedraz}, \& {Tal-Or}}]{2019A&A...623A..44S}
{Sch{\"o}fer}, P., {Jeffers}, S.~V., {Reiners}, A., {et~al.} 2019, \aap, 623,
  A44

\bibitem[{{Schweitzer} {et~al.}(2019){Schweitzer}, {Passegger}, {Cifuentes},
  {B{\'e}jar}, {Cort{\'e}s-Contreras}, {Caballero}, {del Burgo}, {Czesla},
  {K{\"u}rster}, {Montes}, {Zapatero Osorio}, {Ribas}, {Reiners},
  {Quirrenbach}, {Amado}, {Aceituno}, {Anglada-Escud{\'e}}, {Bauer},
  {Dreizler}, {Jeffers}, {Guenther}, {Henning}, {Kaminski}, {Lafarga},
  {Marfil}, {Morales}, {Schmitt}, {Seifert}, {Solano}, {Tabernero}, \&
  {Zechmeister}}]{2019A&A...625A..68S}
{Schweitzer}, A., {Passegger}, V.~M., {Cifuentes}, C., {et~al.} 2019, \aap,
  625, A68

\bibitem[{{Shappee} {et~al.}(2014){Shappee}, {Prieto}, {Grupe}, {Kochanek},
  {Stanek}, {De Rosa}, {Mathur}, {Zu}, {Peterson}, {Pogge}, {Komossa}, {Im},
  {Jencson}, {Holoien}, {Basu}, {Beacom}, {Szczygie{\l}}, {Brimacombe},
  {Adams}, {Campillay}, {Choi}, {Contreras}, {Dietrich}, {Dubberley},
  {Elphick}, {Foale}, {Giustini}, {Gonzalez}, {Hawkins}, {Howell}, {Hsiao},
  {Koss}, {Leighly}, {Morrell}, {Mudd}, {Mullins}, {Nugent}, {Parrent},
  {Phillips}, {Pojmanski}, {Rosing}, {Ross}, {Sand}, {Terndrup}, {Valenti},
  {Walker}, \& {Yoon}}]{2014ApJ...788...48S}
{Shappee}, B.~J., {Prieto}, J.~L., {Grupe}, D., {et~al.} 2014, \apj, 788, 48

\bibitem[{{Skrutskie} {et~al.}(2006){Skrutskie}, {Cutri}, {Stiening},
  {Weinberg}, {Schneider}, {Carpenter}, {Beichman}, {Capps}, {Chester},
  {Elias}, {Huchra}, {Liebert}, {Lonsdale}, {Monet}, {Price}, {Seitzer},
  {Jarrett}, {Kirkpatrick}, {Gizis}, {Howard}, {Evans}, {Fowler}, {Fullmer},
  {Hurt}, {Light}, {Kopan}, {Marsh}, {McCallon}, {Tam}, {Van Dyk}, \&
  {Wheelock}}]{2006AJ....131.1163S}
{Skrutskie}, M.~F., {Cutri}, R.~M., {Stiening}, R., {et~al.} 2006, \aj, 131,
  1163

\bibitem[{{Smith} {et~al.}(2012){Smith}, {Stumpe}, {Van Cleve}, {Jenkins},
  {Barclay}, {Fanelli}, {Girouard}, {Kolodziejczak}, {McCauliff}, {Morris}, \&
  {Twicken}}]{2012PASP..124.1000S}
{Smith}, J.~C., {Stumpe}, M.~C., {Van Cleve}, J.~E., {et~al.} 2012, \pasp, 124,
  1000

\bibitem[{{Soto} {et~al.}(2021){Soto}, {Anglada-Escud{\'e}}, {Dreizler},
  {Molaverdikhani}, {Kemmer}, {Rodr{\'\i}guez-L{\'o}pez}, {Lillo-Box},
  {Pall{\'e}}, {Espinoza}, {Caballero}, {Quirrenbach}, {Ribas}, {Reiners},
  {Narita}, {Hirano}, {Amado}, {B{\'e}jar}, {Bluhm}, {Burke}, {Caldwell},
  {Charbonneau}, {Cloutier}, {Collins}, {Cort{\'e}s-Contreras}, {Girardin},
  {Guerra}, {Harakawa}, {Hatzes}, {Irwin}, {Jenkins}, {Jensen}, {Kawauchi},
  {Kotani}, {Kudo}, {Kunimoto}, {Kuzuhara}, {Latham}, {Montes}, {Morales},
  {Mori}, {Nelson}, {Omiya}, {Pedraz}, {Passegger}, {Rackham}, {Rudat},
  {Schlieder}, {Sch{\"o}fer}, {Schweitzer}, {Selezneva}, {Stockdale}, {Tamura},
  {Trifonov}, {Vanderspek}, \& {Watanabe}}]{2021A&A...649A.144S}
{Soto}, M.~G., {Anglada-Escud{\'e}}, G., {Dreizler}, S., {et~al.} 2021, \aap,
  649, A144

\bibitem[{{Stassun} {et~al.}(2018){Stassun}, {Oelkers}, {Pepper}, {Paegert},
  {De Lee}, {Torres}, {Latham}, {Charpinet}, {Dressing}, {Huber}, {Kane},
  {L{\'e}pine}, {Mann}, {Muirhead}, {Rojas-Ayala}, {Silvotti}, {Fleming},
  {Levine}, \& {Plavchan}}]{2018AJ....156..102S}
{Stassun}, K.~G., {Oelkers}, R.~J., {Pepper}, J., {et~al.} 2018, \aj, 156, 102

\bibitem[{{Stock} {et~al.}(2020){Stock}, {Kemmer}, {Reffert}, {Trifonov},
  {Kaminski}, {Dreizler}, {Quirrenbach}, {Caballero}, {Reiners}, {Jeffers},
  {Anglada-Escud{\'e}}, {Ribas}, {Amado}, {Barrado}, {Barnes}, {Bauer},
  {Berdi{\~n}as}, {B{\'e}jar}, {Coleman}, {Cort{\'e}s-Contreras},
  {D{\'\i}ez-Alonso}, {Dom{\'\i}nguez-Fern{\'a}ndez}, {Espinoza}, {Haswell},
  {Hatzes}, {Henning}, {Jenkins}, {Jones}, {Kossakowski}, {K{\"u}rster},
  {Lafarga}, {Lee}, {L{\'o}pez Gonz{\'a}lez}, {Montes}, {Morales}, {Morales},
  {Pall{\'e}}, {Pedraz}, {Rodr{\'\i}guez}, {Rodr{\'\i}guez-L{\'o}pez}, \&
  {Zechmeister}}]{2020A&A...636A.119S}
{Stock}, S., {Kemmer}, J., {Reffert}, S., {et~al.} 2020, \aap, 636, A119

\bibitem[{{Stumpe} {et~al.}(2014){Stumpe}, {Smith}, {Catanzarite}, {Van Cleve},
  {Jenkins}, {Twicken}, \& {Girouard}}]{Stumpe2014}
{Stumpe}, M.~C., {Smith}, J.~C., {Catanzarite}, J.~H., {et~al.} 2014, \pasp,
  126, 100

\bibitem[{{Stumpe} {et~al.}(2012){Stumpe}, {Smith}, {Van Cleve}, {Twicken},
  {Barclay}, {Fanelli}, {Girouard}, {Jenkins}, {Kolodziejczak}, {McCauliff}, \&
  {Morris}}]{Stumpe2012}
{Stumpe}, M.~C., {Smith}, J.~C., {Van Cleve}, J.~E., {et~al.} 2012, \pasp, 124,
  985

\bibitem[{{Su{\'a}rez Mascare{\~n}o} {et~al.}(2021){Su{\'a}rez Mascare{\~n}o},
  {Damasso}, {Lodieu}, {Sozzetti}, {B{\'e}jar}, {Benatti}, {Zapatero Osorio},
  {Micela}, {Rebolo}, {Desidera}, {Murgas}, {Claudi}, {Gonz{\'a}lez
  Hern{\'a}ndez}, {Malavolta}, {del Burgo}, {D'Orazi}, {Amado}, {Locci},
  {Tabernero}, {Marzari}, {Aguado}, {Turrini}, {Cardona Guill{\'e}n},
  {Toledo-Padr{\'o}n}, {Maggio}, {Aceituno}, {Bauer}, {Caballero},
  {Chinchilla}, {Esparza-Borges}, {Gonz{\'a}lez-{\'A}lvarez}, {Granzer},
  {Luque}, {Mart{\'\i}n}, {Nowak}, {Oshagh}, {Pall{\'e}}, {Parviainen},
  {Quirrenbach}, {Reiners}, {Ribas}, {Strassmeier}, {Weber}, \&
  {Mallonn}}]{2021arXiv211109193S}
{Su{\'a}rez Mascare{\~n}o}, A., {Damasso}, M., {Lodieu}, N., {et~al.} 2021,
  arXiv e-prints, arXiv:2111.09193

\bibitem[{{Su{\'a}rez Mascare{\~n}o} {et~al.}(2016){Su{\'a}rez Mascare{\~n}o},
  {Rebolo}, \& {Gonz{\'a}lez Hern{\'a}ndez}}]{2016A&A...595A..12S}
{Su{\'a}rez Mascare{\~n}o}, A., {Rebolo}, R., \& {Gonz{\'a}lez Hern{\'a}ndez},
  J.~I. 2016, \aap, 595, A12

\bibitem[{{Su{\'a}rez Mascare{\~n}o} {et~al.}(2018){Su{\'a}rez Mascare{\~n}o},
  {Rebolo}, {Gonz{\'a}lez Hern{\'a}ndez}, {Toledo-Padr{\'o}n}, {Perger},
  {Ribas}, {Affer}, {Micela}, {Damasso}, {Maldonado}, {Gonz{\'a}lez-Alvarez},
  {Leto}, {Pagano}, {Scandariato}, {Sozzetti}, {Lanza}, {Malavolta}, {Claudi},
  {Cosentino}, {Desidera}, {Giacobbe}, {Maggio}, {Rainer}, {Esposito},
  {Benatti}, {Pedani}, {Morales}, {Herrero}, {Lafarga}, {Rosich}, \&
  {Pinamonti}}]{2018A&A...612A..89S}
{Su{\'a}rez Mascare{\~n}o}, A., {Rebolo}, R., {Gonz{\'a}lez Hern{\'a}ndez},
  J.~I., {et~al.} 2018, \aap, 612, A89

\bibitem[{{Trifonov} {et~al.}(2021){Trifonov}, {Caballero}, {Morales},
  {Seifahrt}, {Ribas}, {Reiners}, {Bean}, {Luque}, {Parviainen}, {Pall{\'e}},
  {Stock}, {Zechmeister}, {Amado}, {Anglada-Escud{\'e}}, {Azzaro}, {Barclay},
  {B{\'e}jar}, {Bluhm}, {Casasayas-Barris}, {Cifuentes}, {Collins}, {Collins},
  {Cort{\'e}s-Contreras}, {de Leon}, {Dreizler}, {Dressing}, {Esparza-Borges},
  {Espinoza}, {Fausnaugh}, {Fukui}, {Hatzes}, {Hellier}, {Henning}, {Henze},
  {Herrero}, {Jeffers}, {Jenkins}, {Jensen}, {Kaminski}, {Kasper},
  {Kossakowski}, {K{\"u}rster}, {Lafarga}, {Latham}, {Mann}, {Molaverdikhani},
  {Montes}, {Montet}, {Murgas}, {Narita}, {Oshagh}, {Passegger}, {Pollacco},
  {Quinn}, {Quirrenbach}, {Ricker}, {Rodr{\'\i}guez L{\'o}pez}, {Sanz-Forcada},
  {Schwarz}, {Schweitzer}, {Seager}, {Shporer}, {Stangret}, {St{\"u}rmer},
  {Tan}, {Tenenbaum}, {Twicken}, {Vanderspek}, \& {Winn}}]{2021Sci...371.1038T}
{Trifonov}, T., {Caballero}, J.~A., {Morales}, J.~C., {et~al.} 2021, Science,
  371, 1038

\bibitem[{{Trifonov} {et~al.}(2020){Trifonov}, {Lee}, {K{\"u}rster}, {Henning},
  {Grishin}, {Stock}, {Tjoa}, {Caballero}, {Wong}, {Bauer}, {Quirrenbach},
  {Zechmeister}, {Ribas}, {Reffert}, {Reiners}, {Amado}, {Kossakowski},
  {Azzaro}, {B{\'e}jar}, {Cort{\'e}s-Contreras}, {Dreizler}, {Hatzes},
  {Jeffers}, {Kaminski}, {Lafarga}, {Montes}, {Morales}, {Pavlov},
  {Rodr{\'\i}guez-L{\'o}pez}, {Schmitt}, {Solano}, \&
  {Barnes}}]{2020A&A...638A..16T}
{Trifonov}, T., {Lee}, M.~H., {K{\"u}rster}, M., {et~al.} 2020, \aap, 638, A16

\bibitem[{{Trotta}(2008)}]{2008ConPh..49...71T}
{Trotta}, R. 2008, Contemporary Physics, 49, 71

\bibitem[{{Twicken} {et~al.}(2018){Twicken}, {Catanzarite}, {Clarke},
  {Girouard}, {Jenkins}, {Klaus}, {Li}, {McCauliff}, {Seader}, {Tenenbaum},
  {Wohler}, {Bryson}, {Burke}, {Caldwell}, {Haas}, {Henze}, \&
  {Sanderfer}}]{Twicken:DVdiagnostics2018}
{Twicken}, J.~D., {Catanzarite}, J.~H., {Clarke}, B.~D., {et~al.} 2018, \pasp,
  130, 064502

\bibitem[{{Twicken} {et~al.}(2010){Twicken}, {Clarke}, {Bryson}, {Tenenbaum},
  {Wu}, {Jenkins}, {Girouard}, \& {Klaus}}]{twicken:PA2010SPIE}
{Twicken}, J.~D., {Clarke}, B.~D., {Bryson}, S.~T., {et~al.} 2010, in
  \procspie, Vol. 7740, Software and Cyberinfrastructure for Astronomy, 774023

\bibitem[{{Uzsoy} {et~al.}(2020){Uzsoy}, {Price}, \&
  {Rogers}}]{2020AAS...23537507U}
{Uzsoy}, A.~M., {Price}, E., \& {Rogers}, L. 2020, in American Astronomical
  Society Meeting Abstracts, Vol. 235, American Astronomical Society Meeting
  Abstracts \#235, 375.07

\bibitem[{{Van Eylen} {et~al.}(2018){Van Eylen}, {Agentoft}, {Lundkvist},
  {Kjeldsen}, {Owen}, {Fulton}, {Petigura}, \& {Snellen}}]{2018MNRAS.479.4786V}
{Van Eylen}, V., {Agentoft}, C., {Lundkvist}, M.~S., {et~al.} 2018, \mnras,
  479, 4786

\bibitem[{{Van Eylen} {et~al.}(2021){Van Eylen}, {Astudillo-Defru}, {Bonfils},
  {Livingston}, {Hirano}, {Luque}, {Lam}, {Justesen}, {Winn}, {Gandolfi},
  {Nowak}, {Palle}, {Albrecht}, {Dai}, {Estrada}, {Owen}, {Foreman-Mackey},
  {Fridlund}, {Korth}, {Mathur}, {Forveille}, {Mikal-Evans}, {Osborne}, {Ho},
  {Almenara}, {Artigau}, {Barrag{\'a}n}, {Barros}, {Bouchy}, {Cabrera},
  {Caldwell}, {Charbonneau}, {Chaturvedi}, {Cochran}, {Csizmadia}, {Damasso},
  {Delfosse}, {De Medeiros}, {D{\'\i}az}, {Doyon}, {Esposito},
  {F{\H{u}}r{\'e}sz}, {Figueira}, {Georgieva}, {Goffo}, {Grziwa}, {Guenther},
  {Hatzes}, {Jenkins}, {Kabath}, {Knudstrup}, {Latham}, {Lavie}, {Lovis},
  {Mennickent}, {Mullally}, {Murgas}, {Narita}, {Pepe}, {Persson}, {Redfield},
  {Ricker}, {Santos}, {Seager}, {Serrano}, {Smith}, {Mascare{\~n}o}, {Subjak},
  {Twicken}, {Udry}, {Vanderspek}, \& {Osorio}}]{2021MNRAS.tmp.2193V}
{Van Eylen}, V., {Astudillo-Defru}, N., {Bonfils}, X., {et~al.} 2021, \mnras
  [\eprint[arXiv]{2101.01593}]

\bibitem[{{Winn} {et~al.}(2018){Winn}, {Sanchis-Ojeda}, \&
  {Rappaport}}]{2018NewAR..83...37W}
{Winn}, J.~N., {Sanchis-Ojeda}, R., \& {Rappaport}, S. 2018, \nar, 83, 37

\bibitem[{{Wright} {et~al.}(2010){Wright}, {Eisenhardt}, {Mainzer}, {Ressler},
  {Cutri}, {Jarrett}, {Kirkpatrick}, {Padgett}, {McMillan}, {Skrutskie},
  {Stanford}, {Cohen}, {Walker}, {Mather}, {Leisawitz}, {Gautier}, {McLean},
  {Benford}, {Lonsdale}, {Blain}, {Mendez}, {Irace}, {Duval}, {Liu}, {Royer},
  {Heinrichsen}, {Howard}, {Shannon}, {Kendall}, {Walsh}, {Larsen}, {Cardon},
  {Schick}, {Schwalm}, {Abid}, {Fabinsky}, {Naes}, \&
  {Tsai}}]{2010AJ....140.1868W}
{Wright}, E.~L., {Eisenhardt}, P. R.~M., {Mainzer}, A.~K., {et~al.} 2010, \aj,
  140, 1868

\bibitem[{{Wright} {et~al.}(2011){Wright}, {Drake}, {Mamajek}, \&
  {Henry}}]{2011ApJ...743...48W}
{Wright}, N.~J., {Drake}, J.~J., {Mamajek}, E.~E., \& {Henry}, G.~W. 2011,
  \apj, 743, 48

\bibitem[{{Zechmeister} \& {K{\"u}rster}(2009)}]{2009A&A...496..577Z}
{Zechmeister}, M. \& {K{\"u}rster}, M. 2009, \aap, 496, 577

\bibitem[{{Zechmeister} {et~al.}(2018){Zechmeister}, {Reiners}, {Amado},
  {Azzaro}, {Bauer}, {B{\'e}jar}, {Caballero}, {Guenther}, {Hagen}, {Jeffers},
  {Kaminski}, {K{\"u}rster}, {Launhardt}, {Montes}, {Morales}, {Quirrenbach},
  {Reffert}, {Ribas}, {Seifert}, {Tal-Or}, \& {Wolthoff}}]{2018A&A...609A..12Z}
{Zechmeister}, M., {Reiners}, A., {Amado}, P.~J., {et~al.} 2018, \aap, 609, A12

\bibitem[{{Zeng} {et~al.}(2016){Zeng}, {Sasselov}, \&
  {Jacobsen}}]{2016ApJ...819..127Z}
{Zeng}, L., {Sasselov}, D.~D., \& {Jacobsen}, S.~B. 2016, \apj, 819, 127

\bibitem[{{Zhu} \& {Wu}(2018)}]{2018AJ....156...92Z}
{Zhu}, W. \& {Wu}, Y. 2018, \aj, 156, 92

\end{thebibliography}


\begin{appendix} 

\section{Figures}

\begin{figure}[!h]
\centering
\includegraphics[width=0.49\textwidth]{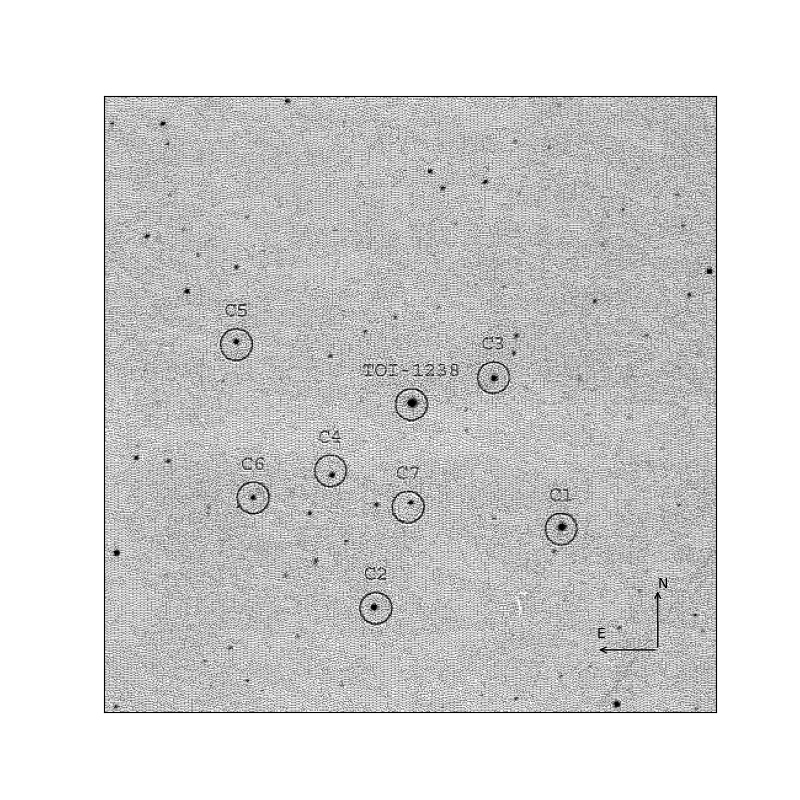}

\caption{SNO $R$-band image of TOI-1238 (in the center). The reference stars used to obtain the differential photometry are marked with circles. The field of view is 13.2\arcmin$\times$13.2\arcmin.}
\label{fig:SNO_field}
\end{figure}

\begin{figure}[!h]
\centering
\includegraphics[width=0.49\textwidth]{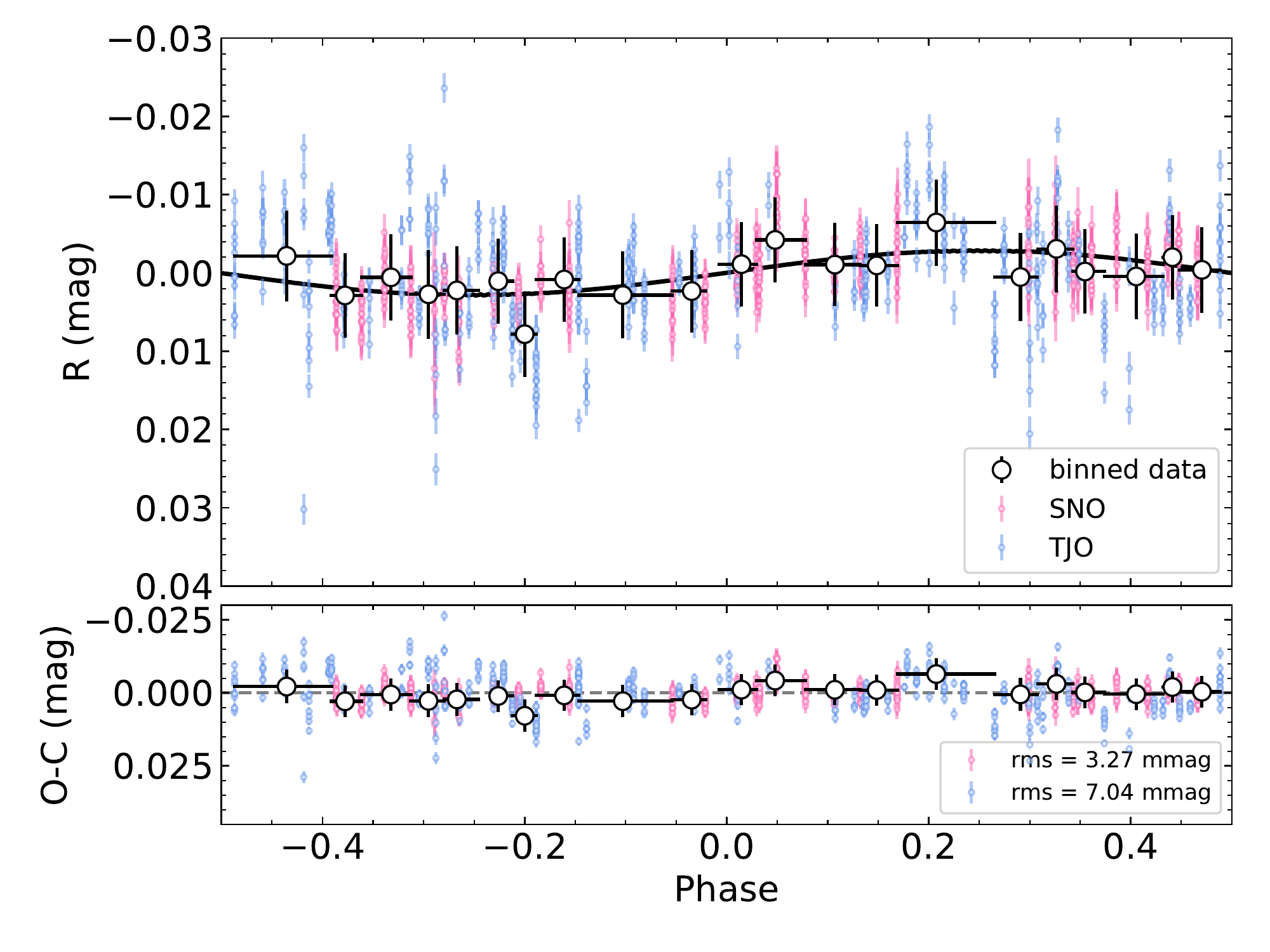}
\caption{TJO and SNO $R$-band LCs folded in phase with the period $P_{\rm rot}$ = 40.3\,d (top panel). Residuals (observations minus model) are shown in the bottom panel. The white dots correspond to the binned data, where the x-error bars illustrate the interval of the bins and the y-error bars take into account the uncertainties of the binned data.}
\label{fig:TJO_SNO_phot_phase}
\end{figure}

\begin{figure}[!h]
        \centering
        \includegraphics[width=\columnwidth]{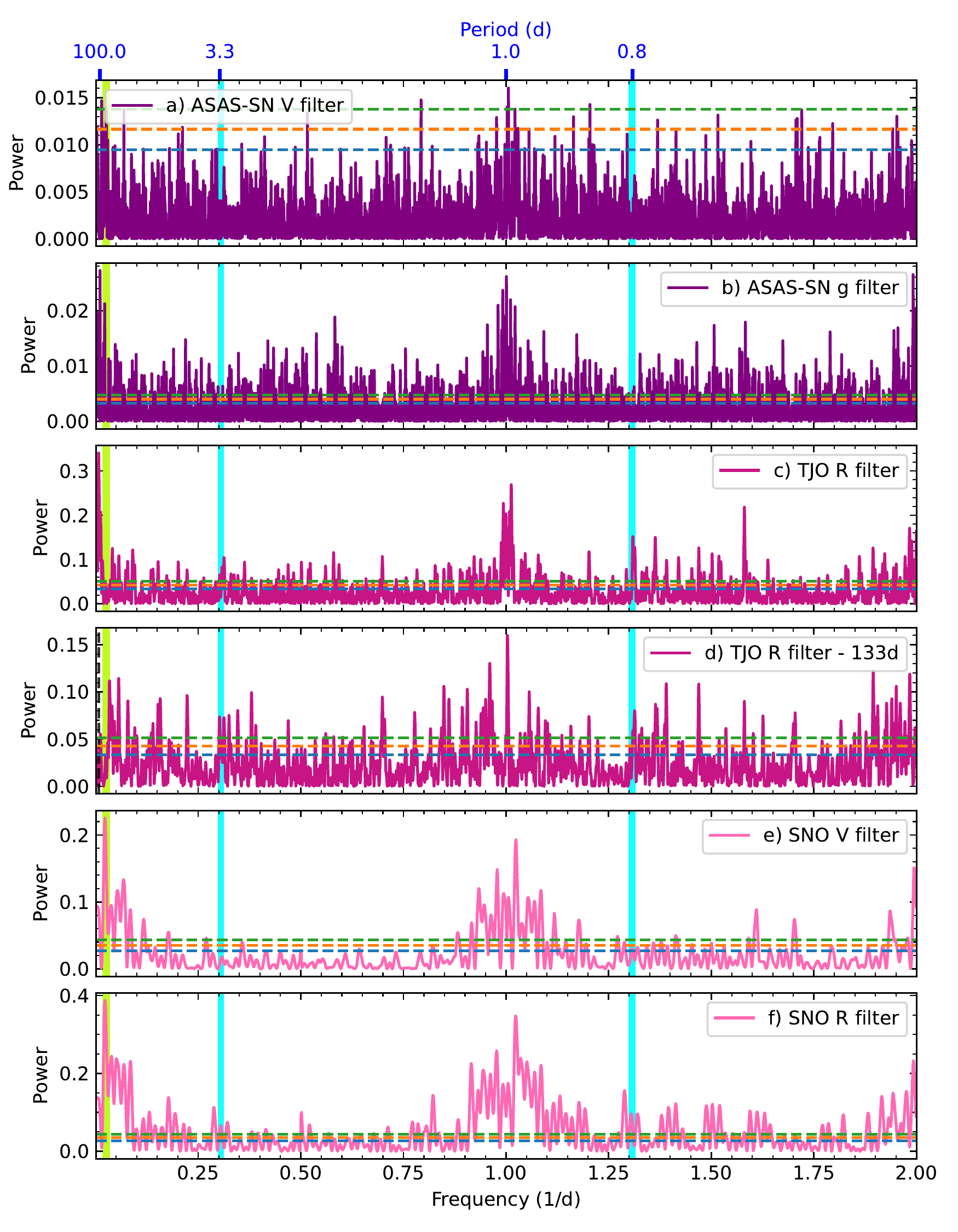}
        \caption{{GLS periodograms of the ASAS-SN $V$ and $g$, TJO $R$, and SNO $V$ and $R$ LCs in the frequency space between 2 and 0.001\,d$^{-1}$. In all panels, the horizontal dashed lines indicate FAP levels of 10\% (blue), 1\% (orange), and 0.1\% (green). The orbital periods of the two transiting planets are marked with vertical blue lines. The greenish band indicates the region where most of the spectroscopic activity indicators have their highest GLS peaks. In the fourth panel, the vertical dashed line indicates the frequency of the signal removed from the data (see text).}}
        \label{fig:GLS_phot_freq_multiplot}
\end{figure}

\begin{figure}[!h]
        \centering
        \includegraphics[width=\columnwidth]{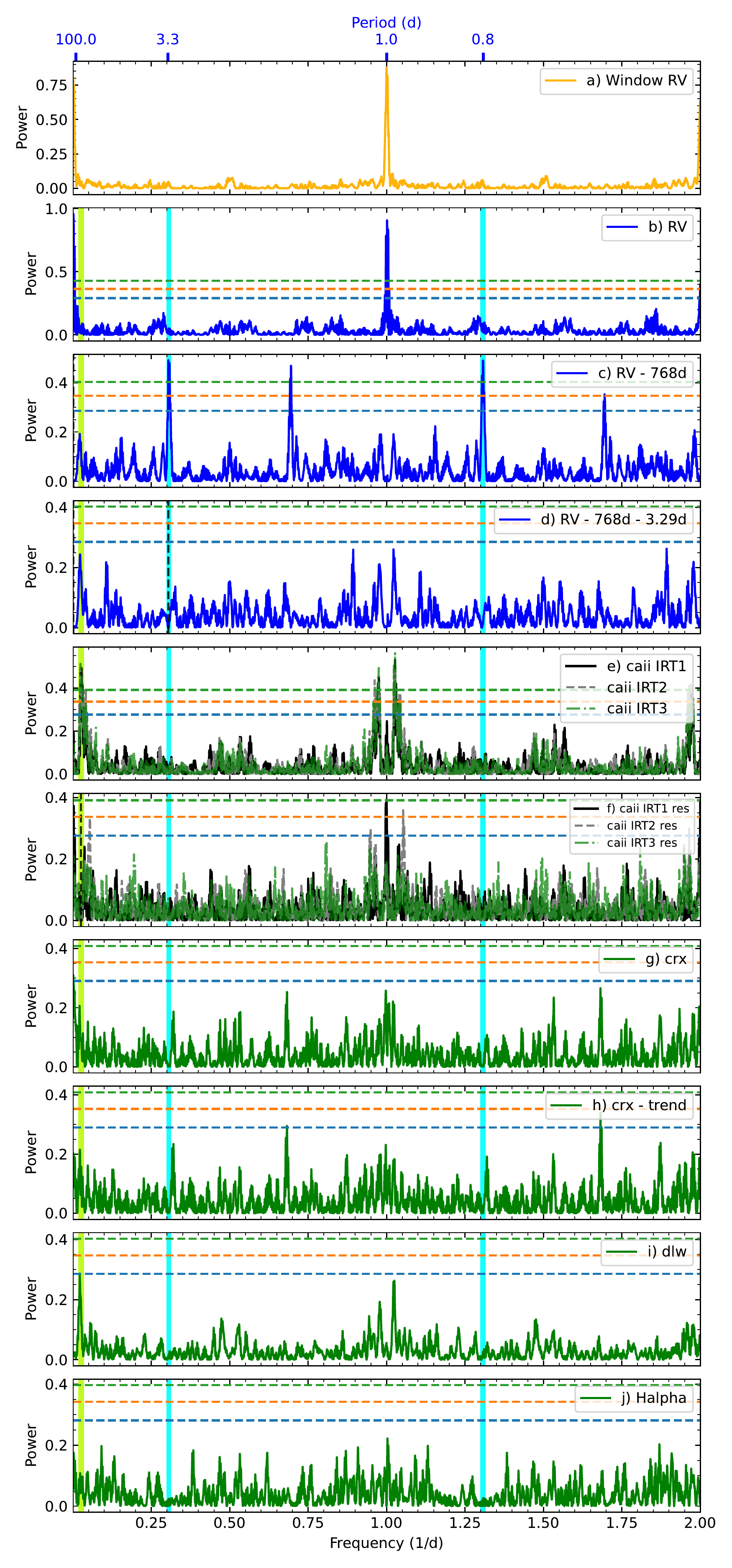}
        \caption{{Window function of the CARMENES data (top panel) and the GLS periodograms of the CARMENES RVs and various spectroscopic indices (panels 2--10) in the frequency space between 2 and 0.001\,d$^{-1}$. The horizontal and vertical lines and bands have the same meanings as in Figs.~\ref{fig:GLS_phot_multiplot} and \ref{fig:GLS_activity_multiplot}.}}
        \label{fig:GLS_activity_multiplot_freq}
\end{figure}

\begin{figure*}[!h]
\centering
\includegraphics[width=\textwidth]{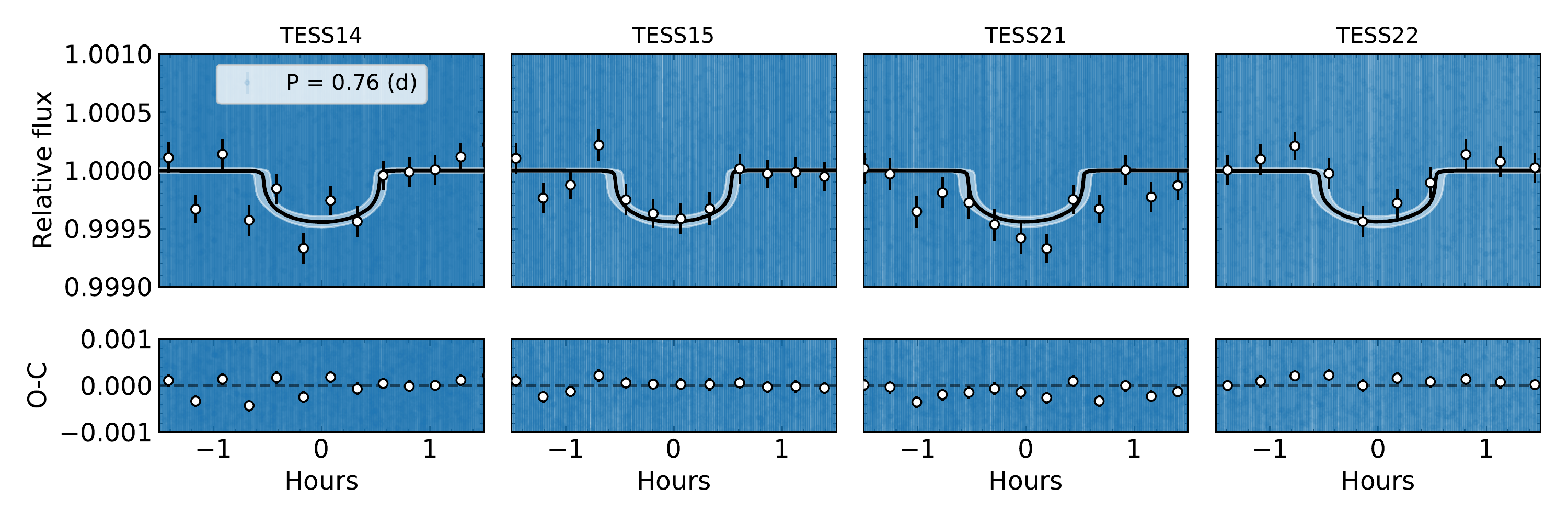}
\includegraphics[width=\textwidth]{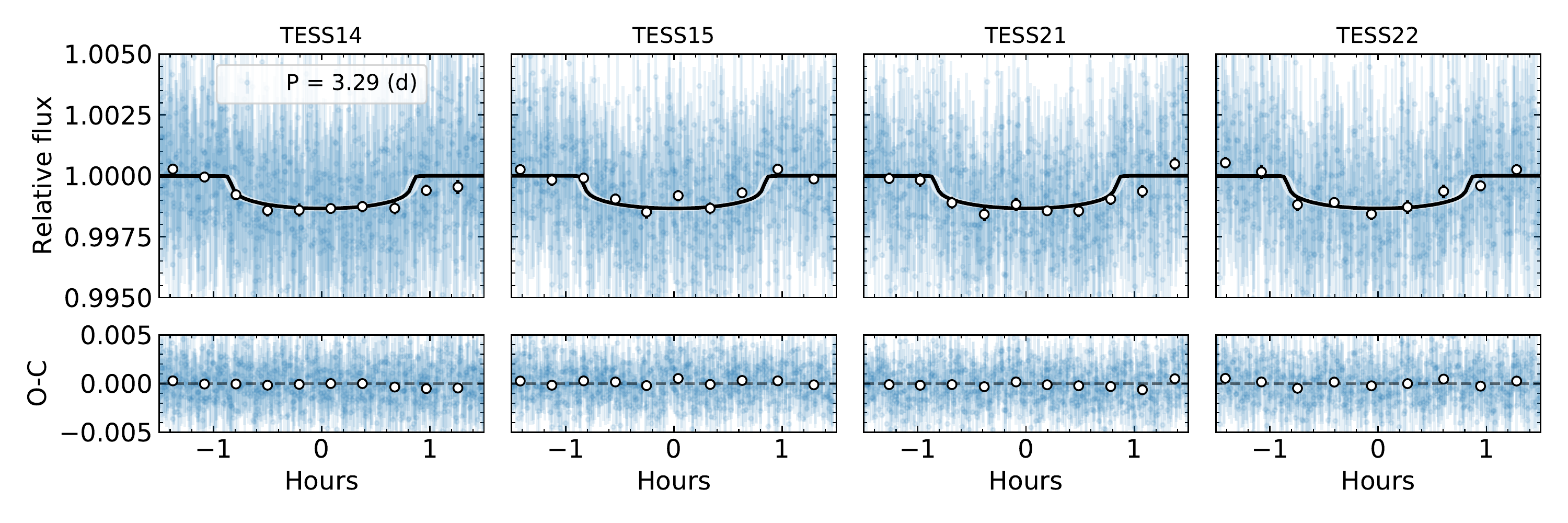}
\caption{\textit{TESS} LCs (blue points) folded in phase with the orbital periods of the transiting planets per sector. The best joint fit solution is plotted as a black line. The white dots correspond to the binned photometric data. The x axis represents the time computed from the mid-transit times as derived from the best joint fit.}
\label{fig:lc_vs_phase}
\end{figure*}

\begin{figure*}[!h]
\centering
\includegraphics[width=\textwidth]{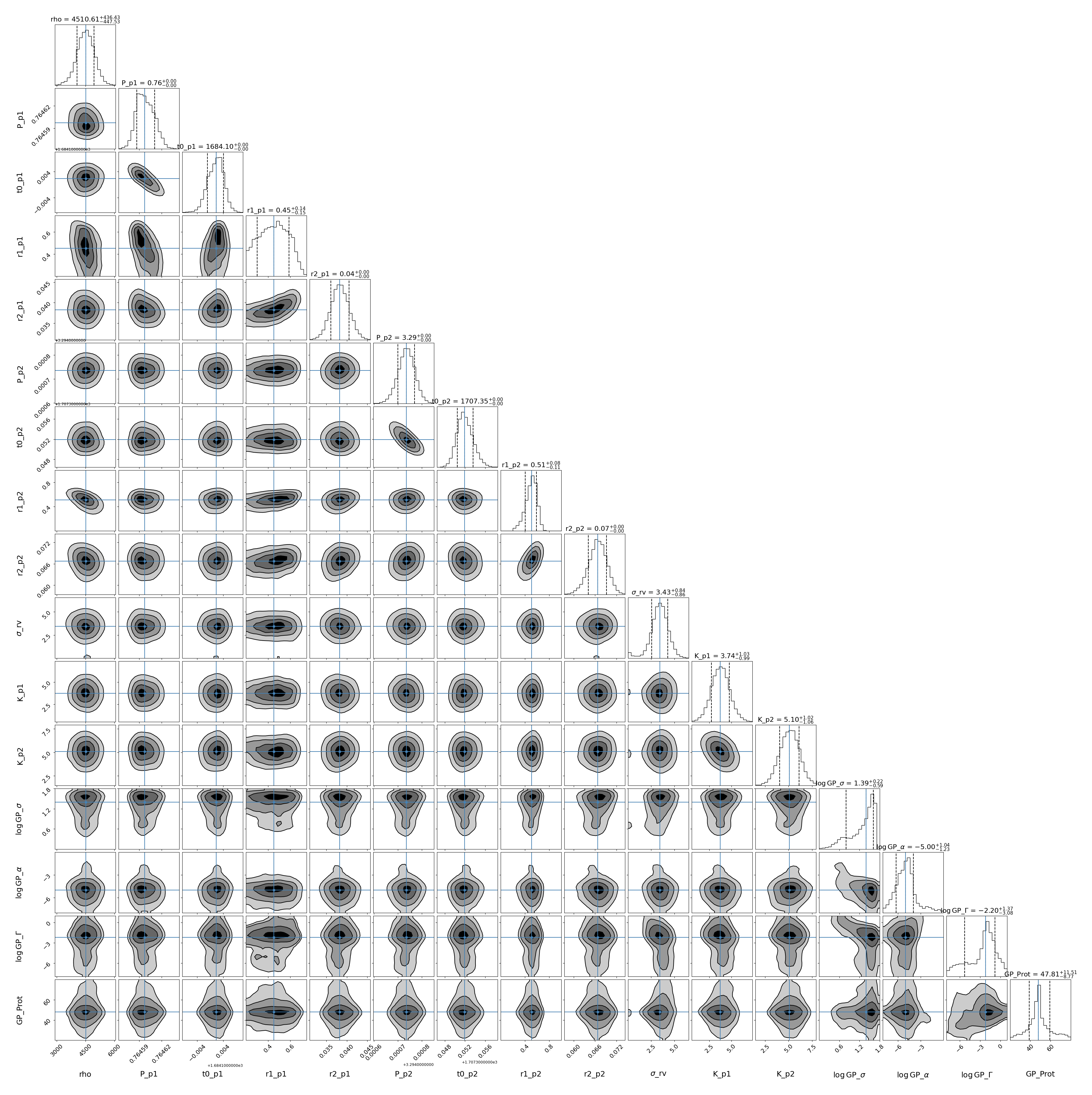}
\caption{Posterior distributions of {some of} the fitted planetary parameters of the TOI-1238 system as obtained from the combined photometric and spectroscopic fit. The vertical dashed lines indicate the 16, 50, and 84\,\%~quantiles that were used to define the optimal values and their associated 1$\sigma$ uncertainty. The blue line stands for the median values of each fitted parameter. 
}
\label{fig:toi1238_cornerplot}
\end{figure*}

\clearpage

\section{Tables}

\begin{table*}[!h]
\centering
\caption{Priors used for the joint LC and RV fit of TOI-1238.}
\label{tab:toi1238_priors_details}
\begin{tabular}{l c c r}

\hline
\hline
\noalign{\smallskip}
Parameter & Prior & Unit & Description \\
\noalign{\smallskip}    
\hline  
\noalign{\smallskip}

\multicolumn{4}{c}{\textit{Stellar parameter}} \\
\noalign{\smallskip}

$\rho_{\star}$                                          & $\mathcal{N}$(4254.2, 578.2) & kg\,m$^{-3}$ & Stellar density\\

\noalign{\smallskip}
\multicolumn{4}{c}{\textit{Photometric parameters}} \\
\noalign{\smallskip}

$\gamma_{\rm TESS, S14}$                                                & $\mathcal{N}$(0, 0.1) & ppm & The offset relative flux for \textit{TESS}\\
$\gamma_{\rm TESS, S15}$                                                & $\mathcal{N}$(0, 0.1) & ppm  & The offset relative flux for \textit{TESS}\\
$\gamma_{\rm TESS, S21}$                                                & $\mathcal{N}$(0, 0.1) & ppm  & The offset relative flux for \textit{TESS}\\
$\gamma_{\rm TESS, S22}$                                                & $\mathcal{N}$(0, 0.1) &  ppm  & The offset relative flux for \textit{TESS}\\

$\sigma_{\rm TESS, S14}$                & $\mathcal{L} \mathcal{U}$(1e$^{-6}$, 0.03)  & ppm & A jitter added in quadrature to the error bars of instrument\\
$\sigma_{\rm TESS, S15}$                & $\mathcal{L} \mathcal{U}$(1e$^{-6}$, 0.02) & ppm     & A jitter added in quadrature to the error bars of instrument\\                 
$\sigma_{\rm TESS, S21}$                & $\mathcal{L} \mathcal{U}$(1e$^{-6}$, 0.02) & ppm & A jitter added in quadrature to the error bars of instrument\\                 
$\sigma_{\rm TESS, S22}$                & $\mathcal{L} \mathcal{U}$(1e$^{-6}$, 0.03) &         ppm & A jitter added in quadrature to the error bars of instrument\\                 

$q1_{\rm TESS}$                                                         & $\mathcal{U}$(0, 1)     &       ...                     & Limb-darkening for photometric instrument\\
$q2_{\rm TESS}$                                                         & $\mathcal{U}$(0, 1) &   ...             & Limb-darkening for photometric instrument\\
$D_{\rm TESS}$                                                          & 1 (fixed)                                       &       ...             & The dilution factor for the photometric instrument.\\

\noalign{\smallskip}
\multicolumn{4}{c}{\textit{RV parameters}} \\
\noalign{\smallskip}

$\gamma$                                                                &  $\mathcal{U}$(-100, 100) & m\,$\rm s^{-1}$ & RV zero point for CARMENES\\
$\sigma$                                                                        & $\mathcal{L} \mathcal{U}$(0.001, 10) & m\,$\rm s^{-1}$ & A jitter added in quadrature \\

\noalign{\smallskip}                                                        
\multicolumn{4}{c}{\textit{GP parameters}} \\
\noalign{\smallskip}            

$\sigma_{\rm GP,RV}$                                & $\mathcal{U}$(0.01, 50)   & m\,$\rm s^{-1}$ & Amplitude of the GP for the RVs\\
$\alpha_{\rm GP,RV}$                                            &  $\mathcal{J}$(1e$^{-8}$, 1) & d$^{-2}$  &  Inverse (squared) length-scale of the external parameter\\
$\Gamma_{\rm GP,RV}$                                    & $\mathcal{J}$(1e$^{-8}$, 10) & ... & Amplitude of the sine-part of the kernel\\
$P_{\rm rot, GP,RV}$                                            & $\mathcal{U}$(20, 80) & d & Period of the quasi-periodic kernel\\

\noalign{\smallskip}                                                        
\multicolumn{4}{c}{\textit{Planet $b$ parameters} }\\   
\noalign{\smallskip}
                                                                                    
$P$                                                                             & $\mathcal{N}$ (0.764, 0.001)    & d             & Period \\
$t_0$ (BJD-2,457,000)           & $\mathcal{N}$ (1684.1, 0.01)          & d               &  Time of periastron passage\\
$e$                                                                             & 0 (fixed)                                                                                       & ...     & Orbital eccentricity \\
$\omega$                                                                &       90 (fixed)                                                                         & deg     & Periastron angle \\
$K$                                                                                     &  $\mathcal{U}$ (0, 8)                           &        m\,$\rm s^{-1}$          & RV semi-amplitude  \\   
$r_1$                                                                           &  $\mathcal{U}$ (0, 1)                                   & ...        & Parameterization for $p$ and $b$  \\   
$r_2$                                                                           &  $\mathcal{U}$ (0, 1)                                   & ...        & Parameterization for $p$ and $b$    \\   
                                                            
\noalign{\smallskip}                                                        
\multicolumn{4}{c}{\textit{Planet $c$ parameters} }\\   
\noalign{\smallskip}                                                        

$P$                                                                             & $\mathcal{N}$ (3.294, 0.001)            & d             & Period \\
$t_0$ (BJD-2,457,000)   & $\mathcal{N}$ (1707.34, 0.01)         & d             &  Time of periastron passage\\
$e$                                                                             & 0 (fixed)                                                                                               & ... & Orbital eccentricity\\
$\omega$                                                                &       90 (fixed)                                                                                 & deg     & Periastron angle\\
$K$                                                                                     &  $\mathcal{U}$ (0, 8)   &         m\,$\rm s^{-1}$        & RV semi-amplitude\\   
$r_1$                                                                           &  $\mathcal{U}$ (0, 1)   &  ...       & Parameterization for $p$ and $b$  \\   
$r_2$                                                                           &  $\mathcal{U}$ (0, 1)   &  ...       & Parameterization for $p$ and $b$    \\   

\noalign{\smallskip}                                                        
\multicolumn{4}{c}{\textit{External companion parameters} }\\   
\noalign{\smallskip}    

$P$                                                                             & $\mathcal{U}$ (400, 1500)               & d             & Period \\
$t_0$ (BJD-2,457,000)   & $\mathcal{U}$ (1900, 3000)            & d             &  Time of periastron passage\\
$e$                                                                             & 0 (fixed)                                                                                               & ... & Orbital eccentricity\\
$\omega$                                                                &       90 (fixed)                                                                                 & deg     & Periastron angle\\
$K$                                                                                     &  $\mathcal{U}$ (0, 300)                                         &         m\,$\rm s^{-1}$         & RV semi-amplitude \\   

\noalign{\smallskip}    
\hline  
\noalign{\smallskip}
\end{tabular}
\tablefoot{The prior labels of $\mathcal{N}$, $\mathcal{U}$,  $\mathcal{J}$, and $\mathcal{L}$ $\mathcal{U}$ represent normal, uniform, Jeffreys, and log-uniform distribution, respectively. The error on the density of the star comes from the stellar mass and radius errors. The upper limit on the photometric jitter term corresponds to the dispersion of the data.}
\end{table*}

\clearpage
\onecolumn
 

\begin{landscape}
\begin{scriptsize}
\begin{longtable}{lcccccccccccccc}
\caption{\label{tab:toi1238_rv_act_data_used} TOI-1238 CARMENES data {included in our analysis}.}\\
\hline\hline
\noalign{\smallskip}
BJD & RV & eRV & CaIRT1 & eCaIRT1 & CaIRT2 & eCaIRT2 & CaIRT3 & eCaIRT3 & CRX & eCRX & dLW & edLW &   H$\alpha$ & eH$\alpha$ \\
(d)  & (m\,s$^{-1}$) &  (m\,s$^{-1}$) &  &  &  &  &  &  & (m\,s$^{-1}$\,Np$^{-1}$) & (m\,s$^{-1}$\,Np$^{-1}$) & (m$^2$\,s$^{-2}$)  & (m$^2$\,s$^{-2}$)  & & \\

 \noalign{\smallskip}
\hline
\noalign{\smallskip}
\endfirsthead
\caption{continued.}\\
\hline\hline
\noalign{\smallskip}
BJD & RV & eRV & CaIRT1 & eCaIRT1 & CaIRT2 & eCaIRT2 & CaIRT3 & eCaIRT3 & CRX & eCRX & dLW & edLW &   H$\alpha$ & eH$\alpha$ \\
(d)  & (m\,s$^{-1}$) &  (m\,s$^{-1}$) &  &  &  &  &  &  & (m\,s$^{-1}$\,Np$^{-1}$) & (m\,s$^{-1}$\,Np$^{-1}$) & (m$^2$\,s$^{-2}$) & (m$^2$\,s$^{-2}$) & & \\

\noalign{\smallskip}
\hline
\endhead
\hline
\endfoot
2458978.6294 & 12.2991 & 6.8233 & 0.5462 & 0.0055 & 0.4282 & 0.0063 & 0.4081 & 0.0060 & -96.3592 & 62.3410 & 33.0130 & 6.4506 & 0.7120 & 0.0054  \\ 
2458980.6420 & 25.4687 & 8.9621 & 0.5283 & 0.0079 & 0.4147 & 0.0096 & 0.3988 & 0.0089 & -39.3722 & 88.8069 & 27.3829 & 8.7092 & 0.7417 & 0.0081  \\ 
2458981.4478 & 23.2731 & 3.1942 & 0.5463 & 0.0028 & 0.4217 & 0.0029 & 0.3982 & 0.0028 & -68.4087 & 28.3277 & 0.7861 & 3.7099 & 0.7038 & 0.0026  \\ 
2458982.5962 & 17.8775 & 4.5214 & 0.5454 & 0.0041 & 0.4155 & 0.0045 & 0.4007 & 0.0043 & -60.1947 & 42.8317 & -5.5279 & 4.8351 & 0.7176 & 0.0039  \\ 
2458988.4656 & 16.6482 & 3.1134 & 0.5567 & 0.0025 & 0.4309 & 0.0026 & 0.4095 & 0.0025 & -22.7424 & 29.1804 & 1.3482 & 2.0761 & 0.7137 & 0.0022  \\ 
2458989.3672 & 22.5039 & 2.6071 & 0.5522 & 0.0026 & 0.4258 & 0.0026 & 0.4119 & 0.0025 & -16.5653 & 23.4493 & 9.3135 & 3.0052 & 0.7069 & 0.0024  \\ 
2458990.4357 & 28.5857 & 3.2883 & 0.5481 & 0.0029 & 0.4323 & 0.0031 & 0.4091 & 0.0030 & -71.0890 & 27.7209 & -3.5179 & 3.6012 & 0.7100 & 0.0027  \\ 
2458992.4290 & 11.3362 & 3.0069 & 0.5522 & 0.0026 & 0.4338 & 0.0027 & 0.4077 & 0.0026 & -7.1100 & 28.1924 & 3.8968 & 2.7798 & 0.7127 & 0.0024  \\ 
2458994.4603 & 7.5435 & 3.3705 & 0.5463 & 0.0027 & 0.4193 & 0.0028 & 0.4070 & 0.0027 & -40.3298 & 31.4931 & 2.8128 & 2.8139 & 0.7148 & 0.0025  \\ 
2458998.4602 & 3.2495 & 4.2965 & 0.5244 & 0.0040 & 0.4132 & 0.0043 & 0.3967 & 0.0042 & -30.2429 & 40.6207 & -22.2695 & 5.4688 & 0.7199 & 0.0040  \\ 
2459000.4160 & 20.3829 & 3.0968 & 0.5326 & 0.0035 & 0.4039 & 0.0037 & 0.3912 & 0.0036 & -23.1747 & 29.4158 & -18.9311 & 4.4068 & 0.7068 & 0.0033  \\ 
2459001.4489 & 11.0096 & 3.2731 & 0.5400 & 0.0033 & 0.4097 & 0.0035 & 0.3867 & 0.0033 & 40.0316 & 30.5645 & -4.3897 & 4.3785 & 0.7225 & 0.0031  \\ 
2459002.3981 & 12.2632 & 2.5005 & 0.5357 & 0.0028 & 0.4026 & 0.0029 & 0.3900 & 0.0028 & -46.8625 & 21.5840 & -12.1010 & 2.9241 & 0.7046 & 0.0026  \\ 
2459003.4168 & 22.2166 & 3.3345 & 0.5345 & 0.0031 & 0.4108 & 0.0032 & 0.3970 & 0.0031 & -37.2941 & 30.8876 & -7.9881 & 2.9240 & 0.7113 & 0.0028  \\ 
2459004.4378 & 14.2590 & 3.2060 & 0.5375 & 0.0032 & 0.4095 & 0.0033 & 0.3928 & 0.0032 & -21.0005 & 30.4122 & -12.3086 & 4.2052 & 0.7071 & 0.0030  \\ 
2459007.4509 & 22.1722 & 7.6640 & 0.5230 & 0.0077 & 0.4274 & 0.0094 & 0.3919 & 0.0088 & ...	     & ...	   & 7.9901 & 8.5497 & 0.7190 & 0.0081  \\ 
2459008.4381 & 12.6887 & 3.1284 & 0.5346 & 0.0033 & 0.4147 & 0.0035 & 0.3997 & 0.0034 & 1.6671 & 30.0904 & -11.4101 & 3.6371 & 0.7110 & 0.0031  \\ 
2459009.4517 & 26.9540 & 3.5163 & 0.5376 & 0.0035 & 0.4199 & 0.0037 & 0.3982 & 0.0036 & -33.8136 & 31.6079 & -2.9220 & 3.5271 & 0.7171 & 0.0034  \\ 
2459010.4533 & 28.7471 & 3.0413 & 0.5464 & 0.0030 & 0.4243 & 0.0032 & 0.4006 & 0.0031 & -31.4311 & 27.5552 & 7.9449 & 2.8730 & 0.7098 & 0.0028  \\ 
2459013.4086 & 18.2862 & 6.7143 & 0.5446 & 0.0064 & 0.4307 & 0.0074 & 0.3920 & 0.0070 & -59.1175 & 63.5833 & 4.2582 & 8.5913 & 0.7071 & 0.0065  \\ 
2459014.3802 & 13.3515 & 2.2812 & 0.5395 & 0.0029 & 0.4286 & 0.0030 & 0.4073 & 0.0029 & 17.8792 & 19.6011 & 7.8211 & 2.6716 & 0.7089 & 0.0026  \\ 
2459015.4371 & 17.2434 & 3.1306 & 0.5446 & 0.0033 & 0.4230 & 0.0035 & 0.4062 & 0.0034 & -25.3464 & 29.1774 & -13.4956 & 4.1748 & 0.7066 & 0.0030  \\ 
2459016.3802 & 20.4550 & 3.1465 & 0.5403 & 0.0033 & 0.4342 & 0.0035 & 0.4080 & 0.0033 & -4.8732 & 24.9033 & 4.3658 & 3.8726 & 0.7266 & 0.0030  \\ 
2459016.4589 & 24.0969 & 3.4353 & 0.5468 & 0.0035 & 0.4278 & 0.0037 & 0.4138 & 0.0035 & 0.7641 & 28.9970 & -5.6826 & 4.6600 & 0.7145 & 0.0032  \\ 
2459016.5412 & ...	   & ...    & ...	 & ...	  & 0.4231 & 0.0191 & ...	 & ...	  & ...	   & ...	  & -3.5007 & 13.4382 & ...	   & ...	   \\ 
2459017.4318 & 22.1507 & 3.1644 & 0.5507 & 0.0037 & 0.4346 & 0.0041 & 0.4045 & 0.0037 & 37.2848 & 28.6356 & -20.8864 & 4.9419 & 0.7171 & 0.0035  \\ 
2459017.5165 & 19.8048 & 5.6736 & 0.5428 & 0.0065 & 0.4280 & 0.0075 & 0.4146 & 0.0071 & 42.6127 & 55.6823 & 30.5198 & 8.3265 & 0.7226 & 0.0061  \\ 
2459018.4115 & 4.9324 & 2.6566 & 0.5503 & 0.0025 & 0.4265 & 0.0026 & 0.4080 & 0.0025 & -5.0073 & 23.0846 & 4.7231 & 2.8462 & 0.7133 & 0.0023  \\ 
2459018.4833 & 12.7496 & 3.6072 & 0.5476 & 0.0037 & 0.4240 & 0.0040 & 0.4031 & 0.0038 & 12.5327 & 33.3872 & 7.9899 & 3.6725 & 0.7138 & 0.0034  \\ 
2459020.4181 & 14.3703 & 2.7739 & 0.5511 & 0.0026 & 0.4287 & 0.0027 & 0.4144 & 0.0026 & -5.5721 & 21.7390 & 5.0660 & 2.5524 & 0.7188 & 0.0023  \\ 
2459024.4360 & 13.9798 & 4.0711 & 0.5516 & 0.0045 & 0.4346 & 0.0049 & 0.4153 & 0.0048 & 26.2544 & 37.6344 & 12.0696 & 4.9312 & 0.7267 & 0.0044  \\ 
2459032.3939 & 15.9407 & 5.1918 & 0.5356 & 0.0052 & 0.4291 & 0.0058 & 0.4113 & 0.0056 & 3.4149 & 49.4701 & 30.0245 & 5.4020 & 0.7290 & 0.0051  \\ 
2459033.3998 & 18.2336 & 4.8145 & 0.5449 & 0.0055 & 0.4203 & 0.0062 & 0.4053 & 0.0060 & -14.2273 & 45.4099 & 27.8730 & 6.9834 & 0.7218 & 0.0056  \\ 
2459034.3708 & -3.3912 & 2.9474 & 0.5389 & 0.0027 & 0.4199 & 0.0027 & 0.3986 & 0.0026 & -52.6128 & 27.2072 & ...	   & ... & 0.7103 & 0.0024  \\ 
2459037.4817 & -1.5446 & 2.8280 & 0.5428 & 0.0031 & 0.4134 & 0.0032 & 0.4026 & 0.0031 & -38.6217 & 26.7669 & -0.4287 & 2.9705 & 0.7201 & 0.0028  \\ 
2459038.4092 & 3.4029 & 3.7206 & 0.5416 & 0.0036 & 0.4212 & 0.0039 & 0.3929 & 0.0037 & 12.5634 & 34.2790 & -5.7973 & 3.9184 & 0.7160 & 0.0033  \\ 
2459040.4244 & -3.8344 & 3.7592 & 0.5404 & 0.0039 & 0.4028 & 0.0042 & 0.3921 & 0.0041 & 30.6914 & 34.2252 & -12.3640 & 4.4489 & 0.7115 & 0.0037  \\ 
2459050.4621 & -12.5590 & 2.8485 & 0.5392 & 0.0030 & 0.4258 & 0.0031 & 0.4005 & 0.0030 & -39.1879 & 25.1942 & -4.3192 & 3.3596 & 0.7128 & 0.0027  \\ 
2459051.3869 & 0.0532 & 3.3114 & 0.5408 & 0.0031 & 0.4173 & 0.0033 & 0.4032 & 0.0032 & -14.2095 & 29.0924 & -3.9970 & 3.5923 & 0.7097 & 0.0028  \\ 
2459053.3966 &   ... &   ... & 0.5479 & 0.0033 & 0.4213 & 0.0035 & 0.4048 & 0.0033 & -6.9152 & 27.2705 & -7.9415 & 3.9786 & 0.7180 & 0.0030  \\ 
2459054.3880 & -4.5984 & 2.9676 & 0.5449 & 0.0027 & 0.4170 & 0.0028 & 0.4005 & 0.0027 & 42.6567 & 26.4337 & 6.0848 & 3.3649 & 0.7158 & 0.0024  \\ 
2459055.4113 & 13.5214 & 3.1001 & 0.5433 & 0.0032 & 0.4216 & 0.0033 & 0.4034 & 0.0032 & 70.0216 & 27.4927 & -7.4510 & 4.0169 & 0.7139 & 0.0029  \\ 
2459057.3751 & -3.0699 & 3.5230 & 0.5405 & 0.0030 & 0.4142 & 0.0031 & 0.4020 & 0.0030 & 15.2184 & 33.5913 & 0.0153 & 3.2978 & 0.7077 & 0.0025  \\ 
2459058.4125 & -2.0478 & 3.0826 & 0.5462 & 0.0035 & 0.4190 & 0.0037 & 0.4036 & 0.0036 & 20.1926 & 30.0287 & -9.4346 & 4.5599 & 0.7104 & 0.0031  \\ 
2459059.3607 & 5.6235 & 3.6988 & 0.5449 & 0.0032 & 0.4213 & 0.0033 & 0.4119 & 0.0032 & 55.8451 & 34.3852 & -1.7411 & 3.7578 & 0.7161 & 0.0029  \\ 
2459061.3752 & 4.2537 & 3.8429 & 0.5537 & 0.0031 & 0.4393 & 0.0033 & 0.4148 & 0.0032 & 24.0274 & 36.5903 & -0.8419 & 3.5789 & 0.7394 & 0.0029  \\ 
2459062.4280 & 6.5352 & 4.5336 & 0.5515 & 0.0043 & 0.4315 & 0.0047 & 0.4166 & 0.0045 & -7.5659 & 44.2782 & 22.1818 & 5.6085 & 0.7274 & 0.0039  \\ 
2459063.4034 & -0.0362 & 3.1169 & 0.5466 & 0.0036 & 0.4239 & 0.0038 & 0.4080 & 0.0037 & 40.8546 & 30.0722 & 25.3997 & 4.9998 & 0.7284 & 0.0031  \\ 
2459237.7532 & -96.0764 & 8.9617 & ...	  & ...	   & 0.4110 & 0.0144 & 0.4241 & 0.0142 & -13.3584 & 87.6544 & 22.3830 & 15.0987 & 0.7325 & 0.0129  \\ 
2459244.7229 & -101.2047 & 4.8675 & 0.5334 & 0.0036 & 0.4051 & 0.0037 & 0.3817 & 0.0037 & 87.7823 & 41.4425 & 1.9888 & 3.9820 & 0.7109 & 0.0035  \\ 
2459248.7167 &   ...	 &   ...  & 0.5338 & 0.0052 & 0.4066 & 0.0057 & 0.3980 & 0.0055 & 66.4999 & 43.4904 & 10.3674 & 4.9137 & 0.7172 & 0.0050  \\ 
2459265.6524 & -104.1914 & 4.4606 & 0.5286 & 0.0046 & 0.4180 & 0.0051 & 0.4055 & 0.0049 & -36.4795 & 42.5400 & 3.4918 & 6.5825 & 0.7183 & 0.0048  \\ 
2459278.6560 & ...       & ...    & 0.5334 & 0.0132 & 0.4158 & 0.0169 & 0.4042 & 0.0157 & ...	   & ...	  & ...	   & ...   & 0.6968 & 0.0151  \\ 
2459295.6677 & -122.2124 & 5.5862 & 0.5209 & 0.0042 & 0.4007 & 0.0045 & 0.3860 & 0.0044 & 86.3146 & 43.5296 & -13.7185 & 6.0405 & 0.7157 & 0.0042  \\ 
2459299.5813 & -102.1456 & 3.3756 & 0.5263 & 0.0032 & 0.4137 & 0.0033 & 0.3902 & 0.0032 & -42.2845 & 31.0441 & 13.0384 & 4.1539 & 0.7035 & 0.0030  \\ 
2459306.5673 & -119.7219 & 3.8087 & 0.5418 & 0.0038 & 0.4236 & 0.0041 & 0.4020 & 0.0039 & 42.1924 & 34.3073 & 8.6188 & 5.2364 & 0.7079 & 0.0037  \\ 
2459310.3896 & -123.7105 & 3.3532 & 0.5422 & 0.0033 & 0.4233 & 0.0035 & 0.3978 & 0.0033 & 25.0196 & 31.0000 & 3.3136 & 4.3916 & 0.7056 & 0.0031  \\ 

\end{longtable}
\end{scriptsize}
\end{landscape}

        \begin{landscape}
                \begin{scriptsize}
                        \begin{longtable}{lcccccccccccccc}
                                \caption{\label{tab:toi1238_rv_act_data_rejected} TOI-1238 CARMENES data {not included in our analysis.}}\\
                                \hline\hline
                                \noalign{\smallskip}
                                BJD & RV & eRV & CaIRT1 & eCaIRT1 & CaIRT2 & eCaIRT2 & CaIRT3 & eCaIRT3 & CRX & eCRX & dLW & edLW &   H$\alpha$ & eH$\alpha$ \\
                                (d)  & (m\,s$^{-1}$) &  (m\,s$^{-1}$) &  &  &  &  &  &  & (m\,s$^{-1}$\,Np$^{-1}$) & (m\,s$^{-1}$\,Np$^{-1}$) & (m$^2$\,s$^{-2}$)  & (m$^2$\,s$^{-2}$)  & & \\
                                
                                \noalign{\smallskip}
                                \hline
                                \noalign{\smallskip}
                                \endfirsthead
                                \caption{continued.}\\
                                \hline\hline
                                \noalign{\smallskip}
                                BJD & RV & eRV & CaIRT1 & eCaIRT1 & CaIRT2 & eCaIRT2 & CaIRT3 & eCaIRT3 & CRX & eCRX & dLW & edLW &   H$\alpha$ & eH$\alpha$ \\
                                (d)  & (m\,s$^{-1}$) &  (m\,s$^{-1}$) &  &  &  &  &  &  & (m\,s$^{-1}$\,Np$^{-1}$) & (m\,s$^{-1}$\,Np$^{-1}$) & (m$^2$\,s$^{-2}$) & (m$^2$\,s$^{-2}$) & & \\

                                \noalign{\smallskip}
                                \hline
                                \endhead
                                \hline
                                \endfoot
                                2459007.4509 &    ...  &    ... &   ...  &   ...  &    ... &   ...  &   ...  &   ...  & -138.3661 & 72.3116  &   ...  &   ...  &   ...  &   ...  \\ 
2459016.5412 & 39.2323 & 11.7379 & 0.5159 & 0.0150 &   ...  &   ... & 0.4366 & 0.0180 & -144.6804 & 117.2962 &    ...  &    ...  & 0.7795 & 0.0143 \\ 
2459034.3708 &    ...  &   ...  &   ...  &   ...  &   ...  &   ...  &   ...  &    ... &    ...    &    ...   & 86.1921 & 6.8763 &   ... &   ...   \\ 
2459237.7532 &    ...  &   ...  & 0.5141  & 0.012  &   ...  &   ...  &   ...  &   ...  &    ...  &    ...  &    ...  &    ...  &    ...  &    ...   \\ 
2459278.6560 & -83.6344 & 13.9066 &   ...  &   ...  &   ...  &   ...  &   ...  &   ...  & -142.0320 & 136.1777 & -120.0847 & 22.7264 &   ...  &   ...   \\ 
                        \end{longtable}
                \end{scriptsize}
        \end{landscape}

\end{appendix}

\end{document}